\documentclass[aps,pra,10pt,final,twocolumn,superscriptaddress,floatfix]{revtex4-2}

\usepackage[utf8]{inputenc}
\usepackage{amssymb,amsmath}
\usepackage{bm}
\usepackage{float}

\usepackage{graphicx, mwe}
\usepackage[normalem]{ulem}
\usepackage{dashbox}
\usepackage{xcolor}
\usepackage{standalone}
\usepackage{tikz}
\usetikzlibrary{arrows.meta, bending}

\usepackage{braket}
\usepackage{comment}

\usepackage{multirow, array}

\newcommand{\kt}[1]{\ensuremath{|#1\rangle}}

\newcommand{\HS}{\mathcal{H}}

\newcommand{\refFig}[2][]{Fig.~\ref{#2}(#1)}
\newcommand{\refEq}[1]{Eq.~(\ref{#1})}
\newcommand{\refEqShort}[1]{(\ref{#1})}
\newcommand{\refSec}[1]{Sec.~\ref{#1}}
\newcommand{\refApp}[1]{App.~\ref{#1}}
\newcommand{\refTab}[1]{Tab.~\ref{#1}}
\newcommand{\citeRef}[1]{Ref.~\cite{#1}}

\newcommand{\tr}{\text{Tr}}

\newcommand{\I}{\mathrm{i}}

\newcommand{\pseudofermionG}{%
\squeeze
    \begin{tabular}{@{}c|c@{}}
    \raisebox{0.35\height}
    {
        \parbox[t]{7ex}{$D_L${\tiny%
             \\ \vspace*{-0.75\baselineskip}$N$ even}}
    }
        &
    \raisebox{0.35\height}
       {
        \parbox[t]{7ex}{$D_{2L}${\tiny%
             \\ \vspace*{-0.75\baselineskip}$N$ odd}}
        }
        \end{tabular}
\squeeze
}

\DeclareMathOperator{\sign}{sign}

\newcommand{\specialS}{$\bigl(\hat{S}\bigr)$}

\newcommand{\BDIAI}{AI $\bigl(\mathrm{BDI}\bigr)$}
\newcommand{\squeeze}{\hspace*{-1.5ex}}

\newcommand{\BDICI}{%
\squeeze
\begin{tabular}{@{}c|c@{}}
        \raisebox{0.35\height}
        {
        \parbox[t]{7ex}{BDI{\tiny%
             \\ \vspace*{-0.75\baselineskip}$N$ even}}
        }
        &
        \raisebox{0.35\height}
        {
            \parbox[t]{7ex}{CI{\tiny%
                 \\ \vspace*{-0.75\baselineskip}$N$ odd}}
        }
        \end{tabular}
\squeeze
    }
\newcommand{\anyonsLabel}{%
\parbox{13ex}{\textbf{anyons\\$0 < |\theta| < \pi$}}}
\newcommand{\bosonsLabel}{\parbox{7 ex}{\textbf{bosons\\$\theta = 0$}}}
\newcommand{\pseudofermionsLabel}{{%
                \parbox{14 ex}{\textbf{pseudo\-fermions%
                \\%
                $\theta = \pi$%
                }}}}

\usepackage{nameref}
\usepackage{hyperref}
\hypersetup{
    colorlinks=true,       
    linkcolor=blue,        
    citecolor=blue,        
    filecolor=magenta,     
    urlcolor=blue,         
    pdfpagemode=FullScreen
}

\begin{document}
\title{Symmetry and integrability in the anyon-Hubbard model}
\author{M. Bonkhoff}
\affiliation{I. Institute for Theoretical Physics, Universit{\"a}t Hamburg, Notkestraße 9, 22607 Hamburg, Germany}
\author{G. Gurney}
\affiliation{Physics Department, American University, Washington, DC 20016, USA}
\author{F. Theel}
\affiliation{Center for Optical Quantum Technologies, Department of Physics, University of Hamburg, Luruper Chaussee 149, 22761 Hamburg Germany }
\author{P. Schmelcher}
\affiliation{Center for Optical Quantum Technologies, Department of Physics, University of Hamburg, Luruper Chaussee 149, 22761 Hamburg Germany }
\affiliation{The Hamburg Centre for Ultrafast Imaging, University of Hamburg, Luruper Chaussee 149, 22761 Hamburg, Germany}
\author{N.L. Harshman}
\affiliation{Physics Department, American University, Washington, DC 20016, USA}
\author{T. Posske}
\affiliation{I. Institute for Theoretical Physics, Universit{\"a}t Hamburg, Notkestraße 9, 22607 Hamburg, Germany}

\begin{abstract}
Recent cold atom experiments have realized one-dimensional anyons and enabled the tuning of 1D~statistics between bosons and fermions.
Here, we analyze the symmetries, integrability, and resulting degeneracies of the underlying
anyon-Hubbard model of finite length.
Our results reveal a switching between symmetry classes AI, BDI, and CI in dependence on system size, particle number, and boundary conditions,
and show that two anyons with periodic boundaries are integrable, while two anyons with open boundary conditions are not. 
We include a comprehensive analysis of all model limits, especially of interacting bosons and pseudofermions and resolve spectral signatures. We additionally reveal an exactly solvable doublon state that hides in the continuum of scattering states and the exact solution of the nullspace of two noninteracting anyons. 
The uncovered symmetries shape the fundamental properties of the one-dimensional anyons at hand, and the predicted states are accessible in state-of-the-art experiments.
\end{abstract}

\maketitle

%

\section{Introduction}
In the last few years, a series of experiments with cold atoms in optical traps have probed the exchange statistics, exclusion statistics, dynamics, and correlations of anyons and anyon-like systems in one dimension~\cite{kwan2024RealizationOnedimensionalAnyons,horowitz1989ExactlySolubleDiffeomorphism,frolian2022Realizing1DTopological,wang2025AnyonizationBosonsOne,zeng2026RealizationFractionalFermi,bakkalihassani2026RevealingPseudoFermionizationChiral,chisholm2022EncodingOnedimensionalTopological}.
In all realizations, the anyon-Hubbard model~\cite{keilmann2011StatisticallyInducedPhase,greschnerDensityDependentSyntheticGauge2014,greschner2015AnyonHubbardModel,strater2016FloquetRealizationSignatures,greschnerProbingExchangeStatistics2018} plays a central role, either realized directly with two particles in a finite system~\cite{kwan2024RealizationOnedimensionalAnyons,bakkalihassani2026RevealingPseudoFermionizationChiral}, in the limit of low statistical parameters \cite{frolian2022Realizing1DTopological,chisholm2022EncodingOnedimensionalTopological}, or relating to the momentum distribution of quasiparticles \cite{dhar2025ObservingAnyonizationBosons,wang2025AnyonizationBosonsOne,zeng2026RealizationFractionalFermi}, revealing anyonic dynamics in quantum walks, anyonic correlations in density functions, the buildup of Friedel oscillations for changing statistical parameter, the existence of chiral bound states, and the relation to fractional statistics \cite{nagies2024BraidStatisticsConstructing}.
The relevance of the anyon-Hubbard model is further enhanced because, properly regularized, its continuum limit coincides with the Kundu model, also called Lieb-Liniger anyons and the chiral BF model~\cite{kundu1999ExactSolutionDouble,bonkhoff2021BosonicContinuumTheory,patu2007CorrelationFunctionsOnedimensional,batchelor2006OneDimensionalInteractingAnyon,aglietti1996AnyonsChiralSolitons,frolian2022Realizing1DTopological,chisholm2022EncodingOnedimensionalTopological} for certain parameter regimes.
Beyond that, a variety of further continuum and lattice models for 1D anyons~\cite{posske2017SecondQuantizationLeinaasMyrheim,harshman2022TopologicalExchangeStatistics},
and multiple extensions and modifications of the anyon-Hubbard model have been proposed \cite{hao2012DynamicalPropertiesHardcore,bonkhoff2025AnyonicPhaseTransitions,harshman2020AnyonsThreebodyHardcore,harshman2022TopologicalExchangeStatistics,nagies2024BraidStatisticsConstructing,greschner2015AnyonHubbardModel,lange2017AnyonicHaldaneInsulator}. Additionally, paraparticles, i.e., mixed representations of the symmetric group, have been discussed~\cite{greenberg1965SelectionRulesParafields, fredenhagen1989SuperselectionSectorsBraid,wang2025ParticleExchangeStatistics}. 
These developments have led to the term anyon being employed beyond its origin in two dimensions \cite{LeinaasMyrheim1977TheoryOfIdenticalParticles,wilczekQuantumMechanicsFractionalSpin1982,kitaevAnyonsExactlySolved2006} or generalized Pauli principles \cite{haldane1991FractionalStatisticsArbitrary}.

One-dimensional anyons enable the unique possibility of tuning the statistical angle $\theta$ in a complete loop from bosons over fermions back to bosons \cite{theel2025ChirallyProtectedState}. The associated chiral-symmetry-protected holonomy of the degenerate zero-modes mimics the topological control found in two-dimensional topologically degenerate anyons \cite{nayak2008NonAbelianAnyonsTopological} and enables holonomic quantum state manipulation. 
However, implementation requires precise understanding of how the energy levels transform under adiabatic manipulation of the model parameters. 
The pivotal role of the anyon-Hubbard model for exploring one-dimensional anyon physics establishes the fundamental interest and practical relevance for resolving and classifying the symmetries, conserved quantities, and integrability properties for all parameter ranges and boundary conditions of the model and for uncovering and cataloging the existence of any exactly solvable states. To this end, a symmetry analysis of the fundamental unitary and antiunitary symmetries and the corresponding universality classes of random matrices with their respective Cartan-Altland-Zirnbauer label \cite{wigner1931GrundlagenQuantenmechanik,cartan1926ClasseRemarquableDespaces,altland1997NonstandardSymmetryClasses,zirnbauer2020ParticleholeSymmetriesCondensed} determines spectral degeneracies to a large extent and reveals how the Hamiltonian decomposes into decoupled sectors. 
Furthermore, the Bethe ansatz techniques provide a framework to investigate integrability and special solutions of one-dimensional models  \cite{betheZurTheorieMetalle1931,essler2010OnedimensionalHubbardModel,baxter2016exactly}. Bethe solutions are rare for bosonic systems with contact interactions beyond the continuum case \cite{lieb1963ExactAnalysisInteracting,haldaneSolidificationSolubleModel1980, choyExactResultsDegenerate1980, choyFailureBetheAnsatzSolutions1982,Kolovsky_2004,Kollath2007,Kollath_2010,Hirsch2020,Fazio2020,Links_2021,Pausch2021,pausch2022OptimalRouteQuantum}, which is why approximate methods are invoked widely \cite{Cazalilla2011,PitaevskiiStringariBook2016}. Yet, for two particles, which describe the dilute limit well \cite{olshanii1998AtomicScatteringPresence,Hofstetter2004,Affleck2004}, exact solutions can be found due to the absence of diffractive scattering \cite{oelkersGroundstatePropertiesAttractive2007,valienteTwoparticleStatesHubbard2008,valienteScatteringResonancesTwoparticle2009,Kollar2013,boschiBoundStatesExpansion2014,longhiTammHubbardSurface2013,poloExactResultsPersistent2020,liTwoparticleStatesOnedimensional2022}.
The two-particle anyon-Hubbard model
was previously investigated in infinite lattices~\cite{greschner2015AnyonHubbardModel,zhang2017GroundstatePropertiesOnedimensional,greschnerProbingExchangeStatistics2018,longhiAnyonicBlochOscillations2012,longhiAnyonsOnedimensionalLattices2012,lauQuantumWalkTwo2022,kwan2024RealizationOnedimensionalAnyons,bakkalihassani2026RevealingPseudoFermionizationChiral,zheng2024NecessityOrthogonalBasis}, revealing 
the stabilization of bound states without external potential \cite{zhang2017GroundstatePropertiesOnedimensional,bakkalihassani2026RevealingPseudoFermionizationChiral} and the transmutation from bosons to pseudofermions at low-energies \cite{greschner2015AnyonHubbardModel,greschnerProbingExchangeStatistics2018}. The dimer case, i.e., two sites, is integrable \cite{bonkhoffCoherencePropertiesRepulsive2023} like the two-site Bose-Hubbard model \cite{Links_2021}. 
Additionally, only few states in the anyon-Hubbard model are exactly solved, among which is the two-anyon bound state in infinite systems \cite{greschner2015AnyonHubbardModel,zhang2017GroundstatePropertiesOnedimensional,kwan2024RealizationOnedimensionalAnyons,bakkalihassani2026RevealingPseudoFermionizationChiral}.


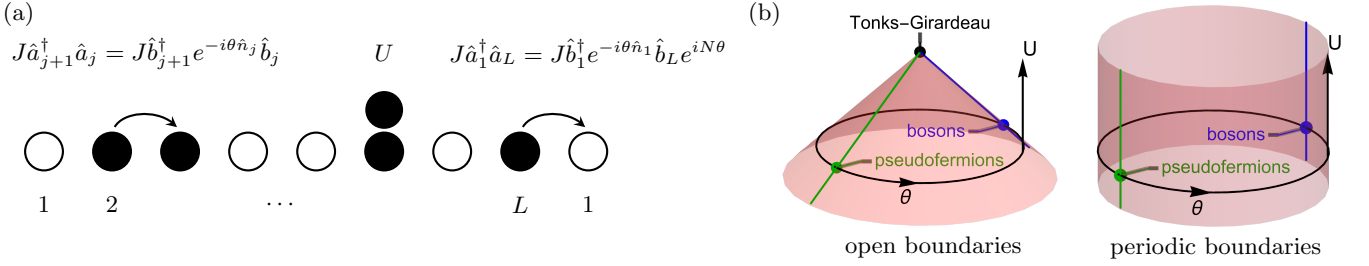
\begin{figure*}
\centering
\parbox[t][10 \baselineskip][t]{0.55\linewidth}{%
(a)\raggedright 
    \quad
    \raisebox{-0.9 \height}{%
    \begin{tikzpicture}[
    scale = 1.0,
    site/.style={circle, draw, thick, minimum size=0.5cm, inner sep=0pt},
    particle/.style={circle, fill=black, minimum size=0.50cm, inner sep=0pt},
    arrow style/.style={->, >=stealth, thick, shorten >=2pt, shorten <=2pt}
]

\def\dx{0.9}
\def\labelHeight{1.3}


\node[site] (s1) at (0, 0) {};

\node[site] (s2) at (\dx, 0) {};
\node[particle] at (s2.center) {};

\node[site] (s3) at (2*\dx, 0) {};
\node[particle] at (s3.center) {};

\node[site] (s4) at (3*\dx, 0) {};
\node[site] (s5) at (4*\dx, 0) {};

\node[site] (s_double) at (5*\dx, 0) {};
\node[particle] at (5*\dx, 0) {};
\node[particle] at (5*\dx, 0.55) {};

\node[site] (s8) at (6*\dx, 0) {};

\node[site] (sL) at (7*\dx, 0) {};
\node[particle] at (sL.center) {};

\node[site] (s1end) at (8*\dx, 0) {};

\node at (0, -0.7) {$1$};
\node at (\dx, -0.7) {$2$};
\node at (3.5*\dx, -0.7) {$\cdots$};
\node at (7*\dx, -0.7) {$L$};
\node at (8*\dx, -0.7) {$1$}; 

\draw[arrow style, bend left=60] (s2.north) to (s3.north);
\draw[arrow style, bend left=60] (sL.north) to (s1end.north);

\node at (1.5*\dx, \labelHeight) {$J \hat{a}^\dagger_{j+1} \hat{a}_j = J \hat{b}^\dagger_{j+1} e^{-i\theta \hat{n}_j} \hat{b}_j$};
\node at (5*\dx, \labelHeight) {$U$}; 
\node at (8.0*\dx, \labelHeight) {$J \hat{a}^\dagger_1 \hat{a}_L = J  \hat{b}^\dagger_1  e^{-i \theta \hat{n}_1} \hat{b}_L e^{i N \theta}$ };
\end{tikzpicture}%
    }
}%
\parbox[t][10 \baselineskip][t]{0.45 \linewidth}{%
(b)\raggedright%
    \raisebox{-0.85\height}{%
        \parbox{0.5\linewidth}{
            \centering
            \includegraphics[width=0.99 \linewidth]{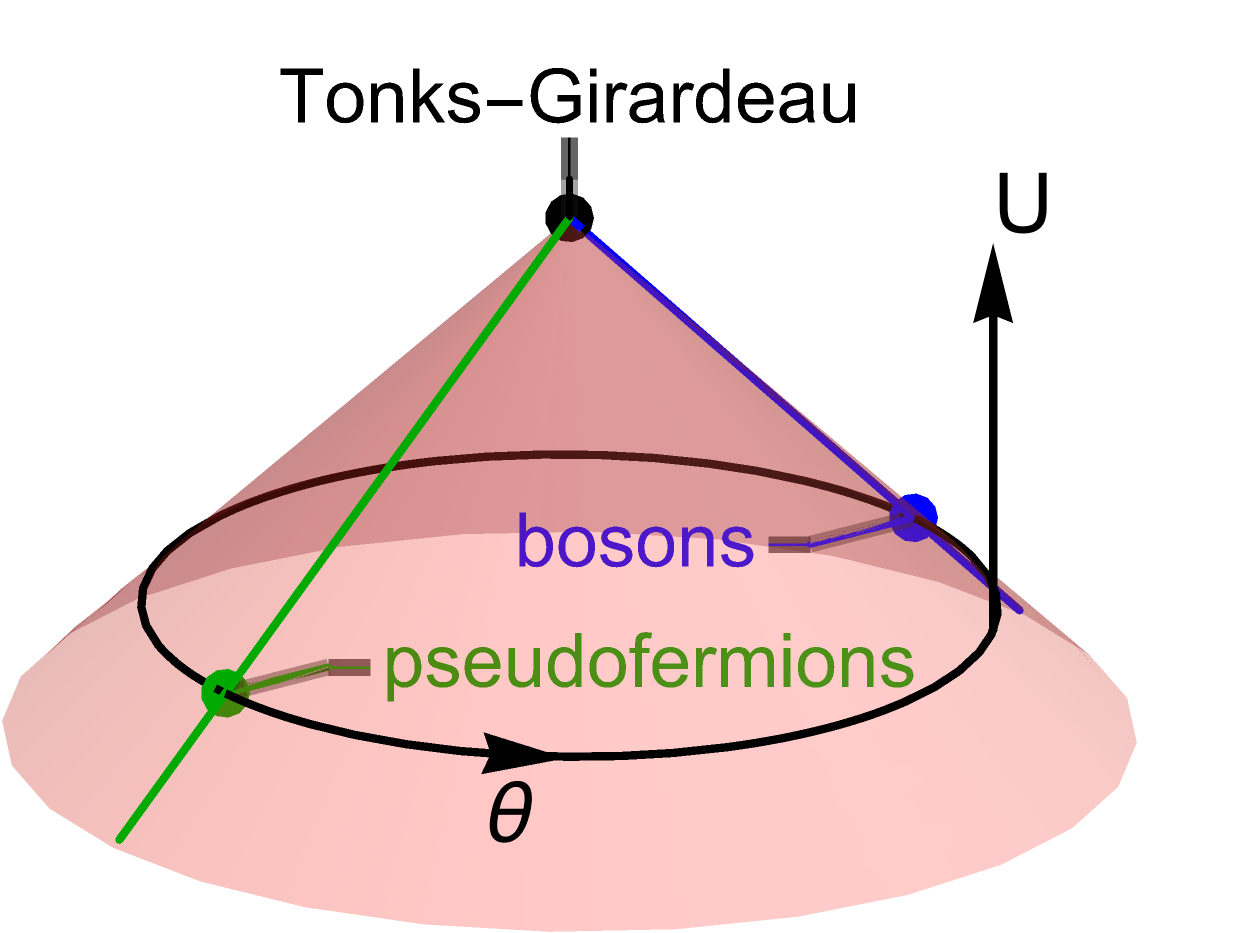}%
            \\open boundaries
        }
    }%
    \raisebox{-0.735\height}{%
        \parbox{0.5\linewidth}{
            \begin{center}
                \includegraphics[width=0.99 \linewidth]{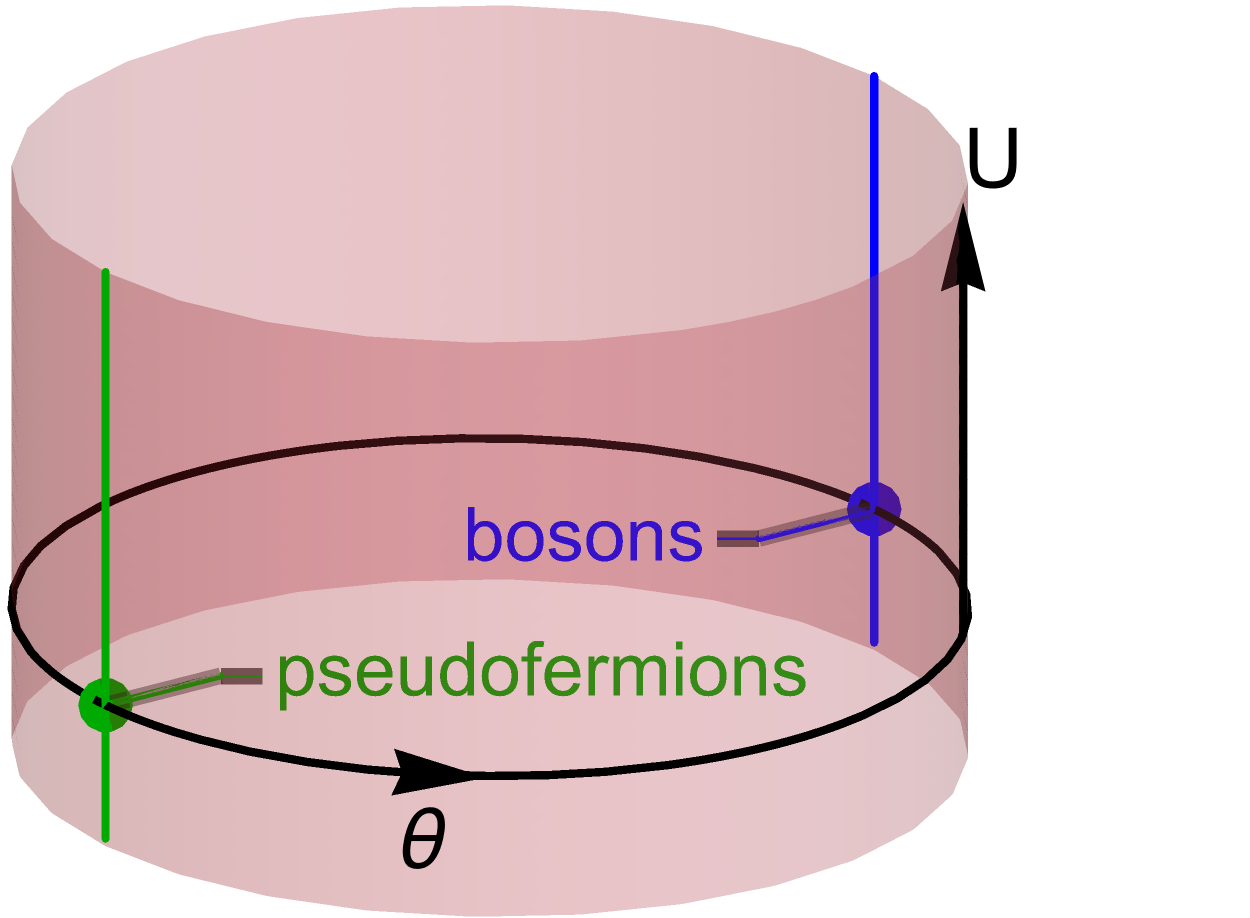}%
                \newline{}
                \hspace*{-2.5em}periodic boundaries
            \end{center}
        }%
    }
}
\caption{\label{fig:ModelSummary}%
(a) The anyon-Hubbard model describes $N$ anyons on $L$ sites with creation  operators $\hat{a}^\dagger_j$, equivalent to interacting bosons with density-dependent Peierls phase. 
(b) Its flat parameter space (red) consists of the Hubbard interaction $U$ and the cyclic statistical parameter $\theta$, realizing bosons for $\theta=0$ and pseudofermions for $\theta=\pi$. For open boundary conditions in the Tonks-Girardeau limit of infinitely strong interactions, all particles become the same (tip of the conic parameter space). Periodic boundary conditions prevent this identification (cylindric parameter space), see \refEq{eq:HBGeneral}.
}
\end{figure*}
 
In this article, we establish the symmetries of the anyon-Hubbard model for generic lattice sizes, particle numbers, statistical parameters, and strengths of Hubbard interaction, and identify integrable regimes.
The symmetries fall into two categories. First, we reveal the point groups, i.e., the unitary symmetries of the model, including their irreducible representations and multiplicities.
Second, we classify the fundamental symmetries of time reversal, charge conjugation, and chirality \cite{cartan1926ClasseRemarquableDespaces,altland1997NonstandardSymmetryClasses,zirnbauerSymmetryClassesRandom2004,zirnbauer2020ParticleholeSymmetriesCondensed}. Previous work had identified an antiunitary symmetry for all $\theta$ \cite{lange2017AnyonicHaldaneInsulator} and chiral symmetry for $U=0$ \cite{theel2025ChirallyProtectedState} for open boundary conditions, and we extend this analysis to periodic boundary conditions and in the bosonic and pseudofermionic limiting cases. All symmetry results are summarized in Tabs.~\ref{tab:intandsym} and \ref{tab:multi}. The methods and results applied here can be extended to related models and future generalizations because multiplicities of irreducible representations are properties of the symmetry operators in the bosonic configuration space, which exists independent of its relevance to a specific model.

Subsequently, we show by the coordinate Bethe ansatz \cite{betheZurTheorieMetalle1931,essler2010OnedimensionalHubbardModel,baxter2016exactly} that the experimentally-relevant case of two anyons with open boundary conditions is only integrable in the bosonic or pseudofermionic limits or in the Tonks-Girardeau limit of infinite Hubbard interactions \cite{tonks1936CompleteEquationState,girardeau1960RelationshipSystemsImpenetrable}.  
This surprising non-integrability for two particles with open boundary conditions, even in the case of no Hubbard interaction, challenges recent findings  \cite{zhangAnyonicBoundStates2023} but is corroborated by the level spacing analysis that we provide.
We also show that, as expected, for periodic boundary conditions, two anyons are integrable irrespective of the model parameters.
Beyond the two-particle case on more than two sites, no integrability can be expected, as is well-established for the simpler Bose-Hubbard model~\cite{Links_2021,essler2005OneDimensionalHubbardModel}.  Constructions of integrable anyon models on the lattice with more than two particles typically rely on additional hardcore conditions to exclude diffractive multi-particle coincidences~\cite{Batchelor_2008,kundu2010QuantumIntegrable1D, patuCorrelationFunctionsOnedimensional2007,batchelorBetheAnsatz1D2007}. 

Finally, we reveal several exact  solutions for the two-particle system. For open boundary conditions, we show that there exists a doublon state whose energy is precisely the same as the Hubbard interaction $E=U$, and we also provide explicit solutions for the zero-energy modes at vanishing Hubbard interactions that are the basis for holonomic quantum state manipulation \cite{theel2025ChirallyProtectedState}. These exactly-solvable states are decoupled from the rest of the spectrum and therefore share features with quantum scars
~\cite{serbyn2021QuantumManybodyScars} and other systems with an extensive number of zero-energy modes~ \cite{banerjeeQuantumScarsZero2021,turnerStableInfinitetemperatureEigenstates2025}. Using the Bethe Ansatz, in App. C we also construct a few additional exactly solvable two-body states for periodic boundary conditions and particular values of $\theta$ corresponding to singular roots of the Bethe ansatz equations.
In total, we establish a guide for understanding symmetries, integrability, and exact solutions in the anyon-Hubbard model and derived correlated hopping models, and provide concrete physical properties in its various parameter regimes.

The article is structured as follows. In \refSec{sec:Model}, we introduce the anyon-Hubbard model and comment on different gauges and boundary conditions. We analyze its symmetries and degeneracies 
\refSec{sec:symm}, especially emphasizing the differences between open and periodic boundary conditions.
This section is the prerequisite for the level statistics and integrability analysis in \refSec{sec:integrability}. In \refSec{sec:SpecialSolutions}, we give analytical solutions for the two-particle anyon-Hubbard model with open boundaries, namely the nullspace of non-interacting anyons and the exactly solved doublon state. We conclude our work in \refSec{sec:Conclusions}.

\begin{table*}
\centering
\renewcommand{\arraystretch}{1.5}

\begin{tabular}{|c|c| *{2}{c|c|c|c|c|c|}}
    \hline

    \multicolumn{2}{|c|}{\multirow{2}{*}{\textbf{OBC}}} & 
    \multicolumn{6}{c|}{\textbf{$L$ even}} & 
    \multicolumn{6}{c|}{\textbf{$L>2$ odd}} \\
    \cline{3-14}

    \multicolumn{2}{|c|}{} & 
    $G$ & {$\mathcal{T}$} & {$\mathcal{C}$} & {$\mathcal{C}^2$} & $\mathcal{S}$ & class & 
    $G$ & {$\mathcal{T}$} & {$\mathcal{C}$} & {$\mathcal{C}^2$} & $\mathcal{S}$ & class \\
    \hline

    \hline
    
    \multirow{2}{*}{\anyonsLabel} & $U = 0$ &
    
    $-$ & $\hat{K}_{\theta}$ & $\hat{C}_{\theta}$ & $(-1)^N$& $\hat{S}$ & \BDICI{} &
   
    $-$ & $\hat{K}_{\theta} $ & $\hat{C}_{\theta}$& 1& $\hat{S}$ & {$\mathrm{BDI}$}
    \\ \cline{2-14} & $U \neq 0$ &
    
    $-$ & $\hat{K}_{\theta}$ &$-$ &$-$ & $-$ & {$\mathrm{AI}$} &
    
    $-$ & $\hat{K}_{\theta}$ & $-$&$-$ & $-$ & {$\mathrm{AI}$} 
    \\ \hline

    \multirow{2}{*}{\bosonsLabel{}} & $U = 0$ &
    ${\mathbb{Z}}_2$ & $\hat{T}$ & $\bigl(\hat{C}_0 \bigr)$&$1$ & $\bigl(\hat{S}\bigr)$ & Integrable &
    
    ${\mathbb{Z}}_2$ & $\hat{T}$ &$\bigl(\hat{C}_0 \bigr)$ &1 & $\bigl(\hat{S}\bigr)$ & Integrable 
    \\ \cline{2-14} & $U \neq 0$ & 
    
    ${\mathbb{Z}}_2$ & $\hat{T}$ & $-$& $-$& $-$ & {$\mathrm{AI}$} &
    
    ${\mathbb{Z}}_2$ & $\hat{T}$ & $-$&$-$ & $-$ & {$\mathrm{AI}$} 
    \\ \hline

    \multirow{2}{*}{\pseudofermionsLabel{}} & $U = 0$ &
    ${\mathbb{Z}}_2$ & $\hat{T}$ & $\bigl(\hat{C}_0 \bigr)$&$1$ & \specialS{} & \BDIAI{}  &
   
    ${\mathbb{Z}}_2$ & $\hat{T}$ & $\hat{C}_{0}$&$1$ & $\hat{S}$ & {$\mathrm{BDI}$} 
    \\ \cline{2-14} & $U \neq 0$ &
   
    ${\mathbb{Z}}_2$ & $\hat{T}$ &$-$ &$-$ & $-$ & {$\mathrm{AI}$} &
    
    ${\mathbb{Z}}_2$ & $\hat{T}$ & $-$&$-$ & $-$ & {$\mathrm{AI}$} 
    \\ \hline

    \multicolumn{2}{|c|}{\multirow{2}{*}{\textbf{PBC}}} & 
    \multicolumn{6}{c|}{\textbf{PBC $L>2$ even}} & 
    \multicolumn{6}{c|}{\textbf{PBC $L>2$ odd}} \\
    \cline{3-14}

    \multicolumn{2}{|c|}{} & 
    $G$ & {$\mathcal{T}$} & {$\mathcal{C}$} & {$\mathcal{C}^2$} & $\mathcal{S}$ & class & 
    $G$ & {$\mathcal{T}$} & {$\mathcal{C}$} & {$\mathcal{C}^2$} & $\mathcal{S}$ & class \\
    \hline

    \hline
    
    \multirow{2}{*}{\anyonsLabel} & $U = 0$ &
    
    ${\mathbb{Z}}_L$ & $\hat{K}_{\theta}$ & $\hat{C}_{\theta}$& $(-1)^N$& $\hat{S}$ &  \BDICI{} &
    
    ${\mathbb{Z}}_L$ & $\hat{K}_{\theta}$ & $-$&$-$ & $-$ & {$\mathrm{AI}$}
    \\ \cline{2-14} & $U \neq 0$ & 
    
    ${\mathbb{Z}}_L$ & $\hat{K}_{\theta}$ &$-$ &$-$ & $-$ & {$\mathrm{AI}$} &
   
    ${\mathbb{Z}}_L$ & $\hat{K}_{\theta}$ & $-$&$-$ & $-$ & {$\mathrm{AI}$} 
    \\ \hline

    \multirow{2}{*}{\bosonsLabel{}} & $U = 0$ &
    
    $D_L$ & $\hat{T}$ &$\bigl(\hat{C}_0 \bigr)$ &$1$ & \specialS{} & Integrable &
   
    $D_L$ & $\hat{T}$ & $-$&$-$ & $-$ & Integrable
    \\ \cline{2-14} & $U \neq 0$ &
   
    $D_L$ & $\hat{T}$ & $-$&$-$ & $-$ & {$\mathrm{AI}$} &
    
    $D_L$ & $\hat{T}$ &$-$ &$-$ & $-$ & {$\mathrm{AI}$} 
    \\ \hline

    \multirow{2}{*}{\pseudofermionsLabel{}} & $U = 0$ &%
    \pseudofermionG{} & $\hat{T}_\pi$ & $\bigl(\hat{C}_0 \bigr)$&$1$ & \specialS{} &\BDIAI{}  &
    
    \pseudofermionG{} & $\hat{T}_\pi$ &$-$ & $-$& $-$ & {$\mathrm{AI}$}
    \\ \cline{2-14} & $U \neq 0$ &
   
    \pseudofermionG{} & $\hat{T}_\pi$ &$-$ & $-$& $-$ & {$\mathrm{AI}$} &
    
    \pseudofermionG{} & $\hat{T}_\pi$ & $-$& $-$& $-$ &  {$\mathrm{AI}$}
    \\ \hline
\end{tabular}

\caption{\label{tab:intandsym}%
    Depending on the parameters and boundary conditions (OBC: open boundary conditions, PBC: periodic boundary conditions), the anyon-Hubbard model assumes different point groups $G$, time reversal symmetries $\mathcal{T}$, and chiral symmetry $\mathcal{S}$,  resulting in a tunable symmetry class $\mathrm{AI}$, $\mathrm{BDI}$, or $\mathrm{CI}$. Time reversal squares to one $\mathcal{T}^2=1$. If $\mathcal{S}$ is a chiral symmetry, there is a charge conjugation symmetry $\mathcal{C} = \mathcal{T}\mathcal{S}$ with $\mathcal{C}^2 = (-1)^{N(L+1)}$ for anyons but $\mathcal{C}^2 =1$ for bosons and pseudo-fermions. Brackets $\left(..\right)$ denote the symmetry is present for the full Hamiltonian in \refEq{eq:Hamiltonian} but mixes symmetry sectors of $G$, cf.~Eqs.~(\ref{eq:obcmix}) and (\ref{eq:pbcmix}). 
    }
\end{table*}

\begin{table*}[t]
\renewcommand{\arraystretch}{1.5}
    \centering
\begin{tabular}{|ccc|c|c|c|c|c|c|}
\hline
\multicolumn{3}{|c|}{\textbf{Parameters}} &
  \textbf{G} &
  \textbf{Irrep Labels} &
  \textbf{Dim} &
  \textbf{Mult} &
  \textbf{Eq.} &
  \textbf{Integrable for $L>2$} \\ \hline

\multicolumn{1}{|c|}{\multirow{2}{*}{\anyonsLabel}} &
  \multicolumn{2}{c|}{OBC} &
  $-$ &
  $-$ &
  1 &
  $1$ &
  $-$ &
  No \\ \cline{2-9} 
\multicolumn{1}{|c|}{} &
  \multicolumn{2}{c|}{PBC} &
  $\langle \hat{{R}}\rangle \sim \mathbb{Z}_L $ &
  $q \in \left\{0,\dots,L-1\right\}$ &
  1 &
  $m^{{\mathbb{Z}}_L}_q$ &
  \refEqShort{eq: subspace_multiplicityintermediatetheta} & 
  \multirow{15}{*}{two-particle} \\ \cline{1-8}

\multicolumn{1}{|c|}{\multirow{5}{*}{\bosonsLabel}} &
  \multicolumn{2}{c|}{OBC} &
  $\langle \hat{P} \rangle \sim \mathbb{Z}_2$ &
  $p \in \{+1, -1\}$ &
  1 &
  $m^{\hat{P}}_{p}$ &
  \refEqShort{tableEq: bosonParityEq} &
   \\ \cline{2-8} 
\multicolumn{1}{|c|}{} &
  \multicolumn{1}{c|}{\multirow{4}{*}{PBC}} &
  \multirow{2}{*}{L odd} &
  \multirow{4}{*}{$\langle \hat{P}, \hat{R} \rangle \sim D_L$} &
  $q =0 \ \& \ p\in\{+1,-1\}$ &
  1 &
  $m^{D_L}_{q,p}$ &
  \refEqShort{tableEq: bosonicOddLTrivialandSign} &  
   \\ \cline{5-8}
\multicolumn{1}{|c|}{} &
  \multicolumn{1}{c|}{} &
   &
   &
  $q \in \{ 1, \ldots,(L-1)/2 \}$ &
  2 &
  $m^{D_L}_{q,q'}$ &
  \refEqShort{tableEq: bosonicOddL2dIrrep} &  
   \\ \cline{3-3} \cline{5-8}
\multicolumn{1}{|c|}{} &
  \multicolumn{1}{c|}{} &
  \multirow{2}{*}{L even} &
   &
  $q \in \{ 0, L/2 \} \ \& \ p\in\{+1,-1\}$ &
  1 &
  $m^{D_L}_{q,p}$ &
  \refEqShort{eq: combined bosonic even L 1D multiplicities} &
   \\ \cline{5-8}
\multicolumn{1}{|c|}{} &
  \multicolumn{1}{c|}{} &
   &
   &
  $q \in \{ 1, \ldots,L/2-1 \}$ &
  2 &
  $m^{D_L}_{q,q'}$ &
  \refEqShort{tableEq: bosonicEvenL2d} &
   \\ \cline{1-8}

\multicolumn{1}{|c|}{\multirow{9}{*}{\pseudofermionsLabel}} &
  \multicolumn{2}{c|}{OBC} &
  $\langle \hat{P}_\pi \rangle \sim \mathbb{Z}_2 $ &
  $p_\pi \in \{+1, -1\}$ &
  1 &
  $m^{\hat{P}_{\pi}}_{p_\pi}$ &
  \refEqShort{eq:multiplicityobcthetapi} &
   \\ \cline{2-8} 
\multicolumn{1}{|c|}{} &
  \multicolumn{1}{c|}{\multirow{4}{*}{\begin{tabular}[c]{@{}c@{}}PBC\\ N even\end{tabular}}} &
  \multirow{2}{*}{L odd} &
  \multirow{4}{*}{$\langle \bar{P}_\pi, \bar{R}_\pi \rangle \sim D_L$} &
  $\bar{q} = 0 \ \& \ \bar{p}_\pi\in\{+1,-1\}$ &
  1 &
  $m^{D_L}_{\bar{q},p_{\pi}}$ &
  \refEqShort{def: fermionic even N even L 1D multiplicity formula} &
   \\ \cline{5-8}
\multicolumn{1}{|c|}{} &
  \multicolumn{1}{c|}{} &
   &
   &
  $\bar{q} \in \{ 1, \ldots,(L-1)/2 \}$ &
  2 &
  $m^{D_L}_{\bar{q},\bar{q}'}$ &
  \refEqShort{eq: Neven2dMult} &
   \\ \cline{3-3} \cline{5-8}
\multicolumn{1}{|c|}{} &
  \multicolumn{1}{c|}{} &
  \multirow{2}{*}{L even} &
   &
  $\bar{q} \in \{ 0, L/2 \} \ \& \ \bar{p}_\pi\in\{+1,-1\}$ &
  1 &
  $m^{D_L}_{\bar{q},p_{\pi}}$ &
  \refEqShort{def: fermionic even N even L 1D multiplicity formula} &
   \\ \cline{5-8}
\multicolumn{1}{|c|}{} &
  \multicolumn{1}{c|}{} &
   &
   &
  $\bar{q} \in \{ 1, \ldots,L/2-1 \}$ &
  2 &
  $m^{D_L}_{\bar{q},\bar{q}'}$ &
  \refEqShort{eq: Neven2dMult} &
   \\ \cline{2-8}
\multicolumn{1}{|c|}{} &
  \multicolumn{1}{c|}{\multirow{4}{*}{\begin{tabular}[c]{@{}c@{}}PBC\\ N odd\end{tabular}}} &
  \multirow{2}{*}{L odd} &
  \multirow{4}{*}{$\langle \bar{P}_\pi, \bar{{R}}_\pi \rangle \sim D_{2L}$} &
  $\bar{q} \in \{ 0, L \} \ \& \ \bar{p}_\pi\in\{+1,-1\}$ &
  1 &
  $m^{D_{2L}}_{\bar{q},p_{\pi}}$ &
  \refEqShort{def: fermionic N odd 1D multiplicities} &
   \\ \cline{5-8}
\multicolumn{1}{|c|}{} &
  \multicolumn{1}{c|}{} &
   &
   &
  $\bar{q} \in \{ 1, \ldots,L-1 \}$ &
  2 &
  $m^{D_{2L}}_{\bar{q},\bar{q}'}$ &
  \refEqShort{apeq: N odd 2d multiplicities} &
   \\ \cline{3-3} \cline{5-8}
\multicolumn{1}{|c|}{} &
  \multicolumn{1}{c|}{} &
  \multirow{2}{*}{L even} &
   &
  $\bar{q} \in \{ 0, L \} \ \& \ \bar{p}_\pi\in\{+1,-1\}$ &
  1 &
  $m^{D_{2L}}_{\bar{q},p_{\pi}}$ &
  \refEqShort{def: fermionic N odd 1D multiplicities} &
   \\ \cline{5-8}
\multicolumn{1}{|c|}{} &
  \multicolumn{1}{c|}{} &
   &
   &
  $\bar{q} \in \{ 1, \ldots,L-1 \}$ &
  2 &
  $m^{D_{2L}}_{\bar{q},\bar{q}'}$ &
  \refEqShort{apeq: N odd 2d multiplicities} &
   \\ \hline
\multicolumn{3}{|c|}{\textbf{all, U=0}} &
  $\langle \hat{S} \rangle \sim \mathbb{Z}_2$ &
  $\chi \in \{+1, -1\}$ &
  1 &
  $m^{\hat{S}}_{\chi}$ &
  \refEqShort{tableEq: bosonParityEq} &
  For $N>2$ only bosons\\ \hline
\end{tabular}

\caption{\label{tab:multi}%
    Each parameter regime and boundary condition (OBC: open boundary conditions, PBC: periodic boundary conditions) comes with unique irreducible representations (irreps) of the respective point groups $G$, generated by the unitary operators between $\langle \cdots \rangle$ in column $G$.  Note that if two sets of parameters have an isomorphic point group, the Hilbert space typically still carries different representations, and hence different multiplicities. Furthermore, we resolve the $Z_2$-group generated by the chiral symmetry $\hat{S}$. The last column summarizes integrability properties; for the non-integrable $N=2$ case with open boundaries there are additionally an extensive number of exactly solvable zero-energy modes, cf.~Sect.~\ref{sec:Nullstates}. 
    }
\end{table*}

\section{The anyon-Hubbard model}
\label{sec:Model}

The anyon-Hubbard model takes the form
\begin{equation}
\label{eq:Hamiltonian}
    \hat{H}= -J \sum_{j=1}^{L} \left( \hat{a}_{j+1}^\dag \hat{a}_j + \mathrm{h.c.}\right) + \frac{U}{2} \sum_{j=1}^L \hat{n}_j (\hat{n}_j - 1),
\end{equation}
where $L$ is the number of sites, $\hat{a}^\dag_j$ the anyon creation operator, $\hat{n}_j = \hat{a}^\dag_j \hat{a}_j$ the anyonic density operator, $J$ the tunneling strength, and $U$ the on-site interaction~\cite{keilmann2011StatisticallyInducedPhase}, see \refFig[a]{fig:ModelSummary}.
We consider open boundary conditions with $\hat{a}_{L+1}\equiv 0=\hat{a}_{0}$, corresponding to the experimental realizations on linear chains \cite{kwan2024RealizationOnedimensionalAnyons, bakkalihassani2026RevealingPseudoFermionizationChiral} and periodic boundary conditions with $\hat{a}_{L+1}=\hat{a}_1$.
The statistical parameter $\theta \in \left[ 0, 2\pi\right)$ classifies the anyonic character of the particles in deformed commutation relations
\begin{eqnarray}
\label{eq:commrel}    
\hat{a}_j \hat{a}_k^\dag - e^{-i\theta \sign(j-k)} \hat{a}^\dag_k \hat{a}_j &=& \delta_{jk},
\nonumber\\
\hat{a}_j \hat{a}_k - e^{i\theta \sign(j-k)} \hat{a}_k \hat{a}_j &=& 0,
\end{eqnarray}
with $\sign(0)=0$ \cite{keilmann2011StatisticallyInducedPhase}. Bosons correspond to $\theta=0$ pseudofermions to $\theta=\pi$, i.e., fermionic anticommutation relations with $i$, $j$ at different sites but bosonic commutation relations at the same site.
The deformed commutation relations in \refEq{eq:commrel} and the Hamiltonian in \refEq{eq:HBGeneral} are invariant under $\theta \to \theta +2\pi$.
In the Tonks-Girardeau limit, $U \to \infty$, all particles behave the same for open boundaries, which is prevented for anyonic periodic boundary conditions, see \refFig[b]{fig:ModelSummary}.

The Hamiltonian in \refEq{eq:Hamiltonian} is transformed into a bosonic model by a fractional Jordan-Wigner transformation. This transformation has a gauge degree of freedom, which we discuss in Appendix~\ref{app:gaugeequiv}. 
We choose the transformation
\begin{equation}
\label{eq:jordanwigner}
    \hat{a}_j=\hat{b}_j e^{i\theta \sum_{k<j}\hat{n}_k}
\end{equation}
introduced in \cite{keilmann2011StatisticallyInducedPhase} and used in \cite{greschner2015AnyonHubbardModel, tangGroundstatePropertiesAnyons2015, langeAnyonicHaldaneInsulator2017, bonkhoff2021BosonicContinuumTheory}, which results in 
\begin{align}
\label{eq:HBGeneral}
    \hat{H} =& -J \sum_{j=1}^{L-1} \left( \hat{b}_{j+1}^\dag e^{-i \theta \hat{n}_{j}}\hat{b}_j +\mathrm{h.c}\right) + \frac{U}{2} \sum_{j=1}^L \hat{n}_j (\hat{n}_j - 1)
    \nonumber \\
     &-J \left( \hat{b}_{L+1}^\dag e^{-i \theta \hat{n}_1}\hat{b}_L e^{i N \theta} +\mathrm{h.c}\right).
\end{align}
Here the kinetic term acquires a density-dependent Peierls phase. For anyonic periodic boundary conditions, the Hamiltonian in \refEq{eq:HBGeneral} has twisted boundary conditions for the bosons with total flux through the ring $N\theta$  ~\cite{patuCorrelationFunctionsOnedimensional2007,liebTwoSolubleModels1961,capelNoteDifferenceAcyclic1975,SeibergShao2024}. We discuss an alternate gauge choice for twisted boundary conditions in \refSec{sec:periodicBoundaries}.
The number operators $\hat{n}_j = \hat{b}^\dag_j \hat{b}_j$ are the same for anyons and bosons.  We denote the basis number states by $\kt{\mathbf{n}} = \kt{n_1,\ldots,n_L}$. The total particle number ${N} = \sum_{j=1}^L \hat{n}_j$ is conserved. The dimension of the Hilbert space $\HS$ for $N$ bosons on $L$ sites is \cite{bose1924PlancksGesetzUnd} 
\begin{equation}
\label{eq:totalHilbertspace}
   D = D(N,L)  = 
   \binom{L+N-1}{N}. 
\end{equation}

\section{Symmetry analysis \label{sec:symm}}

The symmetries of the anyon-Hubbard model depend on the boundary conditions, the statistical parameter $\theta$, the Hubbard parameter $U$, the number of sites $L$, and on the number of particles $N$. In this section, we establish the relevant point groups $G$ and the classification in terms of Cartan-Altland-Zirnbauer classes \cite{cartan1926ClasseRemarquableDespaces,altlandNonstandardSymmetryClasses1997,zirnbauerSymmetryClassesRandom2004}. Our results are summarized in \refTab{tab:intandsym}, including integrability properties discussed in the following \refSec{sec:integrability}.

The point groups $G$ provide conserved quantities that serve as state labels for degenerate energy eigenstates. Further, they provide selection rules that clarify dynamics and simplify calculations. Symmetry allows the decomposition of the Hilbert space into invariant subspaces associated to unitary irreducible representations (irreps) of $G$:
\begin{equation}
    \mathcal{H} = \bigoplus_{\mu\in M^G} \mathcal{H}^G_\mu \sim \bigoplus_{\mu\in M^G}\left( \mathbb{C}^{m^G_\mu} \otimes \mathbb{C}^{d^G_\mu} \right),
\end{equation}
where $M^G$ is the set of  irreps of $G$. For the $D$-dimensional   representations of $G$ on  $\mathcal{H}$ that we consider, the Hilbert subspace $\mathcal{H}^G_\mu$ associated to irrep $\mu$ can be factored into a multiplicity space with dimension $m^G_\mu$ where the Hamiltonian acts non-trivially and the group acts trivially, and an irrep space with dimension $d^G_\mu$, where the Hamiltonian acts trivially and the group acts non-trivially~\cite{hamermeshGroupTheoryIts1989,zirnbauerSymmetryClassesRandom2004}.  For a specific $N$ and $L$, we derive the irrep multiplicities $m^G_\mu$ using character theory for both open and periodic boundary conditions. The key step in character theory requires us to calculate the trace of one symmetry operator for each conjugacy class in the point group~\cite{hamermeshGroupTheoryIts1989}. We calculate these from the explicit action of the symmetry operators on the number states in the bosonic Fock basis using techniques from combinatorics and number theory. For further details, see App.~\ref{app:symm}. In Tab.~\ref{tab:multi}, we summarize results for the relevant point groups $G$, including the irrep labels $\mu$, dimensions $d^G_\mu$, and multiplicities $m^G_\mu$.

Reduction by the unitary commuting symmetries is a prerequisite for determining the Cartan-Altand-Zirnbauer classes for each symmetry sector $\mathcal{H}^G_\mu$. This classification results from analyzing the presence or absence of time reversal, charge conjugation, chiral symmetry, and calculating their squares~\cite{altland1997NonstandardSymmetryClasses,zirnbauerSymmetryClassesRandom2004, ryu2010TopologicalInsulatorsSuperconductors}. Time reversal symmetry corresponds to the existence of an antiunitary operator that commutes with the Hamiltonian on a given symmetry sector, charge conjugation symmetry to an antiunitary anticommuting operator, and chiral symmetry to a unitary anticommuting operator.
For non-integrable symmetry sectors, these classes indicate the expected level spacing 
distributions after the spectral unfolding process~\cite{bohigasChaoticMotionRandom1984, bohigasCharacterizationChaoticQuantum1984, berrySemiclassicalLevelSpacings1984, berrySemiclassicalTheorySpectral1985, berryStatisticsEnergyLevels1986}, which we employ in \refSec{sec:integrability} to verify our analytic results. Although they have a shared mathematical origin in symmetric spaces and random matrices, we emphasize that our results for elemental symmetry classification pertain to the expected asymptotic level statistics of non-integrable systems, and not to the tenfold way classification of non-interacting topological matter~\cite{kitaev2009PeriodicTableTopological,ryu2010TopologicalInsulatorsSuperconductors}.

The physical time reversal operator of spinless bosons is an antiunitary operator $\hat{T}$ such that the bosonic operators  $\hat{T} \hat{b}_j \hat{T} = \hat{b}_j$ and number states $\hat{T}\kt{\mathbf{n}} =\kt{\mathbf{n}}$ remain invariant. This implies that $\hat{T}$ acts like complex conjugation in the bosonic Fock basis and $\hat{T}^2 = 1$. The physical time reversal operator $\hat{T}$ is generally not a symmetry of the Hamiltonian because it inverts $\theta$
\begin{equation}
\label{eq:PhysicalTimeReversalSymmetry}
    \hat{T}\hat{H}(\theta, U)\hat{T} = \hat{H}(-\theta, U).
\end{equation}
Yet, irrespective of the boundary conditions there is a antiunitary generalized time-reversal symmetry~\cite{lange2017AnyonicHaldaneInsulator,robnikFalseTimereversalViolation1986} that commutes with the Hamiltonian for any $\theta$:
\begin{equation}\label{eq:K}
    \hat{K}_\theta = \hat{Q}_\theta \hat{P} \hat{T},
\end{equation}
with $\hat{K}_\theta^2=1$~\cite{langeAnyonicHaldaneInsulator2017}. Here $\hat{Q}_\theta$ is a local gauge transformation of the bosonic number basis
\begin{align}
        \hat{Q}_\theta = \exp\left(i\theta \sum_{j=1}^L \hat{n}_j (\hat{n}_j -1)/2\right),
\end{align}
and $\hat{P}$ is spatial parity $\hat{P}\kt{n_1,\ldots,n_L} = \kt{n_L,\ldots,n_1}$. The operator $\hat{K}_{\theta}$ transforms anyonic operators  as~\cite{bonkhoff2021BosonicContinuumTheory}
\begin{equation}
    \hat{K}_{\theta}\hat{a}_j \hat{K}_{\theta}=\hat{a}_{L-j+1}e^{-i\theta N},
\end{equation}
whereas bosons are transformed non-linearly.

The presence of any antiunitary symmetry that commutes with the Hamiltonian and squares to unity implies that a real representation of the Hamiltonian exists. Further, the spectral statistics must fall into one of the Cartan-Altland-Zirnbauer classes $\mathrm{AI}$, $\mathrm{BDI}$, or $\mathrm{CI}$ for a symmetry sector of a non-integrable case of the anyon-Hubbard model, depending on the existence of chiral symmetry and charge conjugation symmetry in each symmetry sector. Whether those exist depends delicately on the system parameters, as we describe in the next section and summarize in \refTab{tab:intandsym}.

\subsection{Open boundary conditions}\label{sec:openbc}
For open boundary conditions, besides $K_\theta$, we find nontrivial unitary operators that commute with the Hamiltonian only for bosonic ($\theta=0$) and pseudofermionic ($\theta=\pi$) statistical parameters. 

For bosons, physical time reversal $\hat{T}$ and parity $\hat{P}$ become independent symmetries~\cite{Pausch2021,pausch2022OptimalRouteQuantum}. The parity symmetry implies a point group $G \sim {\mathbb{Z}}_2$ that is abelian with one-dimensional irreps labeled by $p = \pm 1$. The multiplicity of energy eigenstates is
    \begin{equation}
        m^{\hat{P}}_\pm = \frac{1}{2}\left( D \pm d_0 \right),
        \label{tableEq: bosonParityEq}
    \end{equation}
where the total Hilbert space dimension $D$ is defined in \refEq{eq:totalHilbertspace}, and one can calculate $d_0$ as the number of palindromic number states. These are states with occupation numbers $n_j = n_{L-j+1}$, which are parity eigenstates with $p=+1$. From combinatorial counting arguments~\cite{theel2025ChirallyProtectedState}, one can show that the number $d_0$ is given by
\begin{equation}\label{eq:d0}
    d_0 = \left\{ \begin{array}{ll}
    0 & \mbox{for $N$ odd, $L$ even,}\\
  
   \binom{\lfloor N/2 \rfloor}{\lceil L/2 \rceil-1}
   & \mbox{else,}
    \end{array}\right.
\end{equation}
where $\lfloor \dots \rfloor$ is the floor and $\lceil \dots \rceil$ the ceiling function.

Pseudofermions with open boundary conditions are also time reversal symmetric with the physical time reversal symmetry $\hat{T}$, see \refEq{eq:PhysicalTimeReversalSymmetry}. Additionally, 
\begin{equation}
\label{def: P hat pi}
    \hat{P}_\pi \equiv \hat{Q}_\pi\hat{P} = \hat{K}_\pi \hat{T}
\end{equation} 
becomes a unitary symmetry, with eigenvalues $p_\pi = \pm 1$.
This is a different realization of $\mathbb{Z}_2$ than that of $\hat{P}$ for bosons. 
For instance, all palindromic states $\hat{P}\kt{\mathbf{n}}=\kt{\mathbf{n}}$ have positive parity $p=+1$ but their $\hat{P}_\pi$ parity is
$
    \hat{P}_\pi \kt{\mathbf{n}} =(-1)^{\nu(\mathbf{n})}\kt{\mathbf{n}},
$ where 
    $\nu(\mathbf{n}) =
            \sum_{j=1}^L 
                n_j \left(
                        n_j -1
                    \right)/2
                \kt{\mathbf{n}}
    $
is the number of colocated pairs in $\kt{\mathbf{n}}$, which is always even for even $L$.
By combinatorics, we find the number of $\hat{P}_\pi$-parity energy eigenstates is
\begin{equation}
\label{eq:multiplicityobcthetapi}
    m^{\hat{P}_\pi}_\pm = \left\{ \begin{array}{l} m^{\hat{P}}_\pm\ \mbox{for $L$ even,}\\
    m^{\hat{P}}_\pm \mp o^\pi\ \mbox{for $L$ odd,}
    \end{array}\right.
\end{equation}
where the number of odd $\hat{P}_\pi$-parity palindromic states is
\begin{equation}\label{eq: opi}
    o^\pi = \frac{1}{2} \sum_{k=0}^{\lfloor N/2\rfloor}\left(1-(-1)^{\mu_k}\right)\binom{\lfloor L/2 \rfloor + k -1}{k},
\end{equation}
where  $\mu_k = (N-2k)(N-2k-1)/2$. Further details and degeneracy multiplicities are discussed in App.~\ref{app:symm}.

In addition to the unitary commuting symmetries, one can define a chiral symmetry operator
    \begin{equation}
        \hat{S} = \exp\left(i\pi \sum_{j=1}^L j \hat{n}_j \right),\label{chiraloperator}
    \end{equation}
that acts like a staggered gauge transformation on the bosonic operators
$ \hat{S} \hat{b}_j \hat{S}= (-1)^j \hat{b}_j$~\cite{damskiMottinsulatorPhaseOnedimensional2006, grusdtTopologicalEdgeStates2013}. 
In general, $\hat{S}$ inverts both energy and the on-site interactions~\cite{yuSymmetryProtectedDynamical2017}
\begin{equation}
    \hat{S}\hat{H}(\theta, U)\hat{S} = -\hat{H}(\theta,- U)
\end{equation}
and $\hat{S}^2=1$.
In the case $U=0$, this unitary operator $\hat{S}$ anticommutes with the Hamiltonian for open boundary conditions and any $L$. One consequence of chiral  symmetry is that a lower bound on the degeneracy of zero-energy modes is given by the discrepancy between the dimensions $m^S_+$ and $m^S_-$ of the subspaces of the Hilbert space with positive  and negative chirality, respectively, where $m^S_+ + m^S_- = D$~\cite{sutherlandLocalizationElectronicWave1986, liebTwoTheoremsHubbard1989}. Perhaps surprisingly, the chiral discrepancy $|m^S_+-m^S_-|=d_0$   is the same as the discrepancy $|m^P_+-m^P_-| = d_0$ between the multiplicity of positive and negative parity states~\cite{theel2025ChirallyProtectedState}. There may be additional zero modes at $E=0$, but these are  accidental or exactly solvable cases; see Sec.~\ref{sec:Nullstates} for examples.

When $\hat{S}$ is a valid chiral symmetry operator for a symmetry sector, then the Cartan-Altland-Zirnbauer class is either $\mathrm{BDI}$ or $\mathrm{CI}$. To establish that distinction, define the antiunitary, anticommuting charge conjugation symmetry operator $\hat{C}_\theta = \hat{K}_\theta\hat{S}$. One can prove the relation
\begin{equation}
    \hat{K}_\theta\hat{S} - (-1)^{N(L+1)}\hat{S}\hat{K}_\theta=0,
\end{equation}
so charge conjugation symmetry satisfies
\begin{equation}
    \hat{C}_\theta^2 = (-1)^{N(L+1)},
\end{equation}
for arbitrary $\theta$ and $U=0$, distinguishing the two cases. Note that for $\theta =0$ and $\theta=\pi$, the correct charge conjugation operator is $\hat{C}_0 = \hat{T}\hat{S}$, which always satisfies $\hat{C}_{0}^2 = 1$.

Note that for the case of non-interacting bosons with $\theta=0$ and $U=0$ or pseudofermions with only statistical interactions $\theta=\pi$ and $U=0$, parity (or gauge parity) and chirality are both symmetries and satisfy \begin{eqnarray}
   \hat{S} \hat{P}\hat{S}&=& (-1)^{N(L+1)}\hat{P}\nonumber
    \\
    \hat{S} \hat{P}_{\pi}\hat{S} &=&    (-1)^{N(L+1)}\hat{P}_{\pi}. \label{eq: commutationparitieschiral} 
\end{eqnarray}
Therefore, they are not simultaneously diagonalizable when $N(L+1)$ is odd, which is precisely the case where $d_0 = 0$ (cf.\ Eq.~\ref{eq:d0}) and there are no zero-energy modes coming from chiral symmetry, although there may be additional zero-energy modes due to `accidental' fine-tuning or quasi-exactly solvable states). By denoting the mutual eigenstates of $\hat{H}(\theta,0)$ and $\hat{P}$ at $\theta = 0$ (or $\hat{P}_\pi$ at $\theta = \pi$) by $\kt{E,p}$ ($\kt{E,p_\pi}$), respectively, this implies
 \begin{eqnarray}\label{eq:obcmix}
     \hat{S}\kt{E,p} &=& \kt{-E, (-1)^{N(L+1)}p}\nonumber\\
      \hat{S}\kt{E,p_\pi} &=& \kt{-E, (-1)^{N(L+1)}p_\pi}, 
 \end{eqnarray}
i.e., the parity of eigenstates with opposite energy is reversed when $N(L+1)$ is odd. Therefore although the Hamiltonian respects chiral symmetry, the symmetry sectors do not if $N(L+1)$ is odd, and the non-integrable pseudofermionic cases  fall into class $\mathrm{AI}$ instead of $\mathrm{BDI}$ as might be expected.

\subsection{Periodic boundary conditions
\label{sec:periodicBoundaries}
}

Periodic boundary conditions require incorporating rotational symmetry into the   point group and also affect the Cartan labels by preventing the existence of chiral symmetry for odd $L$. 
To simplify the analysis, we transform the Hamiltonian in \refEq{eq:Hamiltonian} into the periodic gauge (see App.~\ref{app:gaugeequiv}) where it is manifestly rotation-symmetric by inserting
\begin{align}
\label{eq:jordanwignerPBC1}
    \hat{a}_j=\hat{b}_j e^{i\theta \sum_{k<j}\hat{n}_k} e^{-i\theta N j/L}
\end{align}
into \refEq{eq:Hamiltonian}. This yields
\begin{eqnarray}
\label{eq:HPBC}
    \hat{H}^{\mathrm{P}}(\theta,U) &=& -J \sum_{j=1}^{L} \left( \hat{b}_{j+1}^\dag e^{-i \theta \hat{n}_{j}}\hat{b}_j e^{i \theta N/L}+\mathrm{h.c}\right)
    \nonumber \\
    & &+ \frac{U}{2} \sum_{j=1}^L \hat{n}_j (\hat{n}_j - 1),
\end{eqnarray}
where $\hat{b}_{L+1} \equiv \hat{b}_1$.
The Hamiltonian in \refEq{eq:HPBC} commutes with the rotation/translation operator $\hat{R}$ with $\hat{R}^L = 1$ that has the action~\cite{essler2010OnedimensionalHubbardModel}
\begin{eqnarray}
\hat{R}\kt{n_1,n_2,\ldots,n_L} &=& \kt{n_L, n_1, \ldots, n_{L-1}}\nonumber\\
    \hat{R} \hat{b}_j \hat{R}^\dag &=& \hat{b}_{j-1}.
\end{eqnarray}
The group generated by $\hat{R}$ is isomorphic to the abelian cyclic group $\mathbb{Z}_L$. We decompose the Hilbert space $\mathcal{H}$ into symmetry sectors $\mathcal{H}^{\mathbb{Z}_L}_q$ corresponding to irreps with angular momentum $q \in \{0, \pm  1 , \pm 2, \ldots, \pm (L-1)/2\}$ for odd $L$ and $q \in \{0, \pm 1, \pm 2, \ldots, \pm L/2-1, L/2 \}$ for even $L$. The eigenvalues of $\hat{R}$ are $\omega^q$, where $\omega =e^{2\pi i /L }$. 
The multiplicity of states with a given $q$ is
\begin{equation}
    m^{{\mathbb{Z}}_L}_q = \frac{1}{L} \sum_{h|(N,L)} C_{h}(q)D(N/h,L/h),
    \label{eq: subspace_multiplicityintermediatetheta}
\end{equation}
where the sum over $h|(N,L)$ means the sum over all the divisors $h$ of the greatest common divisor of $N$ and $L$ and $C_{h}(q)$ is Ramanujan's sum, see App.~\ref{app:symm}.
Because they are in different symmetry sectors, note that states with different $q$ may be degenerate at particular values of $\theta$ and $U$.
However, when degeneracies occur within a given symmetry sector, those may be so-called ``accidental degeneracies'', but they often reveal additional symmetries and/or quasi-exact solvability, i.e., exactly solvable states embedded within the spectrum (see examples below in \refSec{sec:Nullstates}).  Such additional degeneracies are known to exist in periodic Hubbard-type tight-binding models and spin chains~\cite{heilmann1971ViolationNoncrossingRule, sriramshastryDecoratedStartriangleRelations1988, grosseSymmetryHubbardModel1989, nicolau2026FragmentationZeroModes}. 

Regarding the Cartan labels, interacting anyons with periodic boundary conditions respect the generalized time reversal symmetry $\hat{K}_\theta$ just like anyons with open boundary conditions. Yet, chiral symmetry is broken if the system has an odd number of sites, leaving the system in symmetry class $\mathrm{AI}$ (for non-integrable systems) even at $U=0$.
For an even number of sites $L$ and $\theta \neq 0, \pi$, chiral symmetry at vanishing Hubbard interactions is restored within each symmetry sector, and the system enters either symmetry classes $\mathrm{BDI}$ and $\mathrm{CI}$ again, for even $N(L+1)$ and odd $N(L+1)$, respectively. For even $L$, the following relation additionally holds

\begin{align}
    &\hat{R}\hat{S}-e^{i\pi N}\hat{S}\hat{R}=0, \label{eq: rotation and chirality commutation relation}
\end{align}
implying that the following energy eigenstates are connected by $\hat{S}$,
\begin{align}
\hat{S}\kt{E,q} = \kt{-E,q+ NL/2}.
\end{align}

The point groups and Cartan classes change for bosons and pseudofermions with periodic boundary conditions.
Interacting bosons are parity $\hat{P}$ and time reversal $\hat{T}$ symmetric. From $\hat{T}\hat{R} = \hat{R}^\dag$ and $\hat{P}\hat{R}\hat{P} = \hat{R}^\dag$, we see that both $\hat{P}$ and  $\hat{T}$ mix sectors with the opposite center of mass angular momentum $q$
\begin{equation}\label{eq:map0}
    \hat{T} \mathcal{H}^{\mathbb{Z}_L}_q = \hat{P} \mathcal{H}^{\mathbb{Z}_L}_q = \mathcal{H}^{\mathbb{Z}_L}_{-q}.
\end{equation}
Adding parity to the point group, now the operators $\hat{P}$ and $\hat{R}$ generate a unitary representation of the dihedral group $D_L \cong \mathbb{Z}_2 \ltimes \mathbb{Z}_L$, $|D_{L}|=2L$ \cite{hamermeshGroupTheoryIts1989,Hirsch2020}:
\begin{equation}
    D_L \sim \langle  s,r \ | r^L=s^2=1, sr^ks^{-1}=r^{L-k}\rangle.
    \label{eq: dihedral presentation}
\end{equation}
Parity mixes ${\mathbb{Z}}_L$ irreps with $q$ and $-q$ into a single two-dimensional irrep of $D_L$ and leaves the irrep $q=0$ and, additionally for even $L$, the irrep $q=L/2$ invariant~\cite{Hirsch2020}.
For odd $L$, $D_L$ has two one-dimensional representations with $q=0$ and $p= \pm 1$ and $(L-1)/2$ two-dimensional representations that combine the $\pm q$ irreps of ${\mathbb{Z}}_L$ for $q \neq 0$. For $L$ even, $D_L$ has four one-dimensional representations with $q=0$ and $q=L/2$ and $p= \pm 1$ and $(L-2)/2$ two-dimensional representations that combine the $\pm q$ irreps of ${\mathbb{Z}}_L$ for $q \neq 0, \pi$. The multiplicities of these irreps for $L$ odd are
\begin{align}
m_{0,\pm 1}^{D_L}=& \ \frac{m^{{\mathbb{Z}}_L}_0 \pm d_0}{2}, \label{tableEq: bosonicOddLTrivialandSign}
\\
    m^{D_L}_{q,-q} =& \ m^{{\mathbb{Z}}_L}_q\ \mbox{for}\ q\neq 0 \label{tableEq: bosonicOddL2dIrrep},
\end{align}
and for $L$ even are
\begin{align}
m^{D_L}_{0,\pm 1} =& \ \frac{1}{2}\left(m^{{\mathbb{Z}}_L}_0\pm \frac{d_0 + d_1}{2}\right), \label{tableEq: bosonicEvenLTrivAndSign}
\\
m^{D_L}_{L/2,\pm 1}=& \ \frac{1}{2}\left(m^{{\mathbb{Z}}_L}_{L/2}\pm \frac{d_0 - d_1}{2}\right), \label{tableEq: bosonicEvenLAltAndAltSign}
\\
m^{D_L}_{q,-q} =& \ m^{{\mathbb{Z}}_L}_q\ \mbox{for}\ q\neq 0, L/2,
\label{tableEq: bosonicEvenL2d}
\end{align}
where 
\begin{equation}
    d_1=\binom{L/2+\lfloor N/2\rfloor}{L/2}+\binom{L/2+\lfloor(N+1)/2\rfloor-1}{L/2},
\end{equation}
see App.~\ref{app:symm} for details.

For pseudofermions, additional subtleties arise in the point group analysis. Consider the gauge transformation 
\begin{equation}
    \hat{{W}}(\alpha) = e^{i \alpha \sum_{j=1}^L j \hat{n}_j},
    \label{eq: W gauge transform}
\end{equation}
which connects the boundary gauge and periodic gauge (cf.~App.~\ref{app:gaugeequiv}). The gauge-transformed time reversal operator 
\begin{equation}
    \hat{T}_\pi \equiv  \hat{W}\!\left(\frac{2\pi N}{L}\right) \hat{T} = \hat{W}\!\left(\frac{\pi N}{L}\right) \hat{T}\ \hat{W}^\dag\!\left(\frac{\pi N}{L}\right), 
\end{equation}
with $\hat{T}_\pi^2=1$,
commutes with the Hamiltonian in \refEq{eq:Hamiltonian}. Using $\hat{K}_\pi$, we define the gauge-transformed parity
\begin{equation}
    \bar{P}_\pi \equiv \omega^{-N^2/2} \hat{T}_\pi \hat{K}_\pi = \omega^{-N^2/2}\ \hat{W}\!\left(\frac{2\pi N}{L}\right) \hat{P}_\pi, 
    \label{def: gauge transformed P pi}
\end{equation}
where the phase factor $\omega^{-N^2/2}$ ensures $\bar{P}_\pi^2 = 1$ and $\bar{P}_\pi = \bar{P}^\dag_\pi$. This parity transforms the rotation operator as
\begin{equation}\label{eq:mixeddihedral}
    \bar{P}_\pi \hat{R} \bar{P}_\pi = \omega^{N^2} \hat{R}^{-1}.
\end{equation}
If we adjust the phase of the rotation operator
\begin{equation}\label{eq:twistC}
    \bar{R}_\pi \equiv  \omega^{-N^2/2}\hat{R},
\end{equation}
we find that $\bar{P}_\pi$ and $\bar{R}_\pi$ satisfy the dihedral generating relation
\begin{subequations}\label{eq:twisteddihedral}
\begin{equation}\label{eq:twisteddihedral1}
    \bar{P}_\pi \bar{R}_\pi \bar{P}_\pi = \bar{R}_\pi^{-1}.
\end{equation}
However, the operator $\bar{R}_\pi$ satisfies
\begin{equation}\label{eq:twisteddihedral2}
    \bar{R}_\pi^L = (-1)^N.
\end{equation}
\end{subequations}
When $N$ is odd, the relations in \refEq{eq:twisteddihedral} therefore generate $D_{2L}$, the double-cover of $D_L$~\cite{hamermeshGroupTheoryIts1989}. This doubling is the pseudofermionic manifestation of the doubling required to describe an odd number of fermions on a ring~\cite{patuCorrelationFunctionsOnedimensional2007,liebTwoSolubleModels1961,capelNoteDifferenceAcyclic1975,SeibergShao2024}.
So, like bosons, there is dihedral symmetry for pseudofermions, but a different realization of $D_L$ for even $N$ and a doubling $D_{2L}$ for odd $N$.

To find the multiplicities of the irreps for these parity-twisted realizations, denote the irreps of $\mathbb{Z}_L \subset D_L$ (for even $N$) and ${\mathbb{Z}}_{2L} \subset D_{2L}$ (for odd $N$) by $\bar{q}$. The relation (\ref{eq:twistC}) implies the following maps from the eigenvalues of $\hat{R}$ to the eigenvalues of $\bar{R}_\pi$:
\begin{align}\label{eq:qmap}
    \bar{q} &\equiv q - N^2/2 \!\!\mod L\ \mbox{for even $N$ and} \\
    \bar{q} &\equiv 2q - N^2 \!\!\mod 2L\ \mbox{for odd $N$}\nonumber.
\end{align}
Furthermore, the relation in \refEq{eq:mixeddihedral}, which determines which $q$ are paired by $\bar{P}_\pi$, can be combined with \refEq{eq:qmap} to confirm $\bar{P}_\pi$ mixes the irreps at $\bar{q}$ and $-\bar{q}$. This allows us to calculate the multiplicities for $m^{D_L}_{\bar{q},\bar{p}_\pi}$ and $m^{D_L}_{\bar{q},-\bar{q}}$ ($N$ even) and $m^{D_{2L}}_{\bar{q},\bar{p}_\pi}$ and $m^{D_{2L}}_{\bar{q},-\bar{q}}$ ($N$ odd), from the corresponding multiplicities of $m^{D_L}_{q,p}$ and $m^{D_L}_{q,-q}$. See App.~\ref{app:symm} for more information on these calculations and their derivation.

A final comment on Cartan labels for bosons and pseudofermions: in analogy to \refEq{eq: commutationparitieschiral}, we have at $\theta=0,\pi$, $U=0$,  and even $L$:\begin{eqnarray}\label{eq:pbcmix}
     \hat{S}\kt{E,p,q} &=& \kt{-E, (-1)^{N}p,q+ NL/2}\\
      \hat{S}\kt{E,p_\pi,q} &=& \kt{-E, (-1)^{N}p_\pi,q+ NL/2}\nonumber.
 \end{eqnarray}
 As with open boundary conditions, this implies that although the Hamiltonian has chiral symmetry, the individual symmetry sectors do not have chiral symmetry when $N$ is odd.
\begin{figure*}

\includegraphics[width=0.999\linewidth]{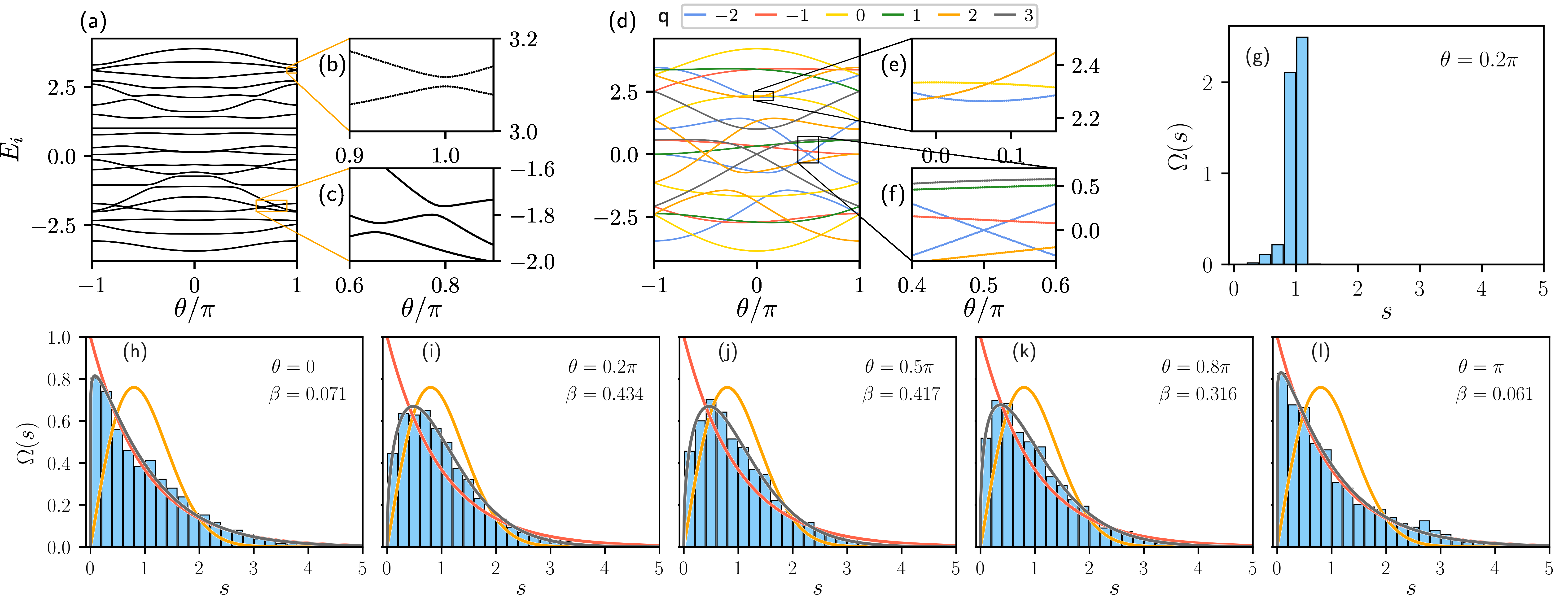}
    \caption{ The two-particle anyon-Hubbard model's integrability depends on its parameters and boundary conditions.
    (a) and (d) depict the energy levels for parameters $N=2$, $L=6$, and $U/J = 1$.
    (a) For open boundaries, the spectrum exhibits avoided level crossings, see insets (b) and (c), a signature of nonintegrability. (d) For periodic boundaries, the line colors indicate angular momenta $q$ according to \refSec{sec:periodicBoundaries}. Crossing between states with different angular momenta is expected (e),  whereas crossings in the same angular momentum sector (f) are indicative of additional exactly-solvable states. 
    (g) and (h-l) depict the symmetry-reduced energy level spacing statistics, see main text for details, for parameters $N=2$ and  $L=100$.
    g) Periodic boundaries: 
    The distribution is sharply peaked around the average level spacing $s=1$. This is characteristic of an effectively one-dimensional integrable system for all $\theta$. (h-l) Open boundaries and $U=0$: For bosons (h) and pseudofermions (l) the fit to the Brody function (\refEq{eq:brody}) with the inset parameter $\beta$ close to $0$ indicates a Poissonian statistics (red curve) characteristic of integrability in a system with more than one effective degree of freedom. For anyons with $0 < \theta < \pi$ (i-k), the level statistics come closer to Wigner-Dyson like (orange curve) with $\beta$ near $0.4$, indicating that two anyons with open boundaries are generally not integrable. 
    \label{fig:levelstatistics} }
\end{figure*}

\section{Integrability}
\label{sec:integrability}

In this section, we address the integrability of the anyon-Hubbard model with two particles, which is the only limit that is expected to be integrable beyond the dimer case. The level variation with $\theta$ depicted in  Figs.~\ref{fig:levelstatistics}(a) and (b) corroborate our symmetry results above and integrability results below~\cite{Neumann1929,Wigner1929,Stepanov2008,braakIntegrabilityWeakDiffraction2014,Kollar2012,Kollar2013,braakIntegrabilityWeakDiffraction2014}. To summarize, in agreement with \refTab{tab:intandsym} and \refTab{tab:multi}, there is no symmetry label for anyons with open boundaries, whereas the states of anyons with periodic boundary conditions are labeled by the angular momentum $q$, see \refEq{eq: subspace_multiplicityintermediatetheta} that each 
have exactly $m^{\mathbb{Z}_L}_q$ states for $0 < \theta < \pi$.
The levels with different $q$ combine at $\theta=0$ and $\theta =\pi$ as predicted by Eqs.~(\ref{eq:map0}) and (\ref{eq:qmap}), respectively, and there are additional crossings within the same $q$-sector at $\theta =0$, $\theta = \pi/2$, and $\theta = \pi$. Such crossing are typical in integrable systems, especially those with exactly-solvable states. In contrast, crossings of different momenta are expected irrespective of integrability because the spectra of different symmetry sectors are uncorrelated. For open boundary conditions on the other hand, there are no symmetry sectors for arbitrary $\theta$ or crossings at individual $\theta$. The individual levels appear repulsive, demonstrating non-integrability.

More insight about integrable regimes is provided by the symmetry-reduced level spacing statistics~\cite{mehta1991random,GUHR1998189,baxter2016exactly,Rabson2004,robnikFalseTimereversalViolation1986}. Expressed in terms of the average spacing $s$ between unfolded energy levels~\cite{brody1981randommatrix, poilblancPoissonVsGOE1993, kieburg2026quantumchaoticsystemsrandommatrix}, a generic non-integrable system has a level spacing distribution determined by the corresponding Cartan-Altand-Zirnbauer class of random matrices. The three relevant classes are summarized in Tab.~\ref{tab:intandsym}, corresponding to Gaussian orthogonal for $\mathrm{AI}$, chiral orthogonal for $\mathrm{BDI}$, and BdG orthogonal for $\mathrm{CI}$ ensembles. The relation to orthogonal rather than unitary ensembles for the anyon-Hubbard is caused
by the role of the anyonic symmetry $\hat{K}_\theta$ as a \lq false\rq \phantom{a} time reversal transformation  \cite{robnikFalseTimereversalViolation1986}. These classes can be distinguished by their symmetry around $E=0$ for $\mathrm{BDI}$ and $\mathrm{CI}$; their statistical spectral properties near $E=0$~\cite{takahasi_energy-level_2001,Evangelou_spectral_2003,rehemanjiang_microwave_2020}; and the existence of chiral symmetry-protected zero modes for $\mathrm{BDI}$~\cite{theel2025ChirallyProtectedState}.

As evidence for the presence or absence of integrability, we  fit the unfolded level statistics to the Brody distribution~\cite{brody1981randommatrix,Fogarty2021}
\begin{equation}\label{eq:brody}
    \Omega^\beta(s) = (\beta +1) b s^\beta \exp(-b s^{\beta+1}),
\end{equation}
where $b = \Gamma\left[(\beta+2)/(\beta+1)\right]^{(\beta +1)}$. This distribution coincides with Poisson statistics for $\beta=0$, with an exponential decay in the average level spacing,
\begin{equation}\label{eq:poisson}
    \Omega^P(s) = e^{-s},
\end{equation}
indicating a generic integrable system with more than one degree of freedom.
For $\beta=1$ on the other hand, \refEq{eq:brody} coincides with the Wigner-Dyson distribution, typical for random matrices in the Gaussian orthogonal class $\mathrm{AI}$ for
\begin{equation}
    \Omega^{WD}(s)= \frac{\pi}{2} s \exp(-\pi s^2/4).
\end{equation}
Increasing $\beta$ reflects increased level repulsion and spectral rigidity. The limit $\beta \to \infty$ leads to a sharp peak at $s=1$ and corresponds to equally-spaced levels~\cite{Berry1977,Garcia-March_2018,Pandey1991,Fogarty2021}. Such distributions are called `picket fence' and are expected for the unfolded spectrum of systems that are effectively one-dimensional after symmetry reduction.

In \refFig[g]{fig:levelstatistics} , we depict the level statistics distribution for $N=2$ and $L=100$ with periodic boundary conditions and $U/J=1$. For $0<\theta<\pi$, all irreps are one-dimensional as the anyonic symmetry $\hat{K}_\theta$ is antiunitary.
There are $100$ symmetry sectors in total, one for each quasimomentum $q$, with either $50$ or $51$ levels (see \refEq{eq: subspace_multiplicityintermediatetheta}). For $\theta=0$ and $\theta=\pi$ on the other hand, $\hat{K}_\theta$ is replaced by the unitary operators $\hat{P}$ and $\hat{P}_{\pi}$ and the corresponding anti-unitary operators $\hat{T}$ and $\hat{T}_\pi$, respectively; see Tab.~\ref{tab:multi}. In these two cases, there are $48$ two-dimensional irreps for $50 > q > 0$ and four one-dimensional irreps for $q=0$ and $q=50$, $ p= \pm 1$ or $p_\pi = \pm 1$, respectively. After separately unfolding and binning all levels, sectors are then averaged (cf.~\cite{turnerStableInfinitetemperatureEigenstates2025}). For all $\theta$, the result is a sharp picket fence distribution, indicating that the system is integrable and has effectively one degree of freedom; in \refFig[g]{fig:levelstatistics} we show only one example for $\theta=\pi/5$. 

In \refFig[h-l]{fig:levelstatistics}, we depict the level statistics distribution open boundary conditions, again for $N=2$ and $L=100$ but for vanishing Hubbard interactions $U=0$. The spectrum is symmetric around $E=0$, which is degenerate with $d_0=50$ zero modes. For $0<\theta<\pi$, there is a single symmetry sector and we consider $E<0$. The parameter $\beta \approx 0.4$ is a clear indication of level repulsion and non-integrability, even though there are only statistical interactions. For $\theta=0$ and $\theta=\pi$, there are two symmetry sectors corresponding to the eigenvalues of the $\hat{P}$ and $\hat{P}_\pi$ operators respectively.
We average their separated, unfolded level statistics as in \refFig[g]{fig:levelstatistics}. The fit values of $\beta$ close to zero indicate this system is integrable with more than one effective degree of freedom. These characteristics persist for finite Hubbard interactions.

We next employ a coordinate Bethe ansatz to investigate the integrability of the two particle anyon-Hubbard model for open and periodic boundary conditions, where previous work investigated infinite systems~\cite{greschner2015AnyonHubbardModel,kwan2024RealizationOnedimensionalAnyons,bakkalihassani2026RevealingPseudoFermionizationChiral}.
We consider,
\begin{align}
	\label{betheansatz}
	 &\vert \psi\rangle \propto \sum_{l,m=-\infty}^\infty\tilde{c}_{l,m}\hat{a}_l^\dagger\hat{a}_m^\dagger \vert 0\rangle,
     \\
     &\tilde{c}_{l,m}=c_{lm}e^{i\theta\sign(l-m)/2},
\end{align} 
where we impose the system size $L$ by the boundary conditions at a later stage, and the $\theta$-dependent phase factor ensures the anyonic statistics of the coefficient $\tilde{c}_{l,m}$, i.e.,  $\tilde{c}_{m,l}=\tilde{c}_{l,m}e^{-i\theta\sign(l-m)}$  \cite{essler2010OnedimensionalHubbardModel,kundu1999ExactSolutionDouble,batchelorBetheAnsatz1D2007}, where the bosonic coefficients obey $c_{lm}=c_{ml}$. Inserting \refEq{betheansatz} into the Hamiltonian \refEq{eq:Hamiltonian}, leads to a recurrence relation for these coefficients
\begin{align}
\sum_{\sigma=\pm1}
&
e^{-\frac{i\sigma\theta}{2}\left(\delta_{l,m}+\delta_{l,m+\sigma}\right)}
c_{l,m+\sigma}
+
e^{-\frac{i\sigma\theta}{2}\left(\delta_{l,m}+\delta_{l+\sigma,m}\right)}
c_{l+\sigma,m}
\nonumber \\
&-
\frac{U\delta_{l,m}-E}{J}c_{l,m}
=0.\label{recurrence}
\end{align}
The basic idea of the coordinate Bethe ansatz is to decompose $c_{lm}$ into a superposition of incoming and outgoing plane waves,
with energy 
\begin{align}	E=-2J\big(\cos(k_1)+\cos(k_2)\big)\label{energydefinition},
\end{align}
where the momenta of the particles are solutions to the transcendental Bethe equations that follow from the respective boundary conditions \cite{betheZurTheorieMetalle1931,essler2010OnedimensionalHubbardModel}.\\ We first prove integrability of two anyons with periodic boundaries and proceed to demonstrate that open boundaries are not integrable. 
For periodic boundary conditions, we
consider the ansatz \cite{essler2010OnedimensionalHubbardModel,boschiBoundStatesExpansion2014,oelkersGroundstatePropertiesAttractive2007,poloExactResultsPersistent2020}, 
\begin{align}
	c_{l,m} = \Bigg\{  \begin{array}{cc} 
		A_{k_1,k_2}e^{i\left(k_1m+k_2l\right)}+A_{k_2,k_1}e^{i\left(k_1l+k_2m\right)} & \hspace{5mm} l<m, \\
		c_{l=m} & \hspace{5mm} l=m, 
    \\
		A_{k_1,k_2}e^{i\left(k_1l+k_2m\right)}+A_{k_2,k_1}e^{i\left(k_1m+k_2l\right)} & \hspace{5mm} l>m. \\
	\end{array} 
    \label{ansatzperiodic}
\end{align}  
Imposing periodic boundary conditions on the first anyon leads to twisted boundary conditions for the second one and vice versa \cite{kundu1999ExactSolutionDouble,patuCorrelationFunctionsOnedimensional2007,batchelorBetheAnsatz1D2007}, e.g.,
\begin{align}
    \tilde{c}_{1,m}=\tilde{c}_{L+1,m},\quad
\label{betheequations2}
	&\tilde{c}_{l,1}=\tilde{c}_{l,L+1}e^{-2 i \theta}.
\end{align} 
This choice leads to the Bethe ansatz equations in logarithmic form,
\begin{align}
    \label{betheequations}
     k_j L = \theta - (-1)^j \arg\left(\frac{A_{k_1,k_2}}{A_{k_2,k_1}}\right) +2\pi n_j
\end{align}
with $j\in\{1,2\}$ and $n_j \in \{1,..,L\}$. Furthermore, these twisted boundary conditions induce persistent currents 
\cite{poloExactResultsPersistent2020,Minguzzi2023} controlled by the statistical parameter $\theta$. We write the Bethe ansatz in \refEq{ansatzperiodic} in center of mass $Q^+=k_1+k_2$ and relative momentum $Q^-=(k_2-k_1)/2$
\cite{essler2010OnedimensionalHubbardModel,valienteTwoparticleStatesHubbard2008,valienteScatteringResonancesTwoparticle2009,greschner2015AnyonHubbardModel,zhang2017GroundstatePropertiesOnedimensional}, 
\begin{align}
    &c_{l,m}= e^{i Q^+ (l+m)/2} c_{\vert l-m \vert},
    \\
    &E=-4J\cos\left(Q^-\right)\cos\left(\frac{Q^+}{2}\right)\label{energydefinitionpbc}
\end{align}
Here, the coefficients for the relative coordinate $k=l-m$ is given by  
\begin{align}
	c_{\vert k\vert}=\left(1-\delta_{k,0}\right)\left(e^{-iq\vert k\vert}+e^{2i\eta(Q,q)}e^{iq\vert k\vert}\right)+c_{0}\delta_{k,0}\label{relativewavefunction},
\end{align}
with the scattering phase shift 
\begin{align}
	\label{scatteringphaseshift}
	&\eta(Q^+,Q^-)=\frac{1}{2}\arg\left(\frac{A_{k_1,k_2}}{A_{k_2,k_1}}\right)=\frac{\pi}{2}-\operatorname{atan2}(y,x)
\phantom{ab}\mathrm{mod}\;\pi
\end{align}
with
\begin{align}
    &x=-\cos\left(\frac{Q^+}{2}\right)
 +\frac{2\left[1+\cos(Q^+-\theta)\right]}{\frac{U}{2J}
    +4\cos(Q^-)\cos(\frac{Q^+}{2})}\cos\left(Q^-\right),\nonumber
	\\
	&y=-\frac{2\left[1+\cos(Q^+-\theta)\right]}{\frac{U}{2J}+4\cos(Q^-)\cos\left(\frac{Q^+}{2}\right)}\sin\left(Q^-\right).\nonumber
\end{align}
The scattering phase is an antisymmetric function in $Q^-$, which ensures that the scattering is unitary.
The coefficient $c_0$ for two particles at the same site
\begin{align}
\label{eq: particlecoincidenceperiodic}
	c_{0}=\frac{\left(e^{-iQ^-}+e^{i2\eta(Q^+,Q^-)}e^{iQ^-}\right)\cos\left(\frac{Q^+-\theta}{2}\right)}{\frac{U}{4J}+\cos\left(Q^-\right)\cos\left(\frac{Q^+}{2}\right)}
\end{align}
ensures the continuity of the solution. The Bethe ansatz equations in 
\refEq{betheequations} then reduce to a single dynamical equation and a quantization condition~\cite{essler2010OnedimensionalHubbardModel,boschiBoundStatesExpansion2014,poloExactResultsPersistent2020}
\begin{align}
&Q^+=\frac{2\pi n_+}{L}+\frac{2\theta}{L}\label{centerofmassquantization},
\\
	&Q^-=\frac{\pi n_-}{L}-\frac{2\eta(Q^+,Q^-)}{L}\label{singlebetheequations},
\end{align}
with 
\begin{align}
    &n_-=n_1-n_2\in \left[0,L-1\right],
    \\
&n_+=n_1+n_2\in\left[1,L\right]\mod L.
\end{align}
Real momenta $Q^-$ describe scattering solutions, while a nonvanishing imaginary part describes bound states \cite{essler2010OnedimensionalHubbardModel,oelkersGroundstatePropertiesAttractive2007,boschiBoundStatesExpansion2014}. The values $Q^-=0,\pm \pi$ are formal solutions of \refEq{singlebetheequations}, but are excluded from the regular scattering sector, as the two plane waves in \refEq{ansatzperiodic} become linearly dependent, or even vanish at $\eta=\pi/2$ 
\cite{izergin1982PauliPrincipleOnedimensional,Alba_2013}. Together with the overall $2\pi$ periodicity of the wavefunction in the relative momentum, this restricts the real roots to one of the intervals $Q^-\in \left(0,\pi\right) $, or $Q^-\in\left(-\pi,0\right) $, respectively. For imaginary solutions on the other hand, the real part of the Bethe momentum is restricted to $\mathrm{Re}(Q^-)=0,\pi$, in order to ensure real valued energies according to \refEq{energydefinitionpbc}. Only half of the imaginary axis has to be considered due to the reflection symmetry of the energy with respect to $Q^-$.

We next address the uniqueness of the Bethe roots and especially discuss non-regular zero energy solutions \cite{essler2010OnedimensionalHubbardModel}.
First, for generic parameters and Bethe roots, the scattering phase and its derivatives $\partial_{Q^-}\eta(Q^+,Q^-)$ are bounded and smooth, whenever the coefficient ratio in \refEq{singlebetheequations} is finite, but nonzero. Therefore, the solutions of \refEq{singlebetheequations} are uniquely labeled by the integers $n_{\pm}$, if 
\begin{align}
 F(Q^+,Q^-)=  \frac{ LQ^-+2\eta(Q^-,Q^+) }{\pi}
\end{align}
is strictly monotone in $Q^-$, and its derivative with respect to $Q^-$ is positive.
According to \refEq{scatteringphaseshift} this is generally not fulfilled if the energy equals the Hubbard interaction, i.e. $U=E$.  
This case includes the nullspace of noninteracting anyons with $U=E=0$, as well as finite Hubbard interactions $U=E\neq0$.
The parametrization of the scattering phase can therefore become ill-defined and multiple solutions exist. According to \refEq{energydefinition}, the zero-energy solutions originate either from a relative momentum of $Q^-=\pi/2$, or a center of mass momentum of $Q^+=\pi$, whereas one explicitly considers the limit $E=U$ in the case of finite Hubbard interactions. 
The factor $1+\cos(Q^+-\theta)$ in \refEq{scatteringphaseshift} dictates whether the respective limits $\lim_{Q^-\rightarrow\pi/2}\eta(Q^+,Q^-)$, $\lim_{Q^+\rightarrow\pi}\eta(Q^+,Q^-)$, or $\lim_{E\rightarrow U}\eta(Q^+,Q^-)$ are well defined.
We classify four different types of zero energy solutions for noninteracting anyons with periodic boundary conditions in \refApp{app:NullspaceBetheRoots}, whereas one class corresponds to the $U=E$ case if one considers finite $U$ instead.

The additional labeling of states by the quantum numbers $n_-$ explains the crossing of energy levels with equal center of mass momentum $n_+$, which cannot be resolved by the symmetry group analysis of the $N$-particle case in \refFig[d]{fig:levelstatistics}.
In the limit $L\rightarrow\infty$, the momenta $Q^+,Q^-$ become continuous and our solution recovers the previous continuum studies \cite{greschner2015AnyonHubbardModel,zhang2017GroundstatePropertiesOnedimensional,kwan2024RealizationOnedimensionalAnyons}.

We proceed with demonstrating how integrability fails for open boundary conditions for all anyonic values of the statistical parameter $\theta$. This emphasizes the special behavior of anyons as compared to bosons and pseudofermions, which remain integrable for two particles, see \refTab{tab:multi}.
For open boundaries, the scattering off the boundary couples the relative and center of mass parts of the coefficients $c_{l,m}$ in \refEq{betheansatz} \cite{batchelor20051DInteractingBose, longhiTammHubbardSurface2013}
\begin{align}
	c_{l,m}=: \begin{cases} 
	c^<_{l,m} & \hspace{5mm} l<m, \\
		c_{l=m} & \hspace{5mm} l=m, \\
		c^>_{l,m} & \hspace{5mm} l>m, \\
	\end{cases} \label{ansatzopen}
\end{align}   
with coefficients that implement bosonic exchange symmetry,
\begin{align}
\label{coefficientopen1}
c^<_{l,m}
&=
\sum_{\sigma,\sigma'=\pm 1}
\alpha_{\sigma,\sigma'}
e^{\I \left(\sigma k_1 l + \sigma' k_2 m\right)}
+
\beta_{\sigma,\sigma'}
e^{\I \left(\sigma k_2 l + \sigma' k_1 m\right)},
\\
\label{coefficientopen2}
c^>_{l,m}
&=
\sum_{\sigma,\sigma'=\pm 1}
\beta_{\sigma',\sigma}
e^{\I \left(\sigma k_1 l + \sigma' k_2 m\right)}
+
\alpha_{\sigma',\sigma}
e^{\I \left(\sigma k_2 l + \sigma' k_1 m\right)}  .
\end{align}
\refEq{energydefinition} remains valid for the energy, yet, there are more choices of $k_1$ and $k_2$ for open boundaries that result in a real energy \cite{Alba_2013}
\begin{align}
\label{momentumvaluesobc}
1) && \mathrm{Im}(k_1) &= 0 = \mathrm{Im}(k_2), \\
2) && \mathrm{Re}(k_1), &= 0 = \mathrm{Re}(k_2), \\
3) && \mathrm{Im}(k_{1,2}), &= 0 = \mathrm{Re}(k_{2,1}), \\
4) && k_1^* &= k_2.
\end{align}
Physically, these choices correspond to different scenarios of bound and dispersive behavior, a distinction that only becomes unambiguous in the thermodynamic limit $L\rightarrow\infty$  \cite{Kollar2012,Alba_2013,longhiTammHubbardSurface2013,Kollar2013,zhang2013BoundStatesOnedimensional,zhang2023AnyonicBoundStates},
where the scattering states become continuous and bound states can emerge at the edges of this continuum \cite{Kollar2012,Kollar2013}.
Inserting  \refEq{coefficientopen1} and \refEq{coefficientopen2} into the recurrence equation in \refEq{recurrence}  yields a set of continuity conditions for the Bethe amplitudes on the defective lines where $l=m$ or $l=m\pm1$. First, we determine the coefficient $c_{l=m}$ for particles at the same site~\cite{greschner2015AnyonHubbardModel,zhang2017GroundstatePropertiesOnedimensional},
\begin{align}
	c_{l=m}=&e^{-i\theta/2}\left(c^<_{m,m+1}+c^>_{m+1,m}\right)\frac{J}{U-E}
 \\ &+e^{i\theta/2}\left(c^>_{m,m-1}+c^<_{m-1,m}\right)\frac{J}{U-E}.\nonumber
\end{align} 
In the next step the continuity conditions at the defective line $l=m-1$ relates the amplitude ratios of the coefficients $\alpha_{\sigma,\sigma'}$ and $\beta_{\sigma,\sigma'}$ \cite{longhiTammHubbardSurface2013}, 
\begin{align}
  A_{\sigma,\sigma'}\alpha_{\sigma,\sigma'}
+
\widetilde A_{\sigma,\sigma'}\beta_{\sigma',\sigma}=0,\quad\sigma,\sigma'=\pm 1,
\end{align}
where we have introduced the shorthand notation,
\begin{align}
&A_{\sigma,\sigma'}
:=
A(\sigma k_1,\sigma' k_2,\theta),
\\
&\widetilde A_{\sigma,\sigma'}
:=
A(\sigma' k_2,\sigma k_1,\theta),
\\
&A(k_1,k_2,\theta)=e^{-2ik_1}+e^{-i(k_1-k_2)}+e^{-ik_1}E/(J)
\nonumber \\
 &\quad\quad\qquad+\frac{2\left(2e^{-ik_1}+e^{i(k_2-\theta)}+e^{i(-k_2-2k_1+\theta)}\right)}{U/J-E/J}.
 \label{Betheamplitude}
\end{align}
Imposing open boundaries $\tilde{c}_{L+1,m}=\tilde{c}_{0,m}=0$ and  $\bar{c}_{l,L+1}=\bar{c}_{l,0}=0$ relates the coefficients
\begin{align}
\alpha_{\sigma,\sigma'}
&=
\sigma
\left(-e^{2ik_2(L+1)}\right)^{\frac{1-\sigma'}{2}}\alpha_{+,+},
\\
\beta_{\sigma,\sigma'}
&=
\sigma
\left(-e^{2ik_1(L+1)}\right)^{\frac{1-\sigma'}{2}}\beta_{+,+},
\end{align}
 yielding a ($4\times 2$) system of linear equations
\begin{align}
\begin{pmatrix}
A_{+,+} & \widetilde A_{+,+}
\\
e^{2ik_2(L+1)}A_{+,-} & \widetilde A_{+,-}
\\
A_{-,+} & e^{2ik_1(L+1)}\widetilde A_{-,+}
\\
e^{2ik_2(L+1)}A_{-,-} & e^{2ik_1(L+1)}\widetilde A_{-,-}
\end{pmatrix}
\begin{pmatrix}
\alpha_{+,+}\\ \beta_{+,+}
\end{pmatrix}
=0.\label{matrixdefinition}
\end{align}
Nonzero solutions require the matrix in
~\refEq{matrixdefinition} to have rank smaller than two.
Equivalently, all of its \(2\times2\) minors must vanish. For generic roots
and finite scattering ratios, this condition is equivalent to the Bethe
equations
\begin{align}
e^{2ik_1(L+1)}
&=
SR
=
\widetilde R\,\widetilde S,\nonumber
\\
\label{openboundarybethe1}
\\
e^{2ik_2(L+1)}
&=
S^{-1}\widetilde R
=
R\,\widetilde S^{-1},
\nonumber
\end{align}
where
\begin{align}
S
&:=
\frac{A(k_2,k_1,\theta)}{A(k_1,k_2,\theta)},
&
R
&:=
\frac{A(-k_1,k_2,\theta)}{A(k_2,-k_1,\theta)},
\\
\widetilde R
&:=
\frac{A(-k_2,k_1,\theta)}{A(k_1,-k_2,\theta)},
&
\widetilde S
&:=
\frac{A(-k_1,-k_2,\theta)}{A(-k_2,-k_1,\theta)}.
\label{scatteringprocesses}
\end{align}
Here, \(S\) describes exchange before reflection, while \(\widetilde S\)
describes exchange after both momenta have been reflected. The factors \(R\)
and \(\widetilde R\) are effective reflection-exchange amplitudes associated
with the channels \(k_1\to -k_1\) and \(k_2\to -k_2\), respectively. Thus, equality of the right-hand sides of \refEq{openboundarybethe1} expresses
path independence of the two physically possible exchange-reflection processes. This path-independence is equivalent to the mathematical compatibility condition of vanishing minors in \refEq{matrixdefinition}, provided that the scattering ratios according to \refEq{Betheamplitude} are finite,
\begin{align}
SR-\widetilde R\,\widetilde S
=
\frac{\mathcal C}
{
A_{+,+}A_{+,-}\widetilde A_{-,+}\widetilde A_{-,-}
}\stackrel{!}{=}0,
\end{align}
with
\begin{align}
\mathcal C
&:=
\widetilde A_{+,+}A_{+,-}A_{-,+}\widetilde A_{-,-}
-
A_{+,+}\widetilde A_{+,-}\widetilde A_{-,+}A_{-,-}.
\label{compatibilityC}
\end{align}
By this, we express the compatibility of path exchanges explicitly as 
\begin{align}
\label{compatibilityfactor}
SR-\widetilde R\,\widetilde S
&=
\frac{
64 i\,\sin\theta\,
\sin k_1\,\sin k_2\,
\left(\cos k_2-\cos k_1\right)
\mathcal Q
}{
\left(\frac{U}{J}-\frac{E}{J}\right)^3
A_{+,+}A_{+,-}\widetilde A_{-,+}\widetilde A_{-,-}
},
\end{align}
\begin{align}
\mathcal Q=
-2\left(\frac{E}{J}\right)^2
+4\frac{E}{J}
\left(\frac{U}{J}-\frac{E}{J}\right)
\sin^2\frac{\theta}{2}
+
32\sin^4\frac{\theta}{2},
\end{align}
which is proportional to $\sin(\theta)$.  Consequently, only bosons ($\theta=0$) and pseudofermions ($\theta=\pi$)
are generally compatible with the Bethe equations.
Therefore, crucially, we find that two anyons with open boundaries are generally not integrable for finite, or even vanishing, Hubbard interactions.
This is consistent with the level spacing analysis in \refFig[c]{fig:ModelSummary} and underlines that recent experiments probe genuine nonintegrable physics.

Let us finally address the structure of the Bethe roots for interacting bosons and pseudofermions \cite{Alba_2013}. 
Here, we remark that \refEq{coefficientopen1} and \refEq{coefficientopen2} are invariant under sign reflections
\(k_j\to -k_j\) and permutations \(k_1\leftrightarrow k_2\). Together with
the \(2\pi\)-periodicity in the Bethe momenta, this allows us to choose a
fundamental representative for each root, where the allowed classes are listed in ~\refEq{momentumvaluesobc}. Concretely, we consider roots with nonzero real part
in the interval \(0<\mathrm{Re}(k_j)<\pi\), while purely imaginary roots are taken with
\(\mathrm{Im}(k_j)>0\) \cite{Alba_2013}. Degenerate roots, i.e., $k_1=k_2$, are excluded from the Bethe ansatz, in analogy to the periodic case. Also. the momenta $k_j=0,\pi$ are inadmissible, which correspond to exactly vanishing wavefunctions according to \refEq{coefficientopen1} and \refEq{coefficientopen2}.

\section{Selected analytic solutions}
\label{sec:SpecialSolutions}

From our analysis emerge two analytic solutions for open boundaries and $U=E$, where the Bethe coefficients in  \refEq{Betheamplitude} diverge.
For $U\neq 0$, we find a unique doublon solution \cite{valiente2008TwoparticleStatesHubbard,boschiBoundStatesExpansion2014}, and for $U=0$ we find the exact classification of the degenerate nullspace \cite{theel2025ChirallyProtectedState}.
These states are submerged in a genuine nonintegrable background, in contrast to the additional special solutions for periodic boundary conditions that we discuss in \refApp{app:NullspaceBetheRoots}.

\subsection{The Doublon State}
\label{sec:Doublon}

Despite the nonintegrability of two anyons with open boundary conditions, we can solve the recurrence relations for the coefficients in \refEq{recurrence} directly by the ansatz
\begin{align}
c_{l,m}\propto\delta_{lm}(-e^{i\theta})^{m-1},\phantom{a}E=U
\end{align}
and find that one doublon state decouples from the rest of the spectrum 
\begin{align}
  &\ket{\Psi}_{D}=\frac{1}{\sqrt{2L}}\sum_{j=1}^L(-e^{i\theta})^{j-1}\left(\hat{a}^{\dagger}_j\right)^2\ket{0}.
  \label{eq:doublonstate}
\end{align}
This is a highly delocalized, low-entangled state, which is embedded in the scattering quasi-continuum whenever the onsite interaction satisfies \cite{valienteTwoparticleStatesHubbard2008,boschiBoundStatesExpansion2014,Kollar2012,Kollar2013,zhang2013BoundStatesOnedimensional} 
$
    |U/J| < 4
$
and is isolated in energy otherwise.
We furthermore find a nontrivial Berry phase $\gamma$ that grows linearly in $L$, as well as a giant quantum metric $g_{\theta\theta}$ \cite{Berry1977,ma2010AbelianNonAbelianQuantum}
\begin{align}
    &\gamma=\int_0^{2\pi} \braket{\Psi\vert\partial_{\theta}\Psi}d\theta=\pi(L-1)\label{eq: berrydoublon},
    \\
    &g_{\theta\theta}=\braket{\partial_{\theta}\Psi\vert\partial_{\theta}\Psi}-\vert \braket{\Psi\vert\partial_{\theta}\Psi}\vert^2
    =\frac{L^2-1}{12}\label{eq: metricdoublon}.
\end{align}
\refEq{eq: berrydoublon} implies that loops in the parameter space $\theta\in[0,2\pi)$ are nontrivial for even $L$. The quantum metric is superextensive and so the state in \refEq{eq:doublonstate} is extremely susceptible to variations of $\theta$. Yet, the $U$ direction of the connection remains perfectly flat, i.e., $g_{U \theta}=g_{U U}=g_{\theta U}=0$.

The low entanglement, how the doublon state decouples from the rest of the Hilbert space, and its thermodynamic fragility according to \refEq{eq: metricdoublon}  are signature properties of scar states \cite{serbyn2021QuantumManybodyScars,Moudgalya_2022,Serbyn2023}. 
More precisely, \refEq{eq:doublonstate} corresponds to a state in the tower of scars found in three-body constrained bosons with Hubbard interactions \cite{kaneko2024QuantumManybodyScars}, up to a local, $\theta$-dependent gauge transformation. These states are known to be robust with respect to certain correlated hopping processes, like by the density-dependent Peierls phase in the anyon-Hubbard model.
The operator that creates \refEq{eq:doublonstate} has an emergent $\mathfrak{su}(2)$ structure
\begin{align}
    &\hat{J}^-_\theta=\frac{1}{\sqrt{2}}\sum_{j=1}^L(-e^{i\theta})^{j-1}\left(\hat{a}^{\dagger}_j\right)^2,
    \\
    &\hat{J}^-_\theta=(\hat{J}^+_\theta)^{\dagger},
    \\
   & \hat{J}^z_\theta=\frac{1}{2}\sum_{j=1}^L\left(\hat{n}_j+1\right),
   \\
   &\left[\hat{J}^+_\theta,\hat{J}^-_\theta\right]=2\hat{J}^z_\theta,\quad\left[\hat{J}^z_\theta,\hat{J}^{\pm}_\theta\right]=\pm\hat{J}^{\pm}_\theta.
\end{align}
These relations exclusively hold in the two-particle sector, or when including an additional three-body hardcore constraint~\cite{kaneko2024QuantumManybodyScars}. 
We therefore in particular predict the state to be visible in the recent two-particle realization of the anyon-Hubbard model~\cite{kwan2024RealizationOnedimensionalAnyons}. 

\subsection{Nullstates \label{sec:Nullstates}}
For zero Hubbard interactions $U=0$, the anyon model with open boundary conditions possesses an emergent chiral symmetry with a degenerate nullspace for any $L$ \cite{theel2025ChirallyProtectedState}. As we have seen in section \ref{sec:integrability}, the two-particle Bethe equations are ill-defined in this case, and the system is not integrable. However, we can still find solutions to \refEq{recurrence} that are a basis of the nullspace. 
We find exactly $\lceil \frac{L}{2}\rceil$ basis states, as predicted by \refEq{eq:d0}. This basis consists of the doublon state in \refEq{eq:doublonstate} at $U=0$, as well as $\lfloor \frac{L-1}{2}\rfloor$ additional states of even chirality with coefficients \begin{align}
    &c^{(n)}_{l,m}=\delta_{l,m}\Delta^{(n)}_{l}+(1-\delta_{l,m})F^{(n)}_{l,m}.\label{eq:nullspacecoefficient}
\end{align}
Here, the diagonal term carries a $\theta$-dependence,
\begin{align}
    \Delta^{(n)}_{l}=\frac{4e^{i\theta/2}}{L}(-1)^l\sin\left(q_n\right)\sum_{p=1}^{l-1}e^{i\theta(l-1-p)}\sin\left(2q_n p\right)\label{eq:nullspacecoefficientdiagonal},
  \end{align}
whereas the off-diagonal part does not depend on $\theta$ and only has support if $l$ and $m$ are on the same sub-lattice,
\begin{align}
\label{eq:nullspacecoefficientoffdiagonal}
F^{(n)}_{l,m}=&
\frac{4}{L}(-1)^l\delta_{\mathrm{Mod}[l-m,2],0} 
\nonumber \\
&\times
\begin{cases}
             \sin\left(q_n l\right)\sin\left(q_n(m-1)\right)\phantom{a} l<m,\\
0\qquad l=m,
             \\
            \sin\left(q_n m\right)\sin\left(q_n(l-1)\right)\phantom{a} l>m. \end{cases}
\end{align}
Here, $q_n$ denotes the standing wave momentum 
  \begin{align}
    q_n=\frac{\pi n}{L},\quad\phantom{a}n=1,\dots\bigg\lfloor \frac{L-1}{2}\bigg\rfloor\label{eq:nullspacequantization},
\end{align}
that labels the nullstates
\begin{align}
& \vert\Psi_n\rangle=\frac{1}{N_n}\sum_{l,m=1}^Le^{i\theta \mathrm{sgn}(l-m)/2}c^{(n)}_{l,m}\hat{a}^{\dagger}_l\hat{a}^{\dagger}_m\vert 0,0\rangle,
\\
&  N_n=\sqrt{1+\frac{1}{2}\sum_{r=1}^L \vert\Delta^{(n)}_{r} \vert^2}.\nonumber
\end{align}
Its untypical quantization arises because the nearest-neighbor contact coefficients $c_{l,l+1}$ in \refEq{recurrence} are required to vanish at $E=U=0$. As we show in \refApp{app:NullspaceDirichlet}, this results in an $(L-1)\times(L-1)$ discrete Laplacian equation away from the defective lines, i.e., $\vert l-m\vert\geq 2$. Nevertheless, the coefficients in \refEq{eq:nullspacecoefficient} obey the correct boundary conditions for an $L$-site system, as well as bosonic exchange symmetry by construction.  Similarly to the pure doublon state in \refSec{sec:Doublon}, the nullstate basis is highly delocalized on the two sublattices, and is weakly entangled. The holonomies of the $E=U=0$ subspace and its applications to quantum computing has been addressed numerically in \citeRef{theel2025ChirallyProtectedState}. 
By this, we provide an exactly solvable subspace of the two-particle anyon-Hubbard model with open boundary conditions, which is especially relevant for holonomic state preparation \cite{theel2025ChirallyProtectedState}.

\section{Conclusion and Outlook}
\label{sec:Conclusions}

With this work, we give the complete symmetry analysis of the anyon-Hubbard model for open and periodic boundary conditions in dependence on the statistical parameter, the Hubbard interaction, the system size, and the particle number. We furthermore show that for two anyons, i.e., two particles parameter $\theta \in (0,\pi)$, only periodic boundaries are integrable.
Recent experiments in cold atoms with density-dependent Peierls phases realize the two-particle anyon-Hubbard model with open boundary conditions \cite{kwan2024RealizationOnedimensionalAnyons,bakkalihassani2026RevealingPseudoFermionizationChiral}. Resolving unique experimental signatures of this integrability crossover remains a formidable task, but would provide further insight in the nature of one-dimensional statistical transmutation. Our results stimulate further analytical investigation of boundary specific effects, as for example persistent currents \cite{poloExactResultsPersistent2020}, localization effects, and finite-size corrections in the vicinity of the hard wall boundaries \cite{batchelor20051DInteractingBose,zhang2023AnyonicBoundStates}.
Moreover, the expansion dynamics of ultra-cold atoms differs drastically between interacting and free models \cite{Bloch2013,liuAsymmetricParticleTransport2018,greschnerProbingExchangeStatistics2018}, where the presence of bound states and their weight in the evolution play a pivotal role \cite{boschiBoundStatesExpansion2014}. Its distinction between scattering and bound solutions predestines the Bethe ansatz to furthermore investigate the creation of bound states in the scattering continuum when approaching infinite system sizes \cite{Kollar2012,Kollar2013,longhiTammHubbardSurface2013,zhang2023AnyonicBoundStates}. An interesting extension of our model would be arbitrarily twisted  boundary conditions for the anyon operators, such as $\hat{a}_{L+1}=\hat{a}_1  e^{-i\Phi}$, which describes fluxes $\Phi$ piercing the ring of $L$ sites. For a nontrivial flux $\Phi$ and statistical parameter $\theta$, nontrivial Chern numbers in could occur when adiabatically manipulating degenerate states, in contrast to the currently predicted nontrivial winding numbers \cite{theel2025ChirallyProtectedState}, with intriguing potential for topological pumps.

\begin{acknowledgments}
The authors thank Frank Göhmann and Andreas Klümper for general discussions on integrability, and Joyce Kwan, Bryce Bakkani-Hassani, Perrin Segura, Yanfei Li, and Annie Zhi for discussions of the two-particle anyon-Hubbard model experiment. Furthermore, the authors thank Maxim Olshanii for software support for the level statistics, Ana Hudomal for discussions on quantum scars, and Joshua Lansky for consultation on Ramanujan's sum.   
M.B.\ and T.P.\ acknowledge
funding by the European Union (ERC, QUANTWIST, project number $101039098$). The views and opinions
expressed are however those of the authors only and do
not necessarily reflect those of the European Union
or the European Research Council, Executive Agency.
T.P.\,  N.L.H. \, and P.S.\ acknowledge support by the Cluster of Excellence “CUI: Advanced Imaging of Matter” of the Deutsche Forschungsgemeinschaft (DFG) – EXC
2056 – project ID $390715994$ at the University of Hamburg. 
Additionally,\ N.L.H. was supported by the Deutscher Akademischer Austauschdienst.
\end{acknowledgments}

\clearpage
\appendix

\section{Gauge Equivalence}
\label{app:gaugeequiv}
\setcounter{figure}{0} 
\renewcommand{\thefigure}{A\arabic{figure}}
\setcounter{equation}{0}
\renewcommand{\theequation}{A\arabic{equation}}

To understand how the bosonic model with open boundary conditions in \refEq{eq:HBGeneral} implements the exchange phase $\theta$ and to gain insight into the gauge equivalence, consider $N=2$ and $L=6$ as depicted  in \refFig[a]{fig:configspace}~\cite{kwan2024RealizationOnedimensionalAnyons, nagies2024BraidStatisticsConstructing}. In this graph, vertices represent bosonic Fock states and edges represent nonzero transition amplitudes between them, see the Hamiltonian in \refEq{eq:HBGeneral} in the main text. Most transitions have real amplitudes (depicted in red), except for the case when the rightmost particle hops left onto another particle, (depicted in blue). A non-trivial exchange process is represented by a square plaquette that includes one of the doubly-occupied vertices. The total phase acquired by traversing the Wilson loop around the plaquette, i.e., the sum of the arguments of the complex phases of the transition amplitudes, can be interpreted as a flux through that loop. From this picture we see that the exchange phase of the anyons is implemented as a dynamical phase acquired by bosons completing a closed path around a flux in configuration space. 

Consider local unitary gauge transformations $\hat{U}$ that can be expressed in the form
\begin{equation}\label{eq:ugauge}
    \hat{U} = \exp\left( i \sum_{j=1}^L f_j(\hat{n}_j) \right),
\end{equation}
where $f_j(\hat{n}_j)$ is an arbitrary real function.
 Note that the operators $\hat{Q}_\theta$ and $\hat{R}(\alpha)$ introduced in the main text are of this form.
Such unitary string operators transform the anyon operators as
\begin{equation}\label{eq:gauge}
   \hat{a}'_j = \hat{U} \hat{a}_j \hat{U}^\dag = \exp\left[i \left\{ f_j (\hat{n}_j) - f_j (\hat{n}_j+1)\right\} \right]\hat{a}_j.
\end{equation}
These operators $\hat{a}'_j$ also satisfy  the anyonic commutation relations in \refEq{eq:commrel} in the main text with the same exchange angle $\theta$. However, each of the transformations $\hat{U}$ leads to a different fractional Jordan Wigner transformation and to a different gauge-equivalent bosonic Hamiltonian that simulates the same fractional exchange statistics.

For open boundary conditions, the gauge transformed Hamiltonian $\hat{H}'(\theta, U) = \hat{Q}^\dag_\theta \hat{H}(\theta, U) \hat{Q}_\theta$
therefore realizes anyons with statistical parameter $\theta$ and is used in Refs.~\cite{kwan2024RealizationOnedimensionalAnyons, nagies2024BraidStatisticsConstructing}. Other gauges are possible, including
$\hat{H}''(\theta, U) =  \hat{Q}^\dag_{\theta/2} \hat{H}(\theta, U) \hat{Q}_{\theta/2} $
which has the convenient property that the commuting antiunitary symmetry is $\hat{P}\hat{T}$, instead of the parameter-dependent $\hat{K}_\theta$.

\begin{figure*}
     \centering
     \raisebox{-\height}{\parbox{0.33\linewidth}{%
     (a)\raggedright\\\includegraphics[width=\linewidth]{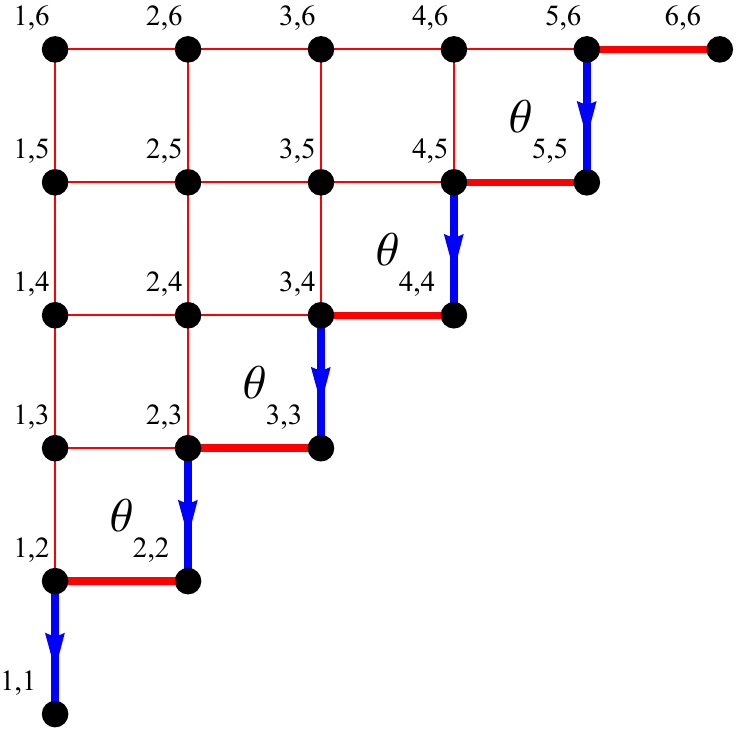}
     }%
     }%
     \raisebox{-\height}{\parbox{0.33\linewidth}
     {%
     (b)\raggedright\\\includegraphics[width=\linewidth]{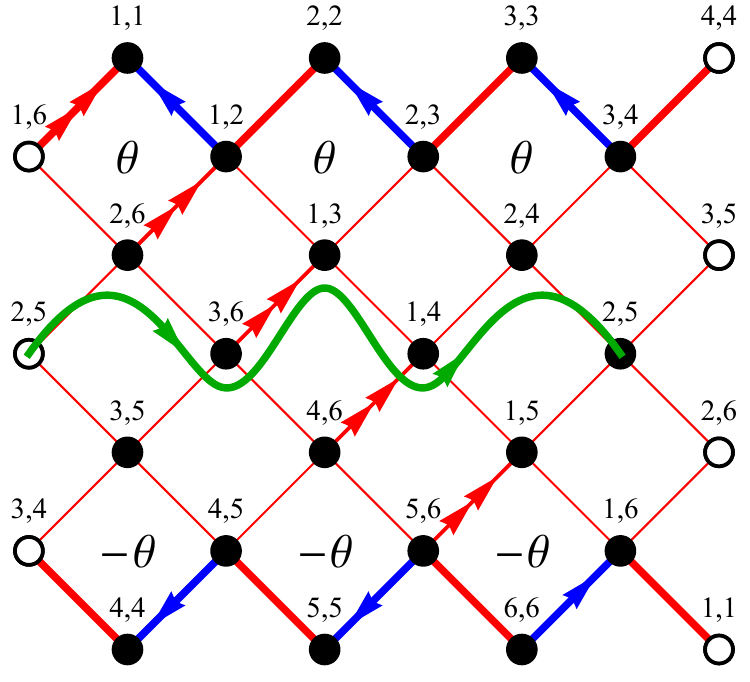}
    }%
    }%
\raisebox{-\height}{\parbox{0.33\linewidth}{%
    (c)\raggedright\\         \includegraphics[width=\linewidth]{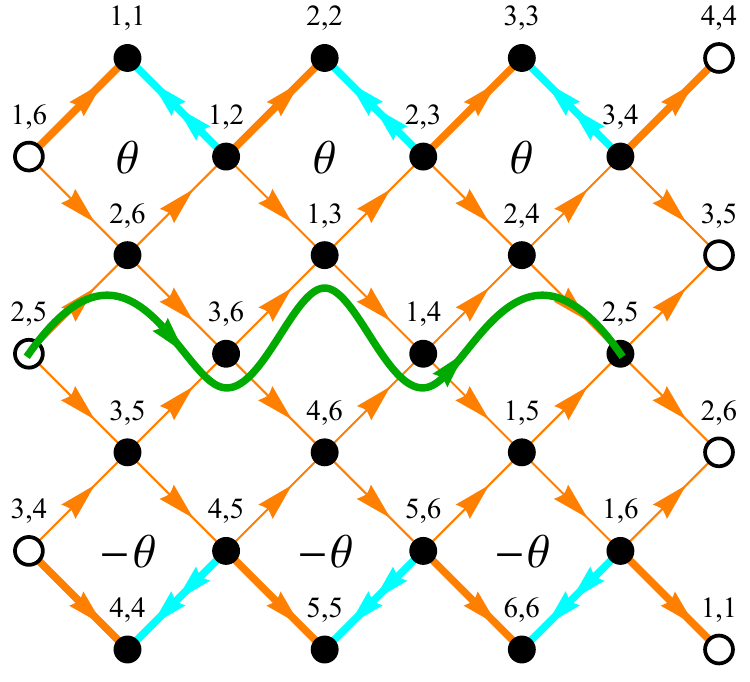}
    }%
    }%
     \caption{For both figures, black dots are labeled by numbers describing the occupation of the two bosons. Subfigure (a) depicts $\hat{H}(\theta,U)$ (\ref{eq:HBGeneral}) for $N=2$ and $L=6$ with open boundaries. Thin red lines indicate transition amplitude $-J$ between connected states; thick red lines indicate amplitude $-\sqrt{2}J$; blue arrows indicate complex amplitude $-\sqrt{2}J e^{i\theta}$ in direction of arrow. Subfigure (b) depicts $\hat{H}(\theta,U)$ (\ref{eq:HBGeneral}) for PBC in the boundary gauge using the same colors and thickness of previous graphs, with the addition of a double red arrow signifying the boundary twist $N \theta$. Open circles represent repeated sites to show that the configuration space has the topology of a M\"obius band. Subfigure (c) depicts bosonic Hamiltonian in the periodic gauge $\hat{H}^{P}(\theta,U)$ (\ref{eq:HPBC}). Thin orange arrows correspond to transition amplitudes $-e^{i \theta/ 3}J$, thick orange arrows to $-\sqrt{2} e^{i \theta/ 3}J$, and thick cyan arrows to $-\sqrt{2} e^{2 i \theta/ 3}J$. For all subfigures, the flux through any Wilson loop in configuration space that is nonzero is labeled. For subfigures (b) and (c), the green arrow represents a loop that a complete cycle around the M\"obius band and encircles a phase $N \theta$.  }
     \label{fig:configspace}
\end{figure*}

When the anyons satisfy periodic boundary conditions, there is an additional complication. The fractional Jordan-Wigner transformation in \refEq{eq:jordanwigner} implements a non-local phase for the operator at site $j$ that depends on the total occupation number for sites $k<j$. For periodic boundaries, the choice of site numbering, and therefore the notion $k<j$, is an additional arbitrary gauge choice. We require the relation $\hat{a}_{L+1}=\hat{a}_1$ to be consistent with the deformed commutation relations in \refEq{eq:commrel} in the main text. These counting ambiguities of the Jordan-Wigner transformation on a circle are known from the fermionization of cyclic spin chains in one dimension and for fermions are resolved by adding a `twist' to the boundary conditions in the bosonic Hamiltonian~\cite{patuCorrelationFunctionsOnedimensional2007,liebTwoSolubleModels1961,capelNoteDifferenceAcyclic1975}.
One way to achieve this uses the fractional Jordan-Wigner transformation in \refEq{eq:jordanwigner}, which results in the bosonic Hamiltonian $H(\theta,U)$ in \refEq{eq:HBGeneral}. 
This boundary gauge is depicted in \refFig[b]{fig:configspace}. Alternatively, one can add a fractional twist to \refEq{eq:jordanwigner} to construct the alternate fractional Jordan-Wigner transformation in \refEq{eq:jordanwignerPBC1} and the corresponding bosonic Hamiltonian $H^P(\theta,U)$ in  \refEq{eq:HPBC} in the main text.
This has been called the periodic gauge for twisted boundary conditions~\cite{linTopologicalInvariantsInteracting2023} and is depicted in \refFig[c]{fig:configspace}. 
Defining $\Theta= \theta N/L$, the gauge transformation $\hat{W}(\Theta)$ (cf.~Eq.~\ref{eq: W gauge transform}) converts from boundary gauge to periodic gauge by \begin{equation}
    \hat{W}(\Theta) \hat{H} (\theta,U) \hat{W}^\dag(\Theta) = \hat{H}^P(\theta,U).
\end{equation}
In either gauge, periodic boundaries for the anyons imply twisted boundary conditions for the bosonic Hamiltonian and a complete Wilson loop around the ring has a nonzero flux through it. The configuration space graph can be understood as embedded in a M\"obius band and in Figs.~\refFig[b]{fig:configspace} and \refFig[c]{fig:configspace} the green arrow traces a Wilson loop around its circumference. The phase acquired by a Wilson loop around the ring is $N\theta$ in both gauges; that is the phase acquired by moving $N$ particles around the ring with flux $\theta$.
For periodic boundaries, the gauge choice determined by \refEq{eq:jordanwignerPBC1} leads to a Hamiltonian in this gauge having $2 M \pi$ periodicity in the statistical angle, where $M = L/\mathrm{gcd}(N,L)$.  This is different to the Hamiltonian for open boundaries in \refEq{eq:HBGeneral}, which has a $2\pi$ periodicity in $\theta$ like the deformed commutation relations.

\section{Details on symmetries and multiplicities of the anyon-Hubbard model}\label{app:symm}
\setcounter{figure}{0} 
\renewcommand{\thefigure}{B\arabic{figure}}
\setcounter{equation}{0}
\renewcommand{\theequation}{B\arabic{equation}}
The representation of the point group $G$ on the Hilbert space $\mathcal{H}$ can be represented by the collection of unitary operators $\{\hat{U}(g) \ | \ g\in G\}$ such that $[\hat{H},\hat{U}(g)]=0 , \ g \in G$. The $D$ dimensional representations of the group elements $g$ can be block-diagonalized according to the set of irreducible representations (irreps) $M^G$ of the group $G$. Each irrep $\mu\in M^G$ may appear with multiplicity $m^{G}_\mu$ in the representation $\hat{U}(G)$. We calculate the multiplicities via the character orthogonality theorem \cite{hamermeshGroupTheoryIts1989}:
\begin{equation}
    m^G_\mu=\frac{1}{|G|}\sum_{g\in G}\chi_\mu^*(g)\chi_{_\mathcal{H}}(g)
\end{equation}
where in general $\chi_{_{\mathcal{H}}}(g)=\text{Tr}(\hat{U}(g))$.

\subsection{Open boundary conditions symmetry multiplicity}

\begin{table}[ht]
\bgroup
\renewcommand{\arraystretch}{1.5}

\begin{tabular}{|c|c|cc|}
\hline
${\mathbb{Z}}_2$  & $I$ & \multicolumn{2}{c|}{$s$} \\ \hline
$\chi_+$ & 1 & \multicolumn{2}{c|}{1} \\ \hline
$\chi_-$ & 1 & \multicolumn{2}{c|}{-1} \\ \hline
$\chi_{\mathcal{H}}$ & $\text{Tr}(\hat{\mathrm{I}})=D$ & \multicolumn{1}{c|}{$\text{Tr}(\hat{P})=d_0$} & $\text{Tr}(\hat{P}_{\pi})=d_0- 2o^{\pi}$ \\ \hline
\end{tabular}
\egroup
\caption{The relevant characters for the two-element group ${\mathbb{Z}}_2$. At $\theta =0$ the representation $U({\mathbb{Z}}_2)$ is $U(I)=\hat{I}_{D}, \ U(s)=\hat{P}$. At $\theta =\pi$, the representation of $U(\mathbb{Z}_2)$ is $U(I)=\hat{I}_{D}, \ U(s)=\hat{Q}_{\pi}\hat{P}=\hat{P}_{\pi}$.}
\label{tab: c2 character table}
\end{table}
In this section, we calculate the multiplicity of parity eigenstates for the bosonic ($\theta=0$) and fermionic ($\theta=\pi$) cases via character theory analysis of the point group ${\mathbb{Z}}_2$ realized by the identity $\hat{\mathrm{I}}$ and $\hat{P}$ (or $\hat{\mathrm{I}}$ and $\hat{P}_{\pi}$ \refEqShort{def: P hat pi}, respectively). The trace of a permutation matrix such as $\hat{P}$ is equal to the number of states invariant under the symmetry. Therefore, $\text{Tr}(\hat{P})=d_0$. The number of palindromic number states and character orthogonality provides
\begin{equation}
    m^{\hat{P}}_{\pm}=\frac{1}{2}\left(\text{Tr}(\hat{\mathrm{I}})\pm \text{Tr}(\hat{P})\right) =\frac{1}{2}\left(D\pm d_0\right).
    \label{apeq: parity multiplicity theta zero}
\end{equation}
The multiplicities for the chiral operator $\hat{S}$ are also given by \refEqShort{apeq: parity multiplicity theta zero} because $\text{Tr}(\hat{S})=d_0$ \cite{theel2025ChirallyProtectedState}.

At $\theta=\pi$, using results summarized in Sec.~\ref{sec:openbc}, we calculate the parity multiplicity at $\theta=\pi$:
\begin{equation}
    m^{\hat{P}_{\pi}}_{\pm}=\frac{\text{Tr}(\hat{\mathrm{I}})\pm \text{Tr}(\hat{P}_{\pi})}{2},
\end{equation}
which simplifies to
\begin{equation}
\label{apeq:multiplicityobcthetapi}
    m^{\hat{P}_\pi}_\pm = \left\{ \begin{array}{l} m^{\hat{P}}_\pm\ \mbox{for $L$ even}\\
    m^{\hat{P}}_\pm \mp o^\pi\ \mbox{for $L$ odd}
    \end{array}\right.,
\end{equation}
because the number of co-located pairs is even for even $L$. 

\subsection{Cyclic Multiplicity}
\begin{table}[ht]
\bgroup
\renewcommand{\arraystretch}{1.5}
\begin{tabular}{|c|cccccc|}
\hline
${\mathbb{Z}}_L$ & \multicolumn{1}{c|}{$I$} & \multicolumn{1}{c|}{$r$} & \multicolumn{1}{c|}{$r^2$} & \multicolumn{1}{c|}{$\dots$} & \multicolumn{1}{c|}{$r^{L-2}$} & $r^{L-1}$ \\ \hline
$\chi_{_1}$ & \multicolumn{1}{c|}{$1$} & \multicolumn{1}{c|}{$\omega$} & \multicolumn{1}{c|}{$\omega^{2}$} & \multicolumn{1}{c|}{$\dots$} & \multicolumn{1}{c|}{$\omega^{(L-2)}$} & $\omega^{(L-1)}$ \\ \hline
$\vdots$ & \multicolumn{6}{c|}{$\vdots$} \\ \hline
$\chi_q$ & \multicolumn{1}{c|}{$1$} & \multicolumn{1}{c|}{$\omega^q$} & \multicolumn{1}{c|}{$\omega^{2q}$} & \multicolumn{1}{c|}{$\dots$} & \multicolumn{1}{c|}{$\omega^{q(L-2)}$} & $\omega^{q(L-1)}$ \\ \hline
$\chi_{_\mathcal{H}}$ & \multicolumn{1}{c|}{$\text{Tr}(\hat{I})$} & \multicolumn{1}{c|}{$\text{Tr}(\hat{{R}})$} & \multicolumn{1}{c|}{$\text{Tr}(\hat{{R}}^2)$} & \multicolumn{1}{c|}{$\dots$} & \multicolumn{1}{c|}{$\text{Tr}(\hat{{R}}^{L-2})$} & $\text{Tr}(\hat{{R}}^{L-1})$ \\ \hline
\end{tabular}
\egroup
\caption{The relevant characters for the order $L$ cyclic group ${\mathbb{Z}}_L$. ${\mathbb{Z}}_L$ has $L$ irreducible representations labeled $\mu_q$, $q=0,\dots,L-1$, with character $\chi_q(r^k)=\omega^{qk}=\exp(2\pi i qk/L)$. The reducible representation $\hat{U}(\mathbb{Z}_L)$ on $\mathcal{H}$ has character $\chi_{_\mathcal{H}}(r^k)=\text{Tr}(\hat{{R}}^k)$.}
\label{tab:cyclicGroupCharacterTable}
\end{table}
We calculate the multiplicity $m^{{\mathbb{Z}}_L}_q$ of eigenstates with momentum $2\pi q/L$ via character orthogonality, using characters from \refTab{tab:cyclicGroupCharacterTable}:
\begin{equation}
    m^{{\mathbb{Z}}_L}_q=\frac{1}{L}\sum_{k=0}^{L-1}\chi^*_q(r^k)\text{Tr}(\hat{{R}}^k).
    \label{apeq: cyclicmultchiform}
\end{equation}
We calculate $\tr(\hat{{R}}^k)$ via the Cyclic Sieving Phenomenon (CSP)~\cite{REINER200417} to be
\begin{equation}
    \text{Tr}(\hat{{R}}^k)=\binom{N+L-1}{N}_{\omega^k}=\frac{[N+L-1]!_{\omega^k}}{[L-1]!_{\omega^k}[N]!_{\omega^k}} \ ,
\end{equation}
the binomial $q$-extension of $D(N,L)$.
Above $[n]!_{\omega}=[n]_{\omega}[n-1]_{\omega}\dots[2]_{\omega}[1]_{\omega}$ and $[n]_{\omega}=1+\omega+\dots+\omega^{n-1}$. The CSP simplifies (\ref{apeq: cyclicmultchiform}) (outlined in~\cite{REINER200417}) to
\begin{equation}
    m^{{\mathbb{Z}}_L}_q = \frac{1}{L} \sum_{h|(N,L)} C_{h}(q)D\left(\frac{N}{h},\frac{L}{h}\right)
    \label{apeq: cyclic multiplicity in Ramanujan form}
\end{equation}
with $h$ indexing over all divisors of the greatest common divisor of $N$ and $L$, ($\text{gcd}(N,L)=(N,L)$), and $C_{h}(q)$ is Ramanujan's sum 
\begin{equation}
    C_q(n)=\sum_{\substack{
p\in[1,q] \\
(p,q)=1\\
}} e^{2\pi i \frac{p}{q}n}
\label{eq: ramanujans_sum}  .
\end{equation}
\subsection{Dihedral Multiplicities at \texorpdfstring{$\theta=0$}{thetaEqualZero}}
\begin{table}[ht]
\bgroup
\renewcommand{\arraystretch}{1.5}
\begin{tabular}{|ccccc|}
\hline
\multicolumn{5}{|c|}{$D_L \; \langle s,r \mid r^L=s^2=1,\, sr^n s^{-1}=r^{-n} \rangle$} \\ \hline
\multicolumn{1}{|c|}{$D_L$} & \multicolumn{1}{c|}{$I$} & \multicolumn{1}{c|}{$r^k$} & \multicolumn{1}{c|}{$sr^{2k}$} & $sr^{2k+1}$ \\ \hline
\multicolumn{1}{|c|}{$\chi_{0,+}$} & \multicolumn{1}{c|}{1} & \multicolumn{1}{c|}{1} & \multicolumn{1}{c|}{1} & 1 \\ \hline
\multicolumn{1}{|c|}{$\chi_{0,-}$} & \multicolumn{1}{c|}{1} & \multicolumn{1}{c|}{1} & \multicolumn{1}{c|}{-1} & -1 \\ \hline
\multicolumn{1}{|c|}{$\chi_{L/2,+}$} & \multicolumn{1}{c|}{1} & \multicolumn{1}{c|}{$(-1)^k$} & \multicolumn{1}{c|}{1} & -1 \\ \hline
\multicolumn{1}{|c|}{$\chi_{L/2,-}$} & \multicolumn{1}{c|}{1} & \multicolumn{1}{c|}{$(-1)^k$} & \multicolumn{1}{c|}{-1} & 1 \\ \hline
\multicolumn{1}{|c|}{$\chi_{q,-q}$} & \multicolumn{1}{c|}{2} & \multicolumn{1}{c|}{$\omega^{qk}+\omega^{-qk}$} & \multicolumn{1}{c|}{0} & 0 \\ \hline
\multicolumn{1}{|c|}{$\chi_{\mathcal{H}}$} & \multicolumn{1}{c|}{Tr$(\hat{I})$} & \multicolumn{1}{c|}{Tr$(\hat{R}^k)$} & \multicolumn{1}{c|}{Tr$(\hat{P})$} & Tr$(\hat{P}\hat{R})$ \\ \hline
\end{tabular}
\egroup
\caption{The first relevant character values for calculating dihedral multiplicities at $\theta=0$ are the 1D irreducible characters $\chi_{0,+}$, $\chi_{0,-}$, $\chi_{L/2,+}$, and $\chi_{L/2,-}$. The parity of $L$ is significant and for odd $L$ only the trivial and sign characters $\chi_{0,+}$ and $\chi_{0,-}$ are relevant, whereas all four 1D characters are relevant when $L$ is even. The character structure for 2D irreps is the same regardless of the parity of $L$ and ranges from $q=1,\dots,\ \lceil L/2 \rceil -1$. The last character is that of the reducible representation furnished by the symmetry operators commuting with $\hat{H}$.}
\label{tab:evenLDihedralCharacterTable}
\end{table}
At $\theta=0$, the rotation and reflection operators $\hat{{R}}$ and $\hat{P}$ realize a representation on $\HS$ of the dihedral group $D_L$ with the presentation
$$\left\langle\hat{P},\hat{R} \ | \ \hat{{R}}^L=\hat{P}^2=\hat{\mathrm{I}}, \ \hat{P}\hat{{R}}\hat{P}^{\dagger}=\hat{{R}}^{\dagger} \right\rangle.$$
We calculate the multiplicities of each dihedral irrep in the permutation representation taking into account the different one-dimensional irrep structures depending on the parity of $L$. The 1D irrep structure of even L elicits two pairs of irreps $\mu_{0,+}$/$\mu_{0,-}$ (the trivial and sign representations) and $\mu_{L/2,+}$/$\mu_{L/2,-}$ (the alternating and sign-alternating representations). The two-dimensional irreducible representations of $D_L$ $\mu_{q,q'}$ are labeled by $q=1,\dots,\ \lceil L/2 \rceil -1$. These 2d irreps can be interpreted as coupling together one-dimensional irreps of the cyclic subgroup ${\mathbb{Z}}_L$.
\subsubsection{One-Dimensional Dihedral Irrep Multiplicities at \texorpdfstring{$\theta=0$}{thetaEqualZero}}
We can combine the first two 1D irreps from \refTab{tab:evenLDihedralCharacterTable} into the character
\begin{equation}
    \chi_{0,\pm}(r^m)=1, \chi_{0,\pm}(sr^m)=\pm1.
    \label{eq: trivial and sign character formula}
\end{equation}
The multiplicities of these irreps in $\HS$ can be rewritten in terms of the reflection and parity cosets
\begin{equation}
    m^{D_L}_{0, \pm}=\frac{1}{2L}\left(\sum_{k=1}^L\tr(\hat{{R}}^k)\pm \sum_{k=1}^L \tr(\hat{P}\hat{{R}}^k)\right).
\end{equation}
For odd L, $\text{Tr}(\hat{P})=\text{Tr}(\hat{P}\hat{{R}})$, simplifying the above to
\begin{equation}
    m^{D_L}_{0,\pm}=\frac{1}{2}(m^{{\mathbb{Z}}_L}_0\pm d_0).
    \label{apeq: odd L triv and sign multiplicity}
\end{equation}
For even L, we write the multiplicity $m^{D_L}_{0, \pm}$ of the trivial/sign irrep in $\HS$ in terms of the reflection and parity cosets as in the case of odd $L$. However, we must separate the parity coset sum into a sum over the even terms $\tr(\hat{P}\hat{{R}}^{2k})$ and odd terms $\tr(\hat{P}\hat{{R}}^{2k+1})$ to account for the trace inequivalency of $\hat{P}$ and $\hat{P}\hat{{R}}$ for even $L$:
\begin{equation}
    \frac{1}{2}\left(m^{{\mathbb{Z}}_L}_{0}\pm\frac{1}{2}\tr(\hat{P}\hat{{R}})\pm\frac{1}{2}\tr(\hat{P})\right)=\frac{m^{{\mathbb{Z}}_L}_{0}}{2}\pm\frac{d_0+d_1}{4}.
\end{equation}
Here $d_1= \text{Tr}(\hat{P}\hat{R})$, and 
\begin{multline}
    \text{Tr}(\hat{P}\hat{R})=\sum_{k=0}^{\lfloor N/2\rfloor}D\left(k,\frac{L}{2}-1\right)D(N-2k,2)\\ =\binom{L/2+\lfloor N/2\rfloor}{L/2}+\binom{L/2+\lfloor(N+1)/2\rfloor-1}{L/2}
\end{multline}
is found combinatorically by taking into account the many possible arrangements of particles on the two reflection-rotation invariant vertices.
Therefore the trivial/sign multiplicity for even L is
\begin{equation}
    m^{D_L}_{0,\pm}=\frac{m^{{\mathbb{Z}}_L}_{0}}{2}\pm\frac{d_0+d_1}{4} \ .
    \label{eq: bosonicTrivandSignMultiplicityEvenL}
\end{equation}
For the one-dimensional multiplicities of an even L dihedral group, we combine the alternating and sign-alternating characters in \refTab{tab:evenLDihedralCharacterTable} into the character
\begin{equation}
    \chi_{_{L/2},\pm}(r^m)=(-1)^m , \ \chi_{_{L/2},\pm}(sr^m)=\pm(-1)^{m} \ .
    \label{eq: alternating and sign-alternating character formula}
\end{equation}
After simplification of the cyclic coset and selecting trace class representatives for $\hat{P}$ and $\hat{P}\hat{{R}}$, we find that
\begin{equation}
    m^{D_L}_{L/2,\pm}=\frac{m^{{\mathbb{Z}}_L}_{L/2}}{2}\pm \frac{d_0}{4}\mp \frac{d_1}{4}
    \label{eq: bosonic even L alt and alt sign multiplicity}
\end{equation}
determines the alternating/sign-alternating multiplicities for even L. The formulae for the trivial/sign and alternating/sign-alternating multiplicities are similar. Here, we combine them into a single formula for the sake of brevity in \refTab{tab:multi}.
\begin{equation}
    m^{D_L}_{q,\pm} = \left\{ \begin{array}{cl}
m^{D_L}_{0,\pm} &  \ q=0, \\
m^{D_L}_{L/2,\pm} &  \ q=L/2,

\end{array} \right.
\label{eq: combined bosonic even L 1D multiplicities}
\end{equation}
for even $L$.
\subsubsection{Two-Dimensional Dihedral Irrep Multiplicities at \texorpdfstring{$\theta=0$}{thetaEqualZero}}
We calculate the multiplicities two-dimensional irreps $\mu_{q,-q}$ of $D_L$ with characters $\chi_{q,-q}(D_L)$, $q=1,\dots,  \lceil L/2 \rceil -1$, given in \refTab{tab:evenLDihedralCharacterTable}. The multiplicity $m^{D_L}_{q,q'}$ is
\begin{gather}
    m^{D_L}_{q,q'}
    =\frac{1}{2L}\sum_{k=1}^L (\omega^{qk}+\omega^{-qk})\tr(\hat{{R}}^k),
\end{gather}
which, after distributing $\tr(\hat{{R}}^k)$ and separating into two summations, simplifies to
\begin{equation}
    m^{D_L}_{q,q'}=\frac{m^{{\mathbb{Z}}_L}_q+m^{{\mathbb{Z}}_L}_{-q}}{2} \ .
    \label{eq: bosonic 2d dihedral irrep formula}
\end{equation}
In the notation for the 2d irrep multiplicity $m^{D_L}_{q,q'}$, $q$ and $q'$ represent the two cyclic subgroup irreps $\mu_q$ and $\mu_{q'}$ that are coupled by the semidirect product structure of $D_L$. In the case of $\theta=0$, we can see that $q$ couples with $-q$, i.e., $q'=-q$.
\subsection{Dihedral Multiplicities at \texorpdfstring{$\theta=\pi$}{thetaEqualPi}}
Let us first provide auxiliary relations that we found necessary to analyze periodic boundary conditions. Concerning the gauge transformation  $\hat{{W}}(\alpha)$ for general $\alpha$ \refEqShort{eq: W gauge transform}, we find
\begin{eqnarray}
\label{eq: conj relations}
    \hat{{W}}(\alpha) \hat{b}_j \hat{{W}}(\alpha)^\dag &= &\exp(-i\alpha j)\hat{b}_j,\\
    \hat{P} \hat{{W}}(\alpha) \hat{P}^\dag &= & \exp(i\alpha (L+1) N)\hat{{W}}^\dag(\alpha),\nonumber\\
    \hat{{R}} \hat{{W}}(\alpha) \hat{{R}}^\dag &= & \exp(i\alpha N)\exp( -i\alpha L \hat{n}_L)\hat{{W}}(\alpha).\nonumber
\end{eqnarray}
The gauge transformation $\hat{Q}_\theta$ commutes with $\hat{P}$ and $\hat{{R}}$.
At $\theta=\pi$, the dihedral representation on $\HS$ is realized by the algebraic relations of $\bar{P}_{\pi}$ \refEqShort{def: gauge transformed P pi} and $\bar{R}_{\pi}$ \refEqShort{eq:twistC}, with presentation 
$$\left\langle\bar{P}_{\pi},\bar{{R}}_{\pi} \ | \ \bar{{R}}_{\pi}^L=(-1)^N \mathrm{I}, \ \bar{P}_{\pi}^2=\mathrm{I}, \ \bar{P}_{\pi}\bar{{R}}_{\pi}\bar{P}_{\pi}^{\dagger}=\bar{{R}}_{\pi}^{\dagger} \right\rangle.$$
For the case of even $N$, the above dihedral presentation is of order $2L$. The only significant difference in this case from the one of $\theta=0$ is that the 2D irreps couple different 1D cyclic irreps. For the case of odd $N$, the above dihedral presentation is of order $4L$ as a consequence of $\bar{{R}}_{\pi}^{2L}=\mathrm{I}$. This projective representation of $D_L$, $\bar{D}_L$, is therefore realized by the double cover of $D_L$ which is the linear representation $D_{2L}$. This case also couples 1D cyclic irreps in 2D dihedral irreps differently than how coupling occurs for $\theta=0$.

Because the operators at $\theta=\pi$ alter the 1D cyclic irrep coupling,  we use the label $\bar{q}$ to where the 1D irrep $q$ is mapped to by the rephasing of the rotation operator  $\hat{{R}}$ that occurs in the definition of $\bar{{R}}_{\pi}$ (\ref{eq:twistC}), i.e.~the mapping (\ref{eq:qmap}). Therefore, for $N$ even the notation $m^{D_L}_{\bar{q},\bar{q}'}$ describes multiplicity of the 2D irrep that couples the mapped 1D cyclic irrep $\bar{q}$ to the mapped irrep $\bar{q}'$. Similarly, the notation $m^{D_L}_{\bar{0},\pm}$ describes the multiplicity of the trivial and sign irreps after mapping induced by $\bar{{R}}_{\pi}$ and $\bar{P}_{\pi}$ and $m^{D_L}_{\overline{L/2},\pm}$ describes the mapping for the alternating and sign-alternating irreps.

\begin{table}[H]

\bgroup
\renewcommand{\arraystretch}{1.5}
\begin{tabular}{|c|cccc|}
\hline
N odd & \multicolumn{4}{c|}{$D_{2L} \sim  \left\langle s,r | r^{2L}=s^2=1, sr^k s^{-1}=r^{-k}\right\rangle$} \\ \hline
$D_{2L}$ & \multicolumn{1}{c|}{$\mathrm{I}$} & \multicolumn{1}{c|}{$r^k$} & \multicolumn{1}{c|}{$sr^{2k}$} & $sr^{2k+1}$ \\ \hline
$\chi_{0,+}$ & \multicolumn{1}{c|}{1} & \multicolumn{1}{c|}{1} & \multicolumn{1}{c|}{1} & 1 \\ \hline
$\chi_{0,-}$ & \multicolumn{1}{c|}{1} & \multicolumn{1}{c|}{1} & \multicolumn{1}{c|}{-1} & -1 \\ \hline
$\chi_{L,+}$ & \multicolumn{1}{c|}{1} & \multicolumn{1}{c|}{$(-1)^k$} & \multicolumn{1}{c|}{1} & -1 \\ \hline
$\chi_{L,-}$ & \multicolumn{1}{c|}{1} & \multicolumn{1}{c|}{$(-1)^k$} & \multicolumn{1}{c|}{-1} & 1 \\ \hline
$\chi_{q,-q}$ & \multicolumn{1}{c|}{2} & \multicolumn{1}{c|}{$\omega^{qk/2}+\omega^{-qk/2}$} & \multicolumn{1}{c|}{0} & 0 \\ \hline
$\chi_{_\mathcal{H}}$ & \multicolumn{1}{c|}{$\text{Tr}(\hat{\mathrm{I}})$} & \multicolumn{1}{c|}{$\text{Tr}(\bar{{R}}_{\pi}^k)$} & \multicolumn{1}{c|}{$\text{Tr}(\bar{P}_{\pi}\bar{{R}}_{\pi}^{2k})$} & $\text{Tr}(\bar{P}_{\pi}\bar{{R}}_{\pi}^{2k+1})$\\ \hline
\end{tabular}
\egroup
\caption{At $\theta=\pi$, $\bar{P}_{\pi}$ and $\bar{R}_{\pi}$ furnish a representation of the dihedral group $D_{2L}$ for odd $N$. Because this is the dihedral group $D_{2L}$, the parity of $L$ is irrelevant and the four 1D irreps are relevant for all $L$.}
\label{tab: theta equals pi odd N dihedral character table}
\end{table}
\subsubsection{Odd \texorpdfstring{$N$}{N} One-Dimensional Multiplicites}
In the case of odd $N$, we require the double covering of the dihedral group, $D_{2L}$. Because we are calculating multiplicities over $|D_{2L}|=4L$, the derivations for even and odd $L$ are similar but distinctions will be made as needed.

First, considering the multiplicity of the trivial/sign irrep in the representation of the double cover $D_{2L}$ on $\HS$, we obtain
\begin{gather}
    \label{eq: N odd 1D mult chi form}
    m^{D_{2L}}_{\bar{0},\pm}=\frac{1}{4L}\left(\sum_{k=1}^{2L} \tr(\hat{{R}}_{\pi}^k)\pm\tr(\bar{P}_{\pi}\hat{{R}}_{\pi}^k)\right).
\end{gather}
We split both the cyclic and parity cosets into pairs of sums, first from $k=1,\dots,L$ and then $k=L+1,\dots,2L$, then use $\hat{{R}}_{\pi}^{L+k}=(-1)^N\hat{{R}}_{\pi}^k$ to simplify (\refEq{eq: N odd 1D mult chi form}) to:
\begin{equation}
    m^{D_{2L}}_{\bar{0},\pm}=\frac{(1+(-1)^N)}{4L}\left(\sum_{k=1}^{L} \tr(\bar{{R}}_{\pi}^k)\pm\tr(\bar{P}_{\pi}\bar{{R}}_{\pi}^k)\right)=0
\end{equation}
for odd $N$, i.e., both even and odd $L$ have trivial and sign multiplicities equal to zero. 

The parity of $L$ is relevant for calculating the multiplicities of the alternating and sign-alternating irreps in the representation of the double cover on $\HS$. In the notation below, recall that in the double-covering case, $\overline{L/2}=L$. Consider the multiplicity of the alternating/sign-alternating irrep:
\begin{gather}
    m^{D_{2L}}_{\overline{L/2},\pm}=\frac{1}{4L}\sum_{k=1}^{2L} (-1)^k\left(\tr(\bar{{R}}_{\pi}^k)\pm\tr(\bar{P}_{\pi}\bar{{R}}_{\pi}^k)\right).
    \label{eq: N odd AltandAltSign First Def}
\end{gather}
For the case of $L$ even, we re-index both coset sums and extract the phases from $\bar{{R}}_{\pi}$. We arrive at
\begin{eqnarray}
    m^{D_{2L}}_{\overline{L/2},\pm}&=&\frac{(1+(-1)^N)}{4L}  \label{eq: N odd AltandSignAlt XNOR form} \\
   &&  \times \sum_{k=1}^L \omega^{\frac{k}{2}(L-N^2)}\left(\tr (\hat{{R}}^k) \pm \tr(\bar{P}_{\pi}\hat{{R}}^k)\right).\nonumber
\end{eqnarray}
For $N$ odd, this formula is zero. Changing $\theta$ from $0 \text{ to } \pi$ maps the two 1D irreps $\mu^{D_L}_{0,\pm}$ of the odd $L$ dihedral group into the alternating and sign-alternating 1D irreps $\mu^{D_{2L}}_{\overline{L/2},\pm}$ of the covering group $D_{2L}$, preserving the expected $\mathbb{Z}_2$ structure of the abelianization of $D_L$. 

For odd $N$ and $L$, we calculate the multiplicity of the alternating and sign-alternating irreps $\mu^{D_{2L}}_{L/2,\pm}$ similarly to the case of even $L$ by simplifying the cyclic coset to the form
\begin{multline}
    m^{D_{2L}}_{\overline{L/2},\pm}=\frac{1}{|D_{2L}|}\sum_{g\in D_{2L}}\chi^*_{L/2,\pm}(g)\chi_{_\mathcal{H}}(g)\\
    =\frac{1}{4L}\sum_{k=1}^L (-1)^k\left(\tr(\hat{{R}}^k)\omega^{-kN^2/2}\pm \tr(\bar{P}_{\pi}\bar{{R}}_{\pi}^k)\right)
\end{multline}
to arrive at
\begin{equation}
\label{eq: odd N theta pi dihedral alt-sign mult intermediate step}
    m^{D_{2L}}_{\overline{L/2},\pm}=\frac{m^{{\mathbb{Z}}_L}_{(N^2-L)/2}}{2}\pm \sum_{k=1}^{2L}\tr(\bar{P}_{\pi}\bar{{R}}_{\pi}^k)(-1)^k
\end{equation}
after rewriting $(-1)^k=\omega^{kL/2}$ and using that $N$ and $L$ are odd. Next we focus on the parity coset. Although our calculation concerns odd $L$, whose dihedral structure places $\{sr^{2m}\}$ and $\{sr^{2m-1}\}$ in the same trace class, because we are working in the double cover $D_{2L}$ we must treat the operators $\left\{\bar{P}_{\pi}\bar{{R}}_{\pi}^{2k}\right\}$ and $\left\{\bar{P}_{\pi}\bar{{R}}_{\pi}^{2k-1}\right\}$ as trace inequivalent. We then later show that they are in fact equivalent. Using our simplification of the cyclic coset \refEqShort{eq: odd N theta pi dihedral alt-sign mult intermediate step} and separating the parity coset into the two trace classes, we arrive at
\begin{equation}
    m^{D_{2L}}_{\overline{L/2},\pm}=\frac{m^{{\mathbb{Z}}_L}_{(N^2-L)/2}}{2}\pm\frac{1}{4}\left( \tr(\bar{P}_{\pi})-\tr(\bar{P}_{\pi}\bar{{R}}_{\pi})\right).
    \label{eq: N odd 1D simplified step}
\end{equation}
We simplify the traces via the conjugation relations \refEq{eq: conj relations} to find that for $N,L$ odd, $\tr(\bar{P}_{\pi})=-\tr(\hat{P}_{\pi})$ and $\tr(\bar{P}_{\pi}\bar{{R}}_{\pi})=-\tr\left((-1)^{\hat{n}_1}\hat{P}_{\pi}\hat{{R}}\right)$. Further simplification of the latter trace readily follows from writing the trace as an inner product over the occupation basis
\begin{equation}
    \tr\left((-1)^{\hat{n}_1}\hat{Q}_{\pi}\hat{P}\hat{{R}}\right)=\sum_{\ket{n}\in\HS}(-1)^{\nu(n)}\bra{n}(-1)^{\hat{n}_1}\hat{P}\hat{{R}}\ket{n}.
\end{equation}
The inner product is only nonzero on reflection-rotation eigenstates for odd $L$. While the reflection $\hat{P}$ reflects about an edge and the vertex $j=(L+1)/2$, the reflection-rotation $\hat{P}\hat{{R}}$ reflects about an edge and the vertex $j=1$. We infer combinatorially that a reflection eigenstate for $N,L$ odd has an odd number of particles on the vertex about which the system is reflected. Therefore $\bra{n}(-1)^{\hat{n}_1}\hat{P}\hat{{R}}\ket{n}=-\bra{n}\hat{P}\hat{{R}}\ket{n}$ and $\tr(\bar{P}_{\pi}\bar{{R}}_{\pi})=-\tr\left((-1)^{\hat{n}_1}\hat{Q}_{\pi}\hat{P}\hat{{R}}\right)=\tr(\hat{P}_{\pi}\hat{{R}})$. Equation (\ref{eq: N odd 1D simplified step}) then simplifies to
\begin{equation}
    m^{D_{2L}}_{\overline{L/2},\pm}=\frac{m^{{\mathbb{Z}}_L}_{(N^2-L)/2}\mp(d_0-2 \ o^{\pi})}{2}.
    \label{eq: N odd L odd Alternating and Sign-Alternating Multiplicity}
\end{equation}
For the sake of brevity in \refTab{tab:multi}, we formulate the above results for even/odd $L$ in the piecewise equation
\begin{equation}
    m^{D_{2L}}_{\overline{L/2},\pm}\equiv
    \begin{cases}
        0  & \ L \text{ even},\\
        \frac{m^{{\mathbb{Z}}_L}_{(N^2-L)/2}\mp(d_0-2 \ o^{\pi})}{2} \ &L \text{ odd} .
    \end{cases}
    \label{eq: combined formula for N odd alt/sign-alt multiplicities}
\end{equation}
Further, we define the most general formulation of $N$ odd 1D multiplicity formulae at $\theta=\pi$,
\begin{equation}
    m^{D_{2L}}_{\bar{q},\pm}\equiv
    \begin{cases}
        0 & \ \bar{q}=\bar{0},\\
        m^{D_{2L}}_{\overline{L/2},\pm} \ &\bar{q}=\overline{L/2} \ .
    \end{cases}
    \label{def: fermionic N odd 1D multiplicities}
\end{equation}
\label{subsec: odd N one-dim multiplicities}
\subsubsection{Odd N Two-Dimensional Multiplicities}
As with the one-dimensional multiplicities in \refSec{subsec: odd N one-dim multiplicities}, calculations of the two-dimensional multiplicities require the double cover of the dihedral group when $N$ is odd. We calculate the multiplicities in the representation of the double cover $D_{2L}$ on $\HS$ of the $\lceil\frac{2L}{2}\rceil-1=L-1$ two-dimensional irreps of $D_{2L}$. Thus the character $\chi_{q,-q}(D_{2L})$ is
\begin{equation}
    \chi_{q,-q}(r^m)=\omega^{qm/2}+\omega^{-qm/2}, \chi_{q,-q}(sr^m)=0,
    \label{eq: double cover 2d character}
\end{equation}
where $ \omega^{\frac{q}{2}}=\exp(2\pi i q/ (2L))$ is a 1D irrep of ${\mathbb{Z}}_{2L}$. The multiplicity of a 2D irrep is calculated via character orthogonality with the characters $\chi_{_{\mathcal{H}}}(D_{2L})$ and $\chi_{q,-q}(D_{2L})$ found in \refTab{tab: theta equals pi odd N dihedral character table}:
\begin{equation}
    m^{D_{2L}}_{\bar{q},\bar{q}'}=\frac{1}{4L}\sum_{k=1}^{2L}(\omega^{qk/2}+\omega^{-qk/2})\tr(\bar{{R}}_{\pi}^k).
\end{equation}
Reindexing into two sums, we expand $\bar{{R}}_{\pi}^{k}=\omega^{-kN^2/2}\hat{{R}}^k$ and arrive at
\begin{multline}
\label{apeq: N odd 2d multiplicities}    m^{D_{2L}}_{\bar{q},\bar{q}'}=\frac{(1+(-1)^{q+1})}{4L}\\\sum_{k=1}^L \omega^{-k\frac{N^2}{2}}\tr(\hat{{R}}^k)\left(\omega^{-k\frac{q}{2}}+\omega^{k\frac{q}{2})}\right)\\ =\frac{(1+(-1)^{q+1})}{4}\left(m^{{\mathbb{Z}}_L}_{(N^2-q)/2}+m^{{\mathbb{Z}}_L}_{(N^2+q)/2}\right),
\end{multline}
for $q=1,\dots,L-1$.
\subsubsection{Even N One-Dimensional Multiplicities}
As is the case for $\theta=0$, at $\theta=\pi$ the dihedral group $D_L$ has different one-dimensional irrep structure depending on the parity of $L$. We treat the two cases separately.
\begin{table}[ht]
\bgroup
\renewcommand{\arraystretch}{1.5}
\begin{tabular}{|c|cccc|}
\hline
N even & \multicolumn{4}{c|}{$D_{L} \sim  \left\langle s,r | r^{L}=s^2=1, sr^k s^{-1}=r^{-k}\right\rangle$} \\ \hline
$D_{L}$ & \multicolumn{1}{c|}{$\mathrm{I}$} & \multicolumn{1}{c|}{$r^k$} & \multicolumn{1}{c|}{$sr^{2k}$} & $sr^{2k+1}$ \\ \hline
$\chi_{0,+}$ & \multicolumn{1}{c|}{1} & \multicolumn{1}{c|}{1} & \multicolumn{1}{c|}{1} & 1 \\ \hline
$\chi_{0,-}$ & \multicolumn{1}{c|}{1} & \multicolumn{1}{c|}{1} & \multicolumn{1}{c|}{-1} & -1 \\ \hline
$\chi_{L/2,+}$ & \multicolumn{1}{c|}{1} & \multicolumn{1}{c|}{$(-1)^k$} & \multicolumn{1}{c|}{1} & -1 \\ \hline
$\chi_{L/2,-}$ & \multicolumn{1}{c|}{1} & \multicolumn{1}{c|}{$(-1)^k$} & \multicolumn{1}{c|}{-1} & 1 \\ \hline
$\chi_{q,-q}$ & \multicolumn{1}{c|}{2} & \multicolumn{1}{c|}{$\omega^{qk}+\omega^{-qk}$} & \multicolumn{1}{c|}{0} & 0 \\ \hline
$\chi_{_\mathcal{H}}$ & \multicolumn{1}{c|}{$\text{Tr}(\hat{\mathrm{I}})$} & \multicolumn{1}{c|}{$\text{Tr}(\bar{{R}}_{\pi}^k)$} & \multicolumn{1}{c|}{$\text{Tr}(\bar{P}_{\pi}\bar{{R}}_{\pi}^{2k})$} & $\text{Tr}(\bar{P}_{\pi}\bar{{R}}_{\pi}^{2k+1})$\\ \hline
\end{tabular}
\egroup
\caption{At $\theta=\pi$, $\bar{P}_{\pi}$ and $\bar{R}_{\pi}$ furnish a representation of the dihedral group $D_{L}$ for even $N$. Because this is the dihedral group $D_{L}$, the parity of $L$ is relevant.}
\label{tab: theta equals pi even N dihedral character table}
\end{table}
We calculate the multiplicities of the two one-dimensional irreps for odd L in \refTab{tab: theta equals pi even N dihedral character table} with character formulae \refEq{eq: trivial and sign character formula} via
\begin{equation}
    m^{D_L}_{\bar{0}, \pm}=\frac{1}{2L}\left(\sum_{k=1}^L\tr(\bar{{R}}_{\pi}^k)\pm \sum_{k=1}^L \tr(\bar{P}_{\pi}\bar{{R}}_{\pi}^k)\right).
\end{equation}
Expanding $\bar{{R}}_{\pi}$ and simplifying the parity coset via trace equivalence,
\begin{equation}
    m^{D_L}_{\bar{0},\pm}=\frac{1}{2L}\sum_{k=1}^L\tr(\hat{{R}}^k)\omega^{-kN^2/2}\pm \frac{\tr(\bar{P}_{\pi})}{2},
\end{equation}
the one-dimensional multiplicities for odd $L$ are
\begin{equation}
    m^{D_L}_{\bar{0}, \pm}=\frac{m^{{\mathbb{Z}}_L}_{N^2/2}\pm (d_0-2 \ o^{\pi})}{2} \ .
    \label{eq: fermionic odd L 1D multiplicities}
\end{equation}
Next, consider the trivial/sign irrep multiplicity for even $L$
\begin{equation}
    m^{D_L}_{\bar{0},\pm}=\frac{1}{2L}\left(\sum_{k=1}^L \tr(\bar{{R}}_{\pi}^k)\pm\sum_{k=1}^L\tr(\bar{P}_{\pi}\bar{{R}}_{\pi}^k)\right).
\end{equation}
Because $L$ is even, separate the parity coset sum into sums over even and odd terms and select trace class representatives
\begin{multline}
    m^{D_L}_{\bar{0},\pm}=\frac{1}{2L}\sum_{k=1}^L \tr(\hat{{R}}^k)\omega^{-kN^2/2}\\ \pm\frac{1}{2L}\sum_{k=1}^{L/2}\tr(\bar{P}_{\pi}\bar{{R}}_{\pi}^{2k-1})\pm\frac{1}{2L}\sum_{k=1}^{L/2}\tr(\bar{P}_{\pi}\bar{{R}}_{\pi}^{2k}),
\end{multline}
simplifying to
\begin{equation}
    m^{D_L}_{\bar{0},\pm}=\frac{m^{{\mathbb{Z}}_L}_{N^2/2}}{2}\pm\frac{\tr(\bar{P}_{\pi}\bar{{R}}_{\pi})}{4}\pm\frac{\tr(\bar{P}_{\pi})}{4}.
\end{equation}
Using the conjugation relations \refEq{eq: conj relations} and $\tr(\bar{P}_{\pi})=d_0$ for $L$ even,
\begin{equation}
    m^{D_L}_{\bar{0},\pm}=\frac{m^{{\mathbb{Z}}_L}_{N^2/2}}{2}\pm\frac{d_0+d_1-2\nu^{\pi}}{4},
    \label{eq: fermionicTrivandSignMultiplicityEvenL}
\end{equation}
where
\begin{equation}
    \nu^{\pi}\equiv \sum_{k=0}^{N/2}D\left(k,\frac{L}{2}-1\right)\left(\sum_{j=0}^{N-2k}\frac{(1-(-1)^{\binom{N-2k-j}{2}+\binom{j}{2}})}{2}\right)
\end{equation}
is calculated via combinatorial arguments analogous to those of \refEq{eq: opi}. We combine the even and odd $L$ trivial/sign multiplicity formulae into
\begin{equation}
    m^{D_{L}}_{\bar{0},\pm}\equiv
    \begin{cases}
        \frac{m^{{\mathbb{Z}}_L}_{N^2/2}\pm (d_0-2 \ o^{\pi})}{2}  & \ L \text{ odd},\\
        \frac{m^{{\mathbb{Z}}_L}_{N^2/2}}{2}\pm\frac{d_0+d_1-2\nu^{\pi}}{4}  \ &L \text{ even}.
    \end{cases}
\end{equation}
Next, consider the multiplicity of the alternating/sign-alternating irrep with character $\chi_{\overline{L/2},\pm}$ from  \refTab{tab: theta equals pi odd N dihedral character table} for even $L$,
\begin{equation}
    \label{apeq: alt-sign mult at pi}
    m^{D_L}_{\overline{L/2},\pm}=\frac{1}{2L}\sum_{k=1}^L (-1)^k\left(\tr(\bar{{R}}_{\pi}^k)\pm\tr(\bar{P}_{\pi}\bar{{R}}_{\pi}^k)\right).
\end{equation}
After simplifying the cyclic coset with  $(-1)^k=\omega^{kL/2}$ and selecting trace class representatives for $\{\bar{P}_{\pi}\bar{{R}}_{\pi}^{2k}\}$ and $\{\bar{P}_{\pi}\bar{{R}}_{\pi}^{2k+1}\}$, \refEq{apeq: alt-sign mult at pi} becomes
\begin{equation}
    m^{D_L}_{\overline{L/2},\pm}=\frac{m^{{\mathbb{Z}}_L}_{(N^2-L)/2}}{2}\pm \left(\frac{1}{4}\tr(\bar{P}_{\pi})-\frac{1}{4}\tr(\bar{P}_{\pi}\bar{{R}}_{\pi})\right).
\end{equation}
Using the conjugation relations \refEq{eq: conj relations},
\begin{equation}
    m^{D_L}_{\overline{L/2},\pm}=\frac{m^{{\mathbb{Z}}_L}_{(N^2-L)/2}}{2}\pm \frac{d_0}{4}\mp \frac{d_1-2\nu^{\pi}}{4} \ 
    \label{eq: fermionic even N alt and alt sign multiplicity}.
\end{equation}
We combine the formulae for even $L$ 1D irrep multiplicities in \refEq{eq: fermionicTrivandSignMultiplicityEvenL} and \refEq{eq: fermionic even N alt and alt sign multiplicity},
\begin{equation}
    m^{D_L}_{\bar{q},\pm}\equiv
    \begin{cases}
        m^{D_L}_{\bar{0},\pm} ,\ & \bar{q}=0\\
        m^{D_L}_{\overline{L/2},\pm}, \ & \bar{q}=\overline{L/2} \ .
    \end{cases}
    \label{def: fermionic even N even L 1D multiplicity formula}
\end{equation}
\subsubsection{Even N Two-Dimensional Multiplicities}
Consider the multiplicity of a two-dimensional irrep with character $\chi_{q,-q}^{D_L}$ in the reducible representation on $\HS$. For $N$ even, the multiplicities are the same for even or odd $L$ and uses the characters in \refTab{tab: theta equals pi even N dihedral character table}:
\begin{equation}
    m^{D_L}_{\bar{q},\bar{q}'}
    =\frac{1}{2L}\sum_{k=1}^L (\omega^{-qk}+\omega^{qk})\omega^{-kN^2/2}\tr(\hat{{R}}^k),
\end{equation}
which simplifies to
\begin{equation}
    m^{D_L}_{\bar{q},\bar{q}'}=\frac{1}{2}\left(m^{{\mathbb{Z}}_L}_{\frac{N^2}{2}-q}+m^{{\mathbb{Z}}_L}_{\frac{N^2}{2}+q}\right).
    \label{eq: Neven2dMult}
\end{equation}

\section{Regular and singular Bethe roots in the nullspace of noninteracting anyons with periodic boundary conditions}
\label{app:NullspaceBetheRoots}
\setcounter{figure}{0} 
\renewcommand{\thefigure}{C\arabic{figure}}
\setcounter{equation}{0}
\renewcommand{\theequation}{C\arabic{equation}}

In the following, we resolve all Bethe roots that appear in the nullspace of two noninteracting anyons with periodic boundaries. We find different classes of solutions, also containing a doublon state. In this case, the state represents a singular eigenstate at $U=E\neq0$ as well.  
Due to the ambiguities in the definition of the scattering phase in this case, we deduce solutions directly from the recurrence relation in \refEq{recurrence} \cite{essler2010OnedimensionalHubbardModel}, while demanding bosonic exchange symmetry for the coefficients and cyclic closure according to \refEq{betheequations2}. We furthermore switch to a basis that is adapted to chiral symmetry 
	\begin{align}
		&\vert Q^+_n,r\rangle=
        \begin{cases}
			\sum_{j=1}^{L-\frac{L}{2}\delta_{r,L/2}}e^{iQ^+_n(j+r/2)}\hat{a}^{\dagger}_j\hat{a}^{\dagger}_{j+r}\vert 0\rangle,\phantom{a} 0<r\leq\frac{L}{2},\\
			\sum_{j=1}^{L}e^{iQ^+_nj}\left(\hat{a}^{\dagger}_j\right)^2\vert 0\rangle,\phantom{a} r=0,\end{cases}
            \\
&\alpha_{Q^+_n}=\frac{\sqrt{2}e^{i\theta/2}\cos(Q^+_n/2)}{\left(e^{iQ^+_n/2}+e^{-i(Q^+_n/2-\theta)}\right)},
	\end{align}
with fixed chirality, as well as manifest bosonic exchange $(r\geq0)$ and translational symmetry
\begin{align}
    &\hat{S}\vert Q^+_n,r\rangle=(-1)^r\vert, Q^+_n,r\rangle
    \\
    &\hat{R}\vert Q^+_n,r\rangle=e^{iQ^+}\vert Q^+_n,r\rangle.
\end{align}
As the lattice is cyclic, $L$ has to be an even number to ensure bi-partiteness, see \refEq{eq:d0}. 
Here $\alpha_{Q^+_n}$ is the doublon weight, i.e., the coefficient of with $r=0$ that describes two-particle coincidences via the contact condition of the recurrence equation in \refEq{recurrence}. 

We find four classes of solutions that we  distinguish by their chiral eigenvalue, by a regular or singular parametrization of Bethe roots, and the underlying mechanism that creates them.
The first class contains solutions with regular Bethe root $Q^-=\pi/2$, whereas neither the bulk recurrence $Q^+\neq\pi$ nor the contact condition $1+\cos(Q^+-\theta)\neq 0$ loses rank. 
The number of these states corresponds to lower-bound in the chiral index theorem for correlated hopping models \cite{Sutherland1986,liebTwoTheoremsHubbard1989,Flach2017,Schomerus2020,theel2025ChirallyProtectedState}, existing for all $\theta$ with positive chirality and a restricted center of mass momentum with \(n\) and \(L/2\) of opposite parity, i.e.,
\(n+L/2\equiv 1 \pmod 2\). The first class consists of only basis states with even particle separations $r$, we find for $L\equiv 2\phantom{a} (\mathrm{mod}\phantom{a} 4)$,
	\begin{align}
		& \vert \Psi_{1}\rangle^{L\equiv 2\phantom{a} (\mathrm{mod}\phantom{a} 4)}_{n+L/2\equiv 1}\propto
        \alpha_{Q^+_n}\vert Q^+_n,0\rangle+\hspace*{-1ex}\sum_{m=1}^{\frac{L/2-1}{2}}(-1)^m\vert Q^+_n,2m\rangle
        \nonumber,
        \\
        &\vert \Psi_{1}\vert^{L\equiv 2\phantom{a} (\mathrm{mod}\phantom{a} 4)}_{n+L/2\equiv 1}=\sqrt{2L\vert \alpha_{Q^+_n}\vert^2+L\left(\frac{L/2-1}{2}\right)}\label{baselinestates},
\end{align}
or if $L$ is divisible by four respectively,
\begin{align}
&\vert \Psi_{1}\rangle^{\mathrm{Mod}[L,4]=0}_{n+L/2\equiv 1}\propto\Bigg(\alpha_{Q^+_n}\vert Q^+_n,0\rangle+
(-1)^{L/4}\vert Q^+_n,L/2\rangle 
\nonumber
\\
&\quad\quad\quad\quad\quad\quad\quad
+\sum_{m=1}^{L/4-1}(-1)^m\vert Q^+_n,2m\rangle
\Bigg),\nonumber
\\
&\vert \Psi_{1}\vert^{\mathrm{Mod}[L,4]=0}_{n+L/2\equiv 1}=\sqrt{2L\vert \alpha_{Q^+_n}\vert^2+L\left(L/4-1\right)+L/2}.
\label{baselinestatesLdivisibleby4}
\end{align}
All other types of solutions are singular Bethe states where the recurrence equation in \refEq{recurrence} loses rank in different ways \cite{essler2010OnedimensionalHubbardModel}. These only exist for finely-tuned values of $\theta$ and center of mass momentum $Q^+$, and embody those states whose presence violates the lower bound of the chiral index theorem \cite{theel2025ChirallyProtectedState}.
\\
\\
The second type of state exists at a center of mass momentum $Q^+=\pi$, where the bulk recurrence equation in \refEq{recurrence}, i.e., away from the defective lines $r=0,1$, becomes exactly zero, 
\begin{align}
    2\cos(Q/2)\left(c_{r+1}+c_{r-1}\right)=0,\quad r\geq2.\label{bulkcontactcondition}
\end{align}
These are states localized in the relative coordinate, that are labeled by the relative separation of the two-particles $r=2,3...,L/2$, besides a chiral eigenvalue of $(-1)^r$ and a translational eigenvalue according to $Q^+=\pi$, 
\begin{align}
		&\vert \Psi_{2}\rangle^{(r)}_{Q^+_n=\pi}=\begin{cases}
\dfrac{1}{\sqrt{L-\frac{L}{2}\delta_{r,L/2}}}
\vert Q^+_n=\pi,r\rangle,
& 2\leq r\leq\frac{L}{2},\\[8pt]
\dfrac{1}{\sqrt{2L}}
\vert Q^+_n=\pi,0\rangle\big|_{\theta=0},
& r=0,\\[8pt]
\dfrac{1}{\sqrt{L}}
\vert Q^+_n=\pi,1\rangle\big|_{\theta=0},
& r=1.
\end{cases}
\label{compactlocalizedstates}
\end{align}
Only in the bosonic limit ($\theta=0$), the indices $r=0,1$ become solutions in the $Q^+=\pi$ sector, i.e.,
\begin{align}
    &2\cos\left(\frac{Q^+}{2}\right)c_2+2\cos\left(\frac{Q^+-\theta}{2}\right)c_0=0,\phantom{a}r=1\nonumber
\\
  &\label{contactconditionnullstates}  
    \\
    &4\cos\left(\frac{Q^+-\theta}{2}\right)c_1=0,\phantom{a}r=0.\nonumber
\end{align}
The third and fourth types of non-regular Bethe solutions arise at a center of mass momentum of $Q^+-\theta=\pi$, exactly where the contact conditions in \refEq{contactconditionnullstates} vanish, not the bulk recurrence. We find a state of negative chirality with support exclusively on opposite sub-lattices, for general even $L$
\begin{align}
		& \vert \Psi_{3}\rangle^{L\equiv 2\phantom{a} (\mathrm{mod}\phantom{a} 4)}_{Q^+_n=\pi+\theta}\propto\Bigg(\sum_{m=0}^{\frac{L/2-3}{2}}(-1)^m\vert \pi+\theta,2m+1\rangle\nonumber
        \\
        &\quad\quad\quad\quad\quad\quad\quad+(-1)^{\frac{L/2-1}{2}}\vert \pi+\theta,L/2\rangle\Bigg),\label{negativechirality}
        \\
        &\vert \Psi_{3}\vert^{L\equiv 2\phantom{a} (\mathrm{mod}\phantom{a} 4)}_{Q^+_n=\pi+\theta}=\sqrt{L\left(\frac{L/2-1}{2}\right)+L/2},\nonumber
\end{align}
or if $L$ is divisible by four respectively,
\begin{align}
&\vert \Psi_{3}\rangle^{\mathrm{Mod}[L,4]=0}_{Q^+_n=\pi+\theta}=\frac{2}{L}\sum_{m=0}^{L/4-1}(-1)^m\vert \pi+\theta,2m+1\rangle.
\label{negativechiralityLdivisibleby4}
\end{align}   
Finally, a complete analysis reveals a doublon state with positive chirality, when the contact condition in \refEq{contactconditionnullstates} vanishes. Similar to the doublon state discussed in \refSec{sec:Doublon}, this state is embedded in the middle of the scattering solutions \cite{valienteTwoparticleStatesHubbard2008,boschiBoundStatesExpansion2014,zhang2013BoundStatesOnedimensional}, 
\begin{align}
		& \vert \Psi_{4}\rangle_{Q^+_n=\pi+\theta}=\frac{1}{\sqrt{2L}}\vert \pi+\theta,0 \rangle\label{resonancedoublon}
	\end{align}
and annihilated by the kinetic energy due to destructive interference with zero weight in the scattering sub-space~\cite{kaneko2024QuantumManybodyScars}. It should be noted, that this state also exists for $U\neq 0$, as a singular Bethe solution, and is embedded in the scattering states depending on the magnitude of the Hubbard interaction. 
\\
\\
We list exemplarily the nullstates for different $\theta$ values and system lengths in \refTab{tab: nullspacepbcL6}, \refTab{tab: nullspacepbcL8} and \refTab{tab:nullspacepbcL10}. It should be noted, that the null states are generally orthogonal for different momenta, chiralities, or fixed relative separations for the bulk singular states in \refEq{compactlocalizedstates}. However, the classification of states 
is generally not unique if the singular conditions in \refEq{bulkcontactcondition} and  \refEq{contactconditionnullstates} coincide in a single momentum sector, or if one of the momenta labeling the regular states in  \refEq{baselinestates} or \refEq{baselinestatesLdivisibleby4} hits these singularities.

\begin{table}[ht]
\bgroup
\renewcommand{\arraystretch}{1.5}
\begin{tabular}{|c|cccc|}
\hline
$L=6$ & \multicolumn{4}{c|}{Nullspace Periodic} \\ \hline
$\theta$ & \multicolumn{1}{c|}{$0$} & \multicolumn{1}{c|}{$0.2\pi$} & \multicolumn{1}{c|}{$\frac{\pi}{2}$} & $\pi$ \\ \hline
$d_0$ & \multicolumn{1}{c|}{5} & \multicolumn{1}{c|}{3} & \multicolumn{1}{c|}{5} & 5 \\ \hline
$\vert\Psi_1\rangle$ & \multicolumn{1}{c|}{3} & \multicolumn{1}{c|}{3} & \multicolumn{1}{c|}{3} & 3 \\ \hline
$\vert\Psi_2\rangle$ & \multicolumn{1}{c|}{0} & \multicolumn{1}{c|}{0} & \multicolumn{1}{c|}{2} & 0 \\ \hline
$\vert\Psi_3\rangle$ & \multicolumn{1}{c|}{1} & \multicolumn{1}{c|}{0} & \multicolumn{1}{c|}{0} & 1 \\ \hline
$\vert\Psi_4\rangle$ & \multicolumn{1}{c|}{1} & \multicolumn{1}{c|}{0} & \multicolumn{1}{c|}{0} & 1 \\ \hline
\end{tabular}
\egroup
\caption{Constituents of the nullspace for different $\theta$, $L=6$ and periodic boundary conditions. Besides the case incommensurate case $\theta=0.2\pi$, all other statistical parameters lead to a nullspace dimension larger than the lower bound of the chiral index theorem in \refEq{eq:d0} \cite{theel2025ChirallyProtectedState}. For $\theta=0$, the choice of $\vert \Psi_4\rangle$ vs. $\vert \Psi_2\rangle$ is arbitrary, as both singular conditions appear at a momentum of the baseline states in  $\vert \Psi_1\rangle$. The cases $\theta=0.2\pi$ and $\theta=\frac{\pi}{2}$ are unique, up to arbitrary rotations, whereas a base-line state at $\theta=\pi$ hits a bulk singularity and so the distinction becomes ambiguous.   }
\label{tab: nullspacepbcL6}
\end{table}

\begin{table}[ht]
\bgroup
\renewcommand{\arraystretch}{1.5}
\begin{tabular}{|c|cccc|}
\hline
$L=8$ & \multicolumn{4}{c|}{Nullspace Periodic} \\ \hline
$\theta$ & \multicolumn{1}{c|}{$0$} & \multicolumn{1}{c|}{$0.2\pi$} & \multicolumn{1}{c|}{$\frac{\pi}{2}$} & $\pi$ \\ \hline
$d_0$ & \multicolumn{1}{c|}{8} & \multicolumn{1}{c|}{4} & \multicolumn{1}{c|}{8} & 6 \\ \hline
$\vert\Psi_1\rangle$ & \multicolumn{1}{c|}{4} & \multicolumn{1}{c|}{4} & \multicolumn{1}{c|}{4} & 4 \\ \hline
$\vert\Psi_2\rangle$ & \multicolumn{1}{c|}{2} & \multicolumn{1}{c|}{0} & \multicolumn{1}{c|}{2} & 2 \\ \hline
$\vert\Psi_3\rangle$ & \multicolumn{1}{c|}{1} & \multicolumn{1}{c|}{0} & \multicolumn{1}{c|}{1} & 0 \\ \hline
$\vert\Psi_4\rangle$ & \multicolumn{1}{c|}{1} & \multicolumn{1}{c|}{0} & \multicolumn{1}{c|}{1} & 0 \\ \hline
\end{tabular}
\egroup
\caption{Constituents of the nullspace for different $\theta$, $L=8$ and periodic boundary conditions. Analogous to \refTab{tab: nullspacepbcL6}, all cases except $\theta=0.2\pi$ exceed the lower bound of the chiral index theorem. Likewise, for the free boson and pseudo-fermion cases, i.e. $\theta=0$ and $\theta=\pi$ respectively, the singularity conditions appear at a baseline momentum, whereas they are not contained in $n+\frac{L}{2}\equiv 1\phantom{a}\mathrm{mod}\phantom{a}2$, or do not exist at all for the other two cases.    }
\label{tab: nullspacepbcL8}
\end{table}

\begin{table}[ht]
\bgroup
\renewcommand{\arraystretch}{1.5}
\renewcommand{\arraystretch}{1.5}
\begin{tabular}{|c|cccc|}
\hline
$L=10$ & \multicolumn{4}{c|}{Nullspace Periodic} \\ \hline
$\theta$ & \multicolumn{1}{c|}{$0$} & \multicolumn{1}{c|}{$0.2\pi$} & \multicolumn{1}{c|}{$\frac{\pi}{2}$} & $\pi$ \\ \hline
$d_0$ & \multicolumn{1}{c|}{9} & \multicolumn{1}{c|}{5} & \multicolumn{1}{c|}{9} & 9 \\ \hline
$\vert\Psi_1\rangle$ & \multicolumn{1}{c|}{5} & \multicolumn{1}{c|}{5} & \multicolumn{1}{c|}{5} & 5 \\ \hline
$\vert\Psi_2\rangle$ & \multicolumn{1}{c|}{2} & \multicolumn{1}{c|}{0} & \multicolumn{1}{c|}{4} & 2 \\ \hline
$\vert\Psi_3\rangle$ & \multicolumn{1}{c|}{1} & \multicolumn{1}{c|}{0} & \multicolumn{1}{c|}{0} & 1 \\ \hline
$\vert\Psi_4\rangle$ & \multicolumn{1}{c|}{1} & \multicolumn{1}{c|}{0} & \multicolumn{1}{c|}{0} & 1 \\ \hline
\end{tabular}
\egroup
\caption{Constituents of the nullspace for different $\theta$, $L=10$ and periodic boundary conditions. As for $L=6,8$ in \refTab{tab: nullspacepbcL6} and \refTab{tab: nullspacepbcL8} respectively, only for statistical parameter of $\theta=0.2\pi$, the chiral sub-matrix of the cyclic Hamiltonian has zero rank. For the extremal cases, $\theta=0,\pi$, baseline momenta hit the singularity mechanisms, whereas the other cases are unambiguous beyond arbitrary rotations in the respective subspaces.  }
\label{tab:nullspacepbcL10}
\end{table}

\section{Nullstates for Open boundary conditions}
\label{app:NullspaceDirichlet}
\setcounter{figure}{0} 
\renewcommand{\thefigure}{D\arabic{figure}}
\setcounter{equation}{0}
\renewcommand{\theequation}{D\arabic{equation}}
In the following we solve the recurrence relation \refEq{recurrence} for $U=E=0$, open boundary conditions, while demanding bosonic exchange symmetry for the coefficients. As the recurrence is in-homogeneous, it is advisory to evaluate \refEq{recurrence} first at the defective line $l=m\equiv r$,
\begin{align}
   & e^{-i\theta/2}c_{r,r+1}+ e^{i\theta/2}c_{r,r-1}+ e^{-i\theta/2}c_{r+1,r}+ e^{i\theta/2}c_{r-1,r}=0\label{eq:nearestneighborconstraintrecurrencedirichlet}
   \\
   &\Rightarrow c_{r,r+1}=- e^{i\theta}c_{r-1,r},\quad c_{0,1}=0\nonumber
   \\
   &\Rightarrow c_{r,r+1}=0\qquad r=1,\dots,L-1 \nonumber,
\end{align}
implying that the nearest-neighbor coefficients have to vanish exactly. Subsequently, we consider the next defective line $l=m+1$ whereas we separate diagonal coefficients $d_m\equiv c_{m,m}$ notationally from the off-diagonal ones,
\begin{align}
D_{m+1}=&-e^{i\theta}D_m-e^{i\theta/2}S_m\nonumber
   \\
   D_m=&(-e^{i\theta})^{m-1}D_1-e^{i\theta/2}\sum_{p=1}^{m-1}(-e^{i\theta})^{m-1-p}S_p\label{eq:innerrecurrencedirichlet}
   \\
S_p\equiv&\left(c_{p,p+2}+c_{p-1,p+1}\right)\nonumber
\end{align}
\refEq{eq:innerrecurrencedirichlet} can be easily evaluated once the solution for the off-diagonal coefficient $S_p$ is known, with the free parameter $D_1$. To this end, we consider $m\geq l+2$ without the loss of generality and define $x\equiv l$ and $y\equiv m-1$,
\begin{align}
    c_{x,y+2}+c_{x,y}+c_{x+1,y+1}+c_{x-1,y-1}=0,\quad m\geq l+2.
\end{align}
In order to re-formulate the nearest-neighbor constraint in \refEq{eq:nearestneighborconstraintrecurrencedirichlet} locally, we furthermore define
\begin{align}
&g_{x,y}\equiv c_{x,y+1},
\\
    &g_{x,y+1}+g_{x,y-1}+g_{x+1,y}+g_{x-1,y}=0,\quad  x
    <y\label{eq:2dlaplaciandirichletnullspace}
\end{align}
yielding a discrete Laplacian with open boundary conditions and Pauli-like constraint on a triangular domain,
\begin{align}
    &g_{x,x}=0\label{eq:PauliconstraintDirichletsolution}
    \\
    &g_{0,y}=g_{x,L}=0.
\end{align}
\refEq{eq:2dlaplaciandirichletnullspace} can be solved by a separation of variables, after extending the solution anti-symmetrically to the complementary triangular region $x>y$, i.e.
\begin{align}
    g^{(n)}_{x,y}=-g^{(n)}_{y,x}\label{eq:antisymmetrizationnullspacedirichlet}
\end{align}
This choice fulfills the constraint in \refEq{eq:PauliconstraintDirichletsolution} and fixes the other open boundary conditions
\begin{align}
     g_{L,y}=g_{x,0}=0.
\end{align}
The eigenvalues of the harmonics in both dimensions exactly cancel each other in analogy to the case of free bosons on the line discussed in \cite{theel2025ChirallyProtectedState}, 
\begin{align}
    g^{(n)}_{x,y}=&\frac{2}{L}\left[\sin(q_n x)\sin(q_{L-n} y)-\sin(q_n y)\sin(q_{L-n} x)\right],\nonumber
\\
  \label{eq:gfunctionrecurrencedirichlet}  
    \\
    q_n=&\frac{\pi n}{L},\nonumber\quad n=1,...\bigg\lfloor \frac{L-1}{2}\bigg\rfloor.
\end{align}
Restoring bosonic exchange symmetry  finally yields the scattering contribution in \refEq{eq:nullspacecoefficientoffdiagonal} in the main text, i.e.
\begin{align}
    F^{(n)}_{l,m}=\begin{cases}
        g^{(n)}_{l,m-1}, \phantom{a}l<m
        \\
        0,\phantom{a}l=m
        \\
        g^{(n)}_{m,l-1} \phantom{a}l>m.
    \end{cases}
\end{align}
Finally, the doublon weight in \refEq{eq:nullspacecoefficientdiagonal} in the main text is obtained by insertion of \refEq{eq:gfunctionrecurrencedirichlet} into the diagonal recurrence relation in \refEq{eq:innerrecurrencedirichlet} and set $D_1=0$. The pure doublon state in \refEq{eq:doublonstate} of the main text, just has to be evaluated at the energy $E=U=0$, or alternatively from  \refEq{eq:innerrecurrencedirichlet} by settin $D_1=1$ and $S_p=0,\phantom{a}\forall p$. Together with the $\lfloor\frac{L-1}{2}\rfloor$ other null states, they span the full nullspace with dimension $d_0$, see \refEq{eq:d0} \cite{theel2025ChirallyProtectedState}.


\begin{thebibliography}{145}%
\makeatletter
\providecommand \@ifxundefined [1]{%
 \@ifx{#1\undefined}
}%
\providecommand \@ifnum [1]{%
 \ifnum #1\expandafter \@firstoftwo
 \else \expandafter \@secondoftwo
 \fi
}%
\providecommand \@ifx [1]{%
 \ifx #1\expandafter \@firstoftwo
 \else \expandafter \@secondoftwo
 \fi
}%
\providecommand \natexlab [1]{#1}%
\providecommand \enquote  [1]{``#1''}%
\providecommand \bibnamefont  [1]{#1}%
\providecommand \bibfnamefont [1]{#1}%
\providecommand \citenamefont [1]{#1}%
\providecommand \href@noop [0]{\@secondoftwo}%
\providecommand \href [0]{\begingroup \@sanitize@url \@href}%
\providecommand \@href[1]{\@@startlink{#1}\@@href}%
\providecommand \@@href[1]{\endgroup#1\@@endlink}%
\providecommand \@sanitize@url [0]{\catcode `\\12\catcode `\$12\catcode `\&12\catcode `\#12\catcode `\^12\catcode `\_12\catcode `\%12\relax}%
\providecommand \@@startlink[1]{}%
\providecommand \@@endlink[0]{}%
\providecommand \url  [0]{\begingroup\@sanitize@url \@url }%
\providecommand \@url [1]{\endgroup\@href {#1}{\urlprefix }}%
\providecommand \urlprefix  [0]{URL }%
\providecommand \Eprint [0]{\href }%
\providecommand \doibase [0]{https://doi.org/}%
\providecommand \selectlanguage [0]{\@gobble}%
\providecommand \bibinfo  [0]{\@secondoftwo}%
\providecommand \bibfield  [0]{\@secondoftwo}%
\providecommand \translation [1]{[#1]}%
\providecommand \BibitemOpen [0]{}%
\providecommand \bibitemStop [0]{}%
\providecommand \bibitemNoStop [0]{.\EOS\space}%
\providecommand \EOS [0]{\spacefactor3000\relax}%
\providecommand \BibitemShut  [1]{\csname bibitem#1\endcsname}%
\let\auto@bib@innerbib\@empty
\bibitem [{\citenamefont {Kwan}\ \emph {et~al.}(2024)\citenamefont {Kwan}, \citenamefont {Segura}, \citenamefont {Li}, \citenamefont {Kim}, \citenamefont {Gorshkov}, \citenamefont {Eckardt}, \citenamefont {{Bakkali-Hassani}},\ and\ \citenamefont {Greiner}}]{kwan2024RealizationOnedimensionalAnyons}%
  \BibitemOpen
  \bibfield  {author} {\bibinfo {author} {\bibfnamefont {J.}~\bibnamefont {Kwan}}, \bibinfo {author} {\bibfnamefont {P.}~\bibnamefont {Segura}}, \bibinfo {author} {\bibfnamefont {Y.}~\bibnamefont {Li}}, \bibinfo {author} {\bibfnamefont {S.}~\bibnamefont {Kim}}, \bibinfo {author} {\bibfnamefont {A.~V.}\ \bibnamefont {Gorshkov}}, \bibinfo {author} {\bibfnamefont {A.}~\bibnamefont {Eckardt}}, \bibinfo {author} {\bibfnamefont {B.}~\bibnamefont {{Bakkali-Hassani}}},\ and\ \bibinfo {author} {\bibfnamefont {M.}~\bibnamefont {Greiner}},\ }\bibfield  {title} {\bibinfo {title} {Realization of one-dimensional anyons with arbitrary statistical phase},\ }\href {https://doi.org/10.1126/science.adi3252} {\bibfield  {journal} {\bibinfo  {journal} {Science}\ }\textbf {\bibinfo {volume} {386}},\ \bibinfo {pages} {1055} (\bibinfo {year} {2024})}\BibitemShut {NoStop}%
\bibitem [{\citenamefont {Horowitz}(1989)}]{horowitz1989ExactlySolubleDiffeomorphism}%
  \BibitemOpen
  \bibfield  {author} {\bibinfo {author} {\bibfnamefont {G.~T.}\ \bibnamefont {Horowitz}},\ }\bibfield  {title} {\bibinfo {title} {Exactly soluble diffeomorphism invariant theories},\ }\href {https://doi.org/10.1007/BF01218410} {\bibfield  {journal} {\bibinfo  {journal} {Commun. Math. Phys.}\ }\textbf {\bibinfo {volume} {125}},\ \bibinfo {pages} {417} (\bibinfo {year} {1989})}\BibitemShut {NoStop}%
\bibitem [{\citenamefont {Fr{\"o}lian}\ \emph {et~al.}(2022)\citenamefont {Fr{\"o}lian}, \citenamefont {Chisholm}, \citenamefont {Neri}, \citenamefont {Cabrera}, \citenamefont {Ramos}, \citenamefont {Celi},\ and\ \citenamefont {Tarruell}}]{frolian2022Realizing1DTopological}%
  \BibitemOpen
  \bibfield  {author} {\bibinfo {author} {\bibfnamefont {A.}~\bibnamefont {Fr{\"o}lian}}, \bibinfo {author} {\bibfnamefont {C.~S.}\ \bibnamefont {Chisholm}}, \bibinfo {author} {\bibfnamefont {E.}~\bibnamefont {Neri}}, \bibinfo {author} {\bibfnamefont {C.~R.}\ \bibnamefont {Cabrera}}, \bibinfo {author} {\bibfnamefont {R.}~\bibnamefont {Ramos}}, \bibinfo {author} {\bibfnamefont {A.}~\bibnamefont {Celi}},\ and\ \bibinfo {author} {\bibfnamefont {L.}~\bibnamefont {Tarruell}},\ }\bibfield  {title} {\bibinfo {title} {Realizing a {{1D}} topological gauge theory in an optically dressed {{BEC}}},\ }\href {https://doi.org/10.1038/s41586-022-04943-3} {\bibfield  {journal} {\bibinfo  {journal} {Nature}\ }\textbf {\bibinfo {volume} {608}},\ \bibinfo {pages} {293} (\bibinfo {year} {2022})}\BibitemShut {NoStop}%
\bibitem [{\citenamefont {Wang}\ \emph {et~al.}(2025)\citenamefont {Wang}, \citenamefont {Vashisht}, \citenamefont {Guo}, \citenamefont {Dhar}, \citenamefont {Landini}, \citenamefont {N{\"a}gerl},\ and\ \citenamefont {Goldman}}]{wang2025AnyonizationBosonsOne}%
  \BibitemOpen
  \bibfield  {author} {\bibinfo {author} {\bibfnamefont {B.}~\bibnamefont {Wang}}, \bibinfo {author} {\bibfnamefont {A.}~\bibnamefont {Vashisht}}, \bibinfo {author} {\bibfnamefont {Y.}~\bibnamefont {Guo}}, \bibinfo {author} {\bibfnamefont {S.}~\bibnamefont {Dhar}}, \bibinfo {author} {\bibfnamefont {M.}~\bibnamefont {Landini}}, \bibinfo {author} {\bibfnamefont {H.-C.}\ \bibnamefont {N{\"a}gerl}},\ and\ \bibinfo {author} {\bibfnamefont {N.}~\bibnamefont {Goldman}},\ }\href {https://doi.org/10.48550/arXiv.2504.21208} {\bibinfo {title} {Anyonization of bosons in one dimension: An effective swap model}} (\bibinfo {year} {2025}),\ \Eprint {https://arxiv.org/abs/2504.21208} {arXiv:2504.21208} \BibitemShut {NoStop}%
\bibitem [{\citenamefont {Zeng}\ \emph {et~al.}(2026)\citenamefont {Zeng}, \citenamefont {Bastianello}, \citenamefont {Dhar}, \citenamefont {Wang}, \citenamefont {Yu}, \citenamefont {Horvath}, \citenamefont {Astrakharchik}, \citenamefont {Guo}, \citenamefont {N{\"a}gerl},\ and\ \citenamefont {Landini}}]{zeng2026RealizationFractionalFermi}%
  \BibitemOpen
  \bibfield  {author} {\bibinfo {author} {\bibfnamefont {Y.}~\bibnamefont {Zeng}}, \bibinfo {author} {\bibfnamefont {A.}~\bibnamefont {Bastianello}}, \bibinfo {author} {\bibfnamefont {S.}~\bibnamefont {Dhar}}, \bibinfo {author} {\bibfnamefont {Z.}~\bibnamefont {Wang}}, \bibinfo {author} {\bibfnamefont {X.}~\bibnamefont {Yu}}, \bibinfo {author} {\bibfnamefont {M.}~\bibnamefont {Horvath}}, \bibinfo {author} {\bibfnamefont {G.~E.}\ \bibnamefont {Astrakharchik}}, \bibinfo {author} {\bibfnamefont {Y.}~\bibnamefont {Guo}}, \bibinfo {author} {\bibfnamefont {H.-C.}\ \bibnamefont {N{\"a}gerl}},\ and\ \bibinfo {author} {\bibfnamefont {M.}~\bibnamefont {Landini}},\ }\href {https://doi.org/10.48550/arXiv.2602.17657} {\bibinfo {title} {Realization of fractional {{Fermi}} seas}} (\bibinfo {year} {2026}),\ \Eprint {https://arxiv.org/abs/2602.17657} {arXiv:2602.17657 [cond-mat.quant-gas]} \BibitemShut {NoStop}%
\bibitem [{\citenamefont {{Bakkali-Hassani}}\ \emph {et~al.}(2026)\citenamefont {{Bakkali-Hassani}}, \citenamefont {Kwan}, \citenamefont {Segura}, \citenamefont {Li}, \citenamefont {Tesfaye}, \citenamefont {{Valent{\'i}-Rojas}}, \citenamefont {Eckardt},\ and\ \citenamefont {Greiner}}]{bakkalihassani2026RevealingPseudoFermionizationChiral}%
  \BibitemOpen
  \bibfield  {author} {\bibinfo {author} {\bibfnamefont {B.}~\bibnamefont {{Bakkali-Hassani}}}, \bibinfo {author} {\bibfnamefont {J.}~\bibnamefont {Kwan}}, \bibinfo {author} {\bibfnamefont {P.}~\bibnamefont {Segura}}, \bibinfo {author} {\bibfnamefont {Y.}~\bibnamefont {Li}}, \bibinfo {author} {\bibfnamefont {I.}~\bibnamefont {Tesfaye}}, \bibinfo {author} {\bibfnamefont {G.}~\bibnamefont {{Valent{\'i}-Rojas}}}, \bibinfo {author} {\bibfnamefont {A.}~\bibnamefont {Eckardt}},\ and\ \bibinfo {author} {\bibfnamefont {M.}~\bibnamefont {Greiner}},\ }\href {https://doi.org/10.48550/arXiv.2602.20421} {\bibinfo {title} {Revealing {{Pseudo-Fermionization}} and {{Chiral Binding}} of {{One-Dimensional Anyons}} using {{Adiabatic State Preparation}}}} (\bibinfo {year} {2026}),\ \Eprint {https://arxiv.org/abs/2602.20421} {arXiv:2602.20421 [cond-mat.quant-gas]} \BibitemShut {NoStop}%
\bibitem [{\citenamefont {Chisholm}\ \emph {et~al.}(2022)\citenamefont {Chisholm}, \citenamefont {Fr{\"o}lian}, \citenamefont {Neri}, \citenamefont {Ramos}, \citenamefont {Tarruell},\ and\ \citenamefont {Celi}}]{chisholm2022EncodingOnedimensionalTopological}%
  \BibitemOpen
  \bibfield  {author} {\bibinfo {author} {\bibfnamefont {C.~S.}\ \bibnamefont {Chisholm}}, \bibinfo {author} {\bibfnamefont {A.}~\bibnamefont {Fr{\"o}lian}}, \bibinfo {author} {\bibfnamefont {E.}~\bibnamefont {Neri}}, \bibinfo {author} {\bibfnamefont {R.}~\bibnamefont {Ramos}}, \bibinfo {author} {\bibfnamefont {L.}~\bibnamefont {Tarruell}},\ and\ \bibinfo {author} {\bibfnamefont {A.}~\bibnamefont {Celi}},\ }\bibfield  {title} {\bibinfo {title} {Encoding a one-dimensional topological gauge theory in a {{Raman-coupled Bose-Einstein}} condensate},\ }\href {https://doi.org/10.1103/PhysRevResearch.4.043088} {\bibfield  {journal} {\bibinfo  {journal} {Phys. Rev. Res.}\ }\textbf {\bibinfo {volume} {4}},\ \bibinfo {pages} {043088} (\bibinfo {year} {2022})}\BibitemShut {NoStop}%
\bibitem [{\citenamefont {Keilmann}\ \emph {et~al.}(2011)\citenamefont {Keilmann}, \citenamefont {Lanzmich}, \citenamefont {McCulloch},\ and\ \citenamefont {Roncaglia}}]{keilmann2011StatisticallyInducedPhase}%
  \BibitemOpen
  \bibfield  {author} {\bibinfo {author} {\bibfnamefont {T.}~\bibnamefont {Keilmann}}, \bibinfo {author} {\bibfnamefont {S.}~\bibnamefont {Lanzmich}}, \bibinfo {author} {\bibfnamefont {I.}~\bibnamefont {McCulloch}},\ and\ \bibinfo {author} {\bibfnamefont {M.}~\bibnamefont {Roncaglia}},\ }\bibfield  {title} {\bibinfo {title} {Statistically induced phase transitions and anyons in {{1D}} optical lattices},\ }\href {https://doi.org/10.1038/ncomms1353} {\bibfield  {journal} {\bibinfo  {journal} {Nat. Commun.}\ }\textbf {\bibinfo {volume} {2}},\ \bibinfo {pages} {361} (\bibinfo {year} {2011})}\BibitemShut {NoStop}%
\bibitem [{\citenamefont {Greschner}\ \emph {et~al.}(2014)\citenamefont {Greschner}, \citenamefont {Sun}, \citenamefont {Poletti},\ and\ \citenamefont {Santos}}]{greschnerDensityDependentSyntheticGauge2014}%
  \BibitemOpen
  \bibfield  {author} {\bibinfo {author} {\bibfnamefont {S.}~\bibnamefont {Greschner}}, \bibinfo {author} {\bibfnamefont {G.}~\bibnamefont {Sun}}, \bibinfo {author} {\bibfnamefont {D.}~\bibnamefont {Poletti}},\ and\ \bibinfo {author} {\bibfnamefont {L.}~\bibnamefont {Santos}},\ }\bibfield  {title} {\bibinfo {title} {Density-{{Dependent Synthetic Gauge Fields Using Periodically Modulated Interactions}}},\ }\href {https://doi.org/10.1103/PhysRevLett.113.215303} {\bibfield  {journal} {\bibinfo  {journal} {Phys. Rev. Lett.}\ }\textbf {\bibinfo {volume} {113}},\ \bibinfo {pages} {215303} (\bibinfo {year} {2014})}\BibitemShut {NoStop}%
\bibitem [{\citenamefont {Greschner}\ and\ \citenamefont {Santos}(2015)}]{greschner2015AnyonHubbardModel}%
  \BibitemOpen
  \bibfield  {author} {\bibinfo {author} {\bibfnamefont {S.}~\bibnamefont {Greschner}}\ and\ \bibinfo {author} {\bibfnamefont {L.}~\bibnamefont {Santos}},\ }\bibfield  {title} {\bibinfo {title} {Anyon {{Hubbard}} model in one-dimensional optical lattices},\ }\href {https://doi.org/10.1103/PhysRevLett.115.053002} {\bibfield  {journal} {\bibinfo  {journal} {Phys. Rev. Lett.}\ }\textbf {\bibinfo {volume} {115}},\ \bibinfo {pages} {053002} (\bibinfo {year} {2015})}\BibitemShut {NoStop}%
\bibitem [{\citenamefont {Str{\"a}ter}\ \emph {et~al.}(2016)\citenamefont {Str{\"a}ter}, \citenamefont {Srivastava},\ and\ \citenamefont {Eckardt}}]{strater2016FloquetRealizationSignatures}%
  \BibitemOpen
  \bibfield  {author} {\bibinfo {author} {\bibfnamefont {C.}~\bibnamefont {Str{\"a}ter}}, \bibinfo {author} {\bibfnamefont {S.~C.~L.}\ \bibnamefont {Srivastava}},\ and\ \bibinfo {author} {\bibfnamefont {A.}~\bibnamefont {Eckardt}},\ }\bibfield  {title} {\bibinfo {title} {Floquet {{Realization}} and {{Signatures}} of {{One-Dimensional Anyons}} in an {{Optical Lattice}}},\ }\href {https://doi.org/10.1103/PhysRevLett.117.205303} {\bibfield  {journal} {\bibinfo  {journal} {Phys. Rev. Lett.}\ }\textbf {\bibinfo {volume} {117}},\ \bibinfo {pages} {205303} (\bibinfo {year} {2016})}\BibitemShut {NoStop}%
\bibitem [{\citenamefont {Greschner}\ \emph {et~al.}(2018)\citenamefont {Greschner}, \citenamefont {Cardarelli},\ and\ \citenamefont {Santos}}]{greschnerProbingExchangeStatistics2018}%
  \BibitemOpen
  \bibfield  {author} {\bibinfo {author} {\bibfnamefont {S.}~\bibnamefont {Greschner}}, \bibinfo {author} {\bibfnamefont {L.}~\bibnamefont {Cardarelli}},\ and\ \bibinfo {author} {\bibfnamefont {L.}~\bibnamefont {Santos}},\ }\bibfield  {title} {\bibinfo {title} {Probing the exchange statistics of one-dimensional anyon models},\ }\href {https://doi.org/10.1103/PhysRevA.97.053605} {\bibfield  {journal} {\bibinfo  {journal} {Phys. Rev. A}\ }\textbf {\bibinfo {volume} {97}},\ \bibinfo {pages} {053605} (\bibinfo {year} {2018})}\BibitemShut {NoStop}%
\bibitem [{\citenamefont {Dhar}\ \emph {et~al.}(2025)\citenamefont {Dhar}, \citenamefont {Wang}, \citenamefont {Horvath}, \citenamefont {Vashisht}, \citenamefont {Zeng}, \citenamefont {Zvonarev}, \citenamefont {Goldman}, \citenamefont {Guo}, \citenamefont {Landini},\ and\ \citenamefont {N{\"a}gerl}}]{dhar2025ObservingAnyonizationBosons}%
  \BibitemOpen
  \bibfield  {author} {\bibinfo {author} {\bibfnamefont {S.}~\bibnamefont {Dhar}}, \bibinfo {author} {\bibfnamefont {B.}~\bibnamefont {Wang}}, \bibinfo {author} {\bibfnamefont {M.}~\bibnamefont {Horvath}}, \bibinfo {author} {\bibfnamefont {A.}~\bibnamefont {Vashisht}}, \bibinfo {author} {\bibfnamefont {Y.}~\bibnamefont {Zeng}}, \bibinfo {author} {\bibfnamefont {M.~B.}\ \bibnamefont {Zvonarev}}, \bibinfo {author} {\bibfnamefont {N.}~\bibnamefont {Goldman}}, \bibinfo {author} {\bibfnamefont {Y.}~\bibnamefont {Guo}}, \bibinfo {author} {\bibfnamefont {M.}~\bibnamefont {Landini}},\ and\ \bibinfo {author} {\bibfnamefont {H.-C.}\ \bibnamefont {N{\"a}gerl}},\ }\bibfield  {title} {\bibinfo {title} {Observing anyonization of bosons in a quantum gas},\ }\href {https://doi.org/10.1038/s41586-025-09016-9} {\bibfield  {journal} {\bibinfo  {journal} {Nature}\ }\textbf {\bibinfo {volume} {642}},\ \bibinfo {pages} {53} (\bibinfo {year} {2025})}\BibitemShut {NoStop}%
\bibitem [{\citenamefont {Nagies}\ \emph {et~al.}(2024)\citenamefont {Nagies}, \citenamefont {Wang}, \citenamefont {Knapp}, \citenamefont {Eckardt},\ and\ \citenamefont {Harshman}}]{nagies2024BraidStatisticsConstructing}%
  \BibitemOpen
  \bibfield  {author} {\bibinfo {author} {\bibfnamefont {S.}~\bibnamefont {Nagies}}, \bibinfo {author} {\bibfnamefont {B.}~\bibnamefont {Wang}}, \bibinfo {author} {\bibfnamefont {A.~C.}\ \bibnamefont {Knapp}}, \bibinfo {author} {\bibfnamefont {A.}~\bibnamefont {Eckardt}},\ and\ \bibinfo {author} {\bibfnamefont {N.~L.}\ \bibnamefont {Harshman}},\ }\bibfield  {title} {\bibinfo {title} {Beyond braid statistics: {{Constructing}} a lattice model for anyons with exchange statistics intrinsic to one dimension},\ }\href {https://doi.org/10.21468/SciPostPhys.16.3.086} {\bibfield  {journal} {\bibinfo  {journal} {SciPost Phys.}\ }\textbf {\bibinfo {volume} {16}},\ \bibinfo {pages} {086} (\bibinfo {year} {2024})}\BibitemShut {NoStop}%
\bibitem [{\citenamefont {Kundu}(1999)}]{kundu1999ExactSolutionDouble}%
  \BibitemOpen
  \bibfield  {author} {\bibinfo {author} {\bibfnamefont {A.}~\bibnamefont {Kundu}},\ }\bibfield  {title} {\bibinfo {title} {Exact {{Solution}} of {{Double}} delta {{Function Bose Gas}} through an {{Interacting Anyon Gas}}},\ }\href {https://doi.org/10.1103/PhysRevLett.83.1275} {\bibfield  {journal} {\bibinfo  {journal} {Phys. Rev. Lett.}\ }\textbf {\bibinfo {volume} {83}},\ \bibinfo {pages} {1275} (\bibinfo {year} {1999})}\BibitemShut {NoStop}%
\bibitem [{\citenamefont {Bonkhoff}\ \emph {et~al.}(2021)\citenamefont {Bonkhoff}, \citenamefont {J{\"a}gering}, \citenamefont {Eggert}, \citenamefont {Pelster}, \citenamefont {Thorwart},\ and\ \citenamefont {Posske}}]{bonkhoff2021BosonicContinuumTheory}%
  \BibitemOpen
  \bibfield  {author} {\bibinfo {author} {\bibfnamefont {M.}~\bibnamefont {Bonkhoff}}, \bibinfo {author} {\bibfnamefont {K.}~\bibnamefont {J{\"a}gering}}, \bibinfo {author} {\bibfnamefont {S.}~\bibnamefont {Eggert}}, \bibinfo {author} {\bibfnamefont {A.}~\bibnamefont {Pelster}}, \bibinfo {author} {\bibfnamefont {M.}~\bibnamefont {Thorwart}},\ and\ \bibinfo {author} {\bibfnamefont {T.}~\bibnamefont {Posske}},\ }\bibfield  {title} {\bibinfo {title} {Bosonic {{Continuum Theory}} of {{One-Dimensional Lattice Anyons}}},\ }\href {https://doi.org/10.1103/PhysRevLett.126.163201} {\bibfield  {journal} {\bibinfo  {journal} {Phys. Rev. Lett.}\ }\textbf {\bibinfo {volume} {126}},\ \bibinfo {pages} {163201} (\bibinfo {year} {2021})}\BibitemShut {NoStop}%
\bibitem [{\citenamefont {Patu}\ \emph {et~al.}(2007)\citenamefont {Patu}, \citenamefont {Korepin},\ and\ \citenamefont {Averin}}]{patu2007CorrelationFunctionsOnedimensional}%
  \BibitemOpen
  \bibfield  {author} {\bibinfo {author} {\bibfnamefont {O.~I.}\ \bibnamefont {Patu}}, \bibinfo {author} {\bibfnamefont {V.~E.}\ \bibnamefont {Korepin}},\ and\ \bibinfo {author} {\bibfnamefont {D.~V.}\ \bibnamefont {Averin}},\ }\bibfield  {title} {\bibinfo {title} {Correlation functions of one-dimensional {{Lieb}}--{{Liniger}} anyons},\ }\href {https://doi.org/10.1088/1751-8113/40/50/004} {\bibfield  {journal} {\bibinfo  {journal} {J. Phys. A: Math. Theor.}\ }\textbf {\bibinfo {volume} {40}},\ \bibinfo {pages} {14963} (\bibinfo {year} {2007})}\BibitemShut {NoStop}%
\bibitem [{\citenamefont {Batchelor}\ \emph {et~al.}(2006)\citenamefont {Batchelor}, \citenamefont {Guan},\ and\ \citenamefont {Oelkers}}]{batchelor2006OneDimensionalInteractingAnyon}%
  \BibitemOpen
  \bibfield  {author} {\bibinfo {author} {\bibfnamefont {M.~T.}\ \bibnamefont {Batchelor}}, \bibinfo {author} {\bibfnamefont {X.-W.}\ \bibnamefont {Guan}},\ and\ \bibinfo {author} {\bibfnamefont {N.}~\bibnamefont {Oelkers}},\ }\bibfield  {title} {\bibinfo {title} {One-{{Dimensional Interacting Anyon Gas}}: {{Low-Energy Properties}} and {{Haldane Exclusion Statistics}}},\ }\href {https://doi.org/10.1103/PhysRevLett.96.210402} {\bibfield  {journal} {\bibinfo  {journal} {Phys. Rev. Lett.}\ }\textbf {\bibinfo {volume} {96}},\ \bibinfo {pages} {210402} (\bibinfo {year} {2006})}\BibitemShut {NoStop}%
\bibitem [{\citenamefont {Aglietti}\ \emph {et~al.}(1996)\citenamefont {Aglietti}, \citenamefont {Griguolo}, \citenamefont {Jackiw}, \citenamefont {Pi},\ and\ \citenamefont {Seminara}}]{aglietti1996AnyonsChiralSolitons}%
  \BibitemOpen
  \bibfield  {author} {\bibinfo {author} {\bibfnamefont {U.}~\bibnamefont {Aglietti}}, \bibinfo {author} {\bibfnamefont {L.}~\bibnamefont {Griguolo}}, \bibinfo {author} {\bibfnamefont {R.}~\bibnamefont {Jackiw}}, \bibinfo {author} {\bibfnamefont {S.-Y.}\ \bibnamefont {Pi}},\ and\ \bibinfo {author} {\bibfnamefont {D.}~\bibnamefont {Seminara}},\ }\bibfield  {title} {\bibinfo {title} {Anyons and chiral solitons on a line},\ }\href {https://doi.org/10.1103/PhysRevLett.77.4406} {\bibfield  {journal} {\bibinfo  {journal} {Phys. Rev. Lett.}\ }\textbf {\bibinfo {volume} {77}},\ \bibinfo {pages} {4406} (\bibinfo {year} {1996})}\BibitemShut {NoStop}%
\bibitem [{\citenamefont {Posske}\ \emph {et~al.}(2017)\citenamefont {Posske}, \citenamefont {Trauzettel},\ and\ \citenamefont {Thorwart}}]{posske2017SecondQuantizationLeinaasMyrheim}%
  \BibitemOpen
  \bibfield  {author} {\bibinfo {author} {\bibfnamefont {T.}~\bibnamefont {Posske}}, \bibinfo {author} {\bibfnamefont {B.}~\bibnamefont {Trauzettel}},\ and\ \bibinfo {author} {\bibfnamefont {M.}~\bibnamefont {Thorwart}},\ }\bibfield  {title} {\bibinfo {title} {Second quantization of {{Leinaas-Myrheim}} anyons in one dimension and their relation to the {{Lieb-Liniger}} model},\ }\href {https://doi.org/10.1103/PhysRevB.96.195422} {\bibfield  {journal} {\bibinfo  {journal} {Phys. Rev. B}\ }\textbf {\bibinfo {volume} {96}},\ \bibinfo {pages} {195422} (\bibinfo {year} {2017})}\BibitemShut {NoStop}%
\bibitem [{\citenamefont {Harshman}\ and\ \citenamefont {Knapp}(2022)}]{harshman2022TopologicalExchangeStatistics}%
  \BibitemOpen
  \bibfield  {author} {\bibinfo {author} {\bibfnamefont {N.~L.}\ \bibnamefont {Harshman}}\ and\ \bibinfo {author} {\bibfnamefont {A.~C.}\ \bibnamefont {Knapp}},\ }\bibfield  {title} {\bibinfo {title} {Topological exchange statistics in one dimension},\ }\href {https://doi.org/10.1103/PhysRevA.105.052214} {\bibfield  {journal} {\bibinfo  {journal} {Phys. Rev. A}\ }\textbf {\bibinfo {volume} {105}},\ \bibinfo {pages} {052214} (\bibinfo {year} {2022})}\BibitemShut {NoStop}%
\bibitem [{\citenamefont {Hao}\ and\ \citenamefont {Chen}(2012)}]{hao2012DynamicalPropertiesHardcore}%
  \BibitemOpen
  \bibfield  {author} {\bibinfo {author} {\bibfnamefont {Y.}~\bibnamefont {Hao}}\ and\ \bibinfo {author} {\bibfnamefont {S.}~\bibnamefont {Chen}},\ }\bibfield  {title} {\bibinfo {title} {Dynamical properties of hard-core anyons in one-dimensional optical lattices},\ }\href {https://doi.org/10.1103/PhysRevA.86.043631} {\bibfield  {journal} {\bibinfo  {journal} {Phys. Rev. A}\ }\textbf {\bibinfo {volume} {86}},\ \bibinfo {pages} {043631} (\bibinfo {year} {2012})}\BibitemShut {NoStop}%
\bibitem [{\citenamefont {Bonkhoff}\ \emph {et~al.}(2025)\citenamefont {Bonkhoff}, \citenamefont {J{\"a}gering}, \citenamefont {Hu}, \citenamefont {Pelster}, \citenamefont {Eggert},\ and\ \citenamefont {Schneider}}]{bonkhoff2025AnyonicPhaseTransitions}%
  \BibitemOpen
  \bibfield  {author} {\bibinfo {author} {\bibfnamefont {M.}~\bibnamefont {Bonkhoff}}, \bibinfo {author} {\bibfnamefont {K.}~\bibnamefont {J{\"a}gering}}, \bibinfo {author} {\bibfnamefont {S.}~\bibnamefont {Hu}}, \bibinfo {author} {\bibfnamefont {A.}~\bibnamefont {Pelster}}, \bibinfo {author} {\bibfnamefont {S.}~\bibnamefont {Eggert}},\ and\ \bibinfo {author} {\bibfnamefont {I.}~\bibnamefont {Schneider}},\ }\bibfield  {title} {\bibinfo {title} {Anyonic {{Phase Transitions}} in the {{1D Extended Hubbard Model}} with {{Fractional Statistics}}},\ }\href {https://doi.org/10.1103/7n1c-vq2p} {\bibfield  {journal} {\bibinfo  {journal} {Phys. Rev. Lett.}\ }\textbf {\bibinfo {volume} {135}},\ \bibinfo {pages} {036601} (\bibinfo {year} {2025})}\BibitemShut {NoStop}%
\bibitem [{\citenamefont {Harshman}\ and\ \citenamefont {Knapp}(2020)}]{harshman2020AnyonsThreebodyHardcore}%
  \BibitemOpen
  \bibfield  {author} {\bibinfo {author} {\bibfnamefont {N.~L.}\ \bibnamefont {Harshman}}\ and\ \bibinfo {author} {\bibfnamefont {A.~C.}\ \bibnamefont {Knapp}},\ }\bibfield  {title} {\bibinfo {title} {Anyons from three-body hard-core interactions in one dimension},\ }\href {https://doi.org/10.1016/j.aop.2019.168003} {\bibfield  {journal} {\bibinfo  {journal} {Ann. Phys.}\ }\textbf {\bibinfo {volume} {412}},\ \bibinfo {pages} {168003} (\bibinfo {year} {2020})}\BibitemShut {NoStop}%
\bibitem [{\citenamefont {Lange}\ \emph {et~al.}(2017{\natexlab{a}})\citenamefont {Lange}, \citenamefont {Ejima},\ and\ \citenamefont {Fehske}}]{lange2017AnyonicHaldaneInsulator}%
  \BibitemOpen
  \bibfield  {author} {\bibinfo {author} {\bibfnamefont {F.}~\bibnamefont {Lange}}, \bibinfo {author} {\bibfnamefont {S.}~\bibnamefont {Ejima}},\ and\ \bibinfo {author} {\bibfnamefont {H.}~\bibnamefont {Fehske}},\ }\bibfield  {title} {\bibinfo {title} {Anyonic {{Haldane Insulator}} in {{One Dimension}}},\ }\href {https://doi.org/10.1103/PhysRevLett.118.120401} {\bibfield  {journal} {\bibinfo  {journal} {Phys. Rev. Lett.}\ }\textbf {\bibinfo {volume} {118}},\ \bibinfo {pages} {120401} (\bibinfo {year} {2017}{\natexlab{a}})}\BibitemShut {NoStop}%
\bibitem [{\citenamefont {Greenberg}\ and\ \citenamefont {Messiah}(1965)}]{greenberg1965SelectionRulesParafields}%
  \BibitemOpen
  \bibfield  {author} {\bibinfo {author} {\bibfnamefont {O.~W.}\ \bibnamefont {Greenberg}}\ and\ \bibinfo {author} {\bibfnamefont {A.~M.~L.}\ \bibnamefont {Messiah}},\ }\bibfield  {title} {\bibinfo {title} {Selection {{Rules}} for {{Parafields}} and the {{Absence}} of {{Para Particles}} in {{Nature}}},\ }\href {https://doi.org/10.1103/PhysRev.138.B1155} {\bibfield  {journal} {\bibinfo  {journal} {Phys. Rev.}\ }\textbf {\bibinfo {volume} {138}},\ \bibinfo {pages} {B1155} (\bibinfo {year} {1965})}\BibitemShut {NoStop}%
\bibitem [{\citenamefont {Fredenhagen}\ \emph {et~al.}(1989)\citenamefont {Fredenhagen}, \citenamefont {Rehren},\ and\ \citenamefont {Schroer}}]{fredenhagen1989SuperselectionSectorsBraid}%
  \BibitemOpen
  \bibfield  {author} {\bibinfo {author} {\bibfnamefont {K.}~\bibnamefont {Fredenhagen}}, \bibinfo {author} {\bibfnamefont {K.~H.}\ \bibnamefont {Rehren}},\ and\ \bibinfo {author} {\bibfnamefont {B.}~\bibnamefont {Schroer}},\ }\bibfield  {title} {\bibinfo {title} {Superselection sectors with braid group statistics and exchange algebras - {{I}}. {{General}} theory},\ }\href {https://doi.org/10.1007/BF01217906} {\bibfield  {journal} {\bibinfo  {journal} {Commun. Math. Phys}\ }\textbf {\bibinfo {volume} {125}},\ \bibinfo {pages} {201} (\bibinfo {year} {1989})}\BibitemShut {NoStop}%
\bibitem [{\citenamefont {Wang}\ and\ \citenamefont {Hazzard}(2025)}]{wang2025ParticleExchangeStatistics}%
  \BibitemOpen
  \bibfield  {author} {\bibinfo {author} {\bibfnamefont {Z.}~\bibnamefont {Wang}}\ and\ \bibinfo {author} {\bibfnamefont {K.~R.~A.}\ \bibnamefont {Hazzard}},\ }\bibfield  {title} {\bibinfo {title} {Particle exchange statistics beyond fermions and bosons},\ }\href {https://doi.org/10.1038/s41586-024-08262-7} {\bibfield  {journal} {\bibinfo  {journal} {Nature}\ }\textbf {\bibinfo {volume} {637}},\ \bibinfo {pages} {314} (\bibinfo {year} {2025})}\BibitemShut {NoStop}%
\bibitem [{\citenamefont {Leinaas}\ and\ \citenamefont {Myrheim}(1977)}]{LeinaasMyrheim1977TheoryOfIdenticalParticles}%
  \BibitemOpen
  \bibfield  {author} {\bibinfo {author} {\bibfnamefont {J.~M.}\ \bibnamefont {Leinaas}}\ and\ \bibinfo {author} {\bibfnamefont {J.}~\bibnamefont {Myrheim}},\ }\bibfield  {title} {\bibinfo {title} {On the theory of identical particles},\ }\href {https://doi.org/10.1007/BF02727953} {\bibfield  {journal} {\bibinfo  {journal} {Nuov. Cim. Soc. Ital. Fis.}\ }\textbf {\bibinfo {volume} {37B}},\ \bibinfo {pages} {1} (\bibinfo {year} {1977})}\BibitemShut {NoStop}%
\bibitem [{\citenamefont {Wilczek}(1982)}]{wilczekQuantumMechanicsFractionalSpin1982}%
  \BibitemOpen
  \bibfield  {author} {\bibinfo {author} {\bibfnamefont {F.}~\bibnamefont {Wilczek}},\ }\bibfield  {title} {\bibinfo {title} {Quantum {{Mechanics}} of {{Fractional-Spin Particles}}},\ }\href {https://doi.org/10.1103/PhysRevLett.49.957} {\bibfield  {journal} {\bibinfo  {journal} {Phys. Rev. Lett.}\ }\textbf {\bibinfo {volume} {49}},\ \bibinfo {pages} {957} (\bibinfo {year} {1982})}\BibitemShut {NoStop}%
\bibitem [{\citenamefont {Kitaev}(2006)}]{kitaevAnyonsExactlySolved2006}%
  \BibitemOpen
  \bibfield  {author} {\bibinfo {author} {\bibfnamefont {A.}~\bibnamefont {Kitaev}},\ }\bibfield  {title} {\bibinfo {title} {Anyons in an exactly solved model and beyond},\ }\href {https://doi.org/10.1016/j.aop.2005.10.005} {\bibfield  {journal} {\bibinfo  {journal} {Ann. Phys.}\ }\textbf {\bibinfo {volume} {321}},\ \bibinfo {pages} {2} (\bibinfo {year} {2006})}\BibitemShut {NoStop}%
\bibitem [{\citenamefont {Haldane}(1991)}]{haldane1991FractionalStatisticsArbitrary}%
  \BibitemOpen
  \bibfield  {author} {\bibinfo {author} {\bibfnamefont {F.~D.~M.}\ \bibnamefont {Haldane}},\ }\bibfield  {title} {\bibinfo {title} {``{{Fractional}} statistics'' in arbitrary dimensions: {{A}} generalization of the {{Pauli}} principle},\ }\href {https://doi.org/10.1103/PhysRevLett.67.937} {\bibfield  {journal} {\bibinfo  {journal} {Phys. Rev. Lett.}\ }\textbf {\bibinfo {volume} {67}},\ \bibinfo {pages} {937} (\bibinfo {year} {1991})}\BibitemShut {NoStop}%
\bibitem [{\citenamefont {Theel}\ \emph {et~al.}(2025)\citenamefont {Theel}, \citenamefont {Bonkhoff}, \citenamefont {Schmelcher}, \citenamefont {Posske},\ and\ \citenamefont {Harshman}}]{theel2025ChirallyProtectedState}%
  \BibitemOpen
  \bibfield  {author} {\bibinfo {author} {\bibfnamefont {F.}~\bibnamefont {Theel}}, \bibinfo {author} {\bibfnamefont {M.}~\bibnamefont {Bonkhoff}}, \bibinfo {author} {\bibfnamefont {P.}~\bibnamefont {Schmelcher}}, \bibinfo {author} {\bibfnamefont {T.}~\bibnamefont {Posske}},\ and\ \bibinfo {author} {\bibfnamefont {N.~L.}\ \bibnamefont {Harshman}},\ }\bibfield  {title} {\bibinfo {title} {Chirally {{Protected State Manipulation}} by {{Tuning One-Dimensional Statistics}}},\ }\href {https://doi.org/10.1103/kzf6-yx24} {\bibfield  {journal} {\bibinfo  {journal} {Phys. Rev. Lett.}\ }\textbf {\bibinfo {volume} {135}},\ \bibinfo {pages} {063401} (\bibinfo {year} {2025})}\BibitemShut {NoStop}%
\bibitem [{\citenamefont {Nayak}\ \emph {et~al.}(2008)\citenamefont {Nayak}, \citenamefont {Simon}, \citenamefont {Stern}, \citenamefont {Freedman},\ and\ \citenamefont {Das~Sarma}}]{nayak2008NonAbelianAnyonsTopological}%
  \BibitemOpen
  \bibfield  {author} {\bibinfo {author} {\bibfnamefont {C.}~\bibnamefont {Nayak}}, \bibinfo {author} {\bibfnamefont {S.~H.}\ \bibnamefont {Simon}}, \bibinfo {author} {\bibfnamefont {A.}~\bibnamefont {Stern}}, \bibinfo {author} {\bibfnamefont {M.}~\bibnamefont {Freedman}},\ and\ \bibinfo {author} {\bibfnamefont {S.}~\bibnamefont {Das~Sarma}},\ }\bibfield  {title} {\bibinfo {title} {Non-{{Abelian}} anyons and topological quantum computation},\ }\href {https://doi.org/10.1103/RevModPhys.80.1083} {\bibfield  {journal} {\bibinfo  {journal} {Rev. Mod. Phys.}\ }\textbf {\bibinfo {volume} {80}},\ \bibinfo {pages} {1083} (\bibinfo {year} {2008})}\BibitemShut {NoStop}%
\bibitem [{\citenamefont {Wigner}(1931)}]{wigner1931GrundlagenQuantenmechanik}%
  \BibitemOpen
  \bibfield  {author} {\bibinfo {author} {\bibfnamefont {E.}~\bibnamefont {Wigner}},\ }\bibfield  {title} {\bibinfo {title} {{Grundlagen der Quantenmechanik}},\ }in\ \href {https://doi.org/10.1007/978-3-663-02555-9_4} {\emph {\bibinfo {booktitle} {{Gruppentheorie und ihre Anwendung auf die Quantenmechanik der Atomspektren}}}},\ \bibinfo {editor} {edited by\ \bibinfo {editor} {\bibfnamefont {E.}~\bibnamefont {Wigner}}}\ (\bibinfo  {publisher} {Vieweg+Teubner Verlag},\ \bibinfo {address} {Wiesbaden},\ \bibinfo {year} {1931})\ pp.\ \bibinfo {pages} {34--43}\BibitemShut {NoStop}%
\bibitem [{\citenamefont {Cartan}(1926)}]{cartan1926ClasseRemarquableDespaces}%
  \BibitemOpen
  \bibfield  {author} {\bibinfo {author} {\bibfnamefont {E.}~\bibnamefont {Cartan}},\ }\bibfield  {title} {\bibinfo {title} {Sur une classe remarquable d'espaces de {{Riemann}}},\ }\href {https://doi.org/10.24033/bsmf.1105} {\bibfield  {journal} {\bibinfo  {journal} {Bull. Soc. Math. Fr.}\ }\textbf {\bibinfo {volume} {54}},\ \bibinfo {pages} {214} (\bibinfo {year} {1926})}\BibitemShut {NoStop}%
\bibitem [{\citenamefont {Altland}\ and\ \citenamefont {Zirnbauer}(1997{\natexlab{a}})}]{altland1997NonstandardSymmetryClasses}%
  \BibitemOpen
  \bibfield  {author} {\bibinfo {author} {\bibfnamefont {A.}~\bibnamefont {Altland}}\ and\ \bibinfo {author} {\bibfnamefont {M.~R.}\ \bibnamefont {Zirnbauer}},\ }\bibfield  {title} {\bibinfo {title} {Nonstandard symmetry classes in mesoscopic normal-superconducting hybrid structures},\ }\href {https://doi.org/10.1103/PhysRevB.55.1142} {\bibfield  {journal} {\bibinfo  {journal} {Phys. Rev. B}\ }\textbf {\bibinfo {volume} {55}},\ \bibinfo {pages} {1142} (\bibinfo {year} {1997}{\natexlab{a}})}\BibitemShut {NoStop}%
\bibitem [{\citenamefont {Zirnbauer}(2020)}]{zirnbauer2020ParticleholeSymmetriesCondensed}%
  \BibitemOpen
  \bibfield  {author} {\bibinfo {author} {\bibfnamefont {M.~R.}\ \bibnamefont {Zirnbauer}},\ }\bibfield  {title} {\bibinfo {title} {Particle-hole symmetries in condensed matter},\ }\bibfield  {journal} {\bibinfo  {journal} {J. Math. Phys.}\ }\textbf {\bibinfo {volume} {62}},\ \href {https://doi.org/10.1063/5.0035358} {10.1063/5.0035358} (\bibinfo {year} {2020})\BibitemShut {NoStop}%
\bibitem [{\citenamefont {Bethe}(1931)}]{betheZurTheorieMetalle1931}%
  \BibitemOpen
  \bibfield  {author} {\bibinfo {author} {\bibfnamefont {H.}~\bibnamefont {Bethe}},\ }\bibfield  {title} {\bibinfo {title} {{Zur Theorie der Metalle: I. Eigenwerte und Eigenfunktionen der linearen Atomkette}},\ }\href {https://doi.org/10.1007/BF01341708} {\bibfield  {journal} {\bibinfo  {journal} {Z. Phys.}\ }\textbf {\bibinfo {volume} {71}},\ \bibinfo {pages} {205} (\bibinfo {year} {1931})}\BibitemShut {NoStop}%
\bibitem [{\citenamefont {Essler}\ \emph {et~al.}(2010)\citenamefont {Essler}, \citenamefont {Frahm}, \citenamefont {G{\"o}hmann}, \citenamefont {Kl{\"u}mper},\ and\ \citenamefont {Korepin}}]{essler2010OnedimensionalHubbardModel}%
  \BibitemOpen
  \bibfield  {author} {\bibinfo {author} {\bibfnamefont {F.}~\bibnamefont {Essler}}, \bibinfo {author} {\bibfnamefont {H.}~\bibnamefont {Frahm}}, \bibinfo {author} {\bibfnamefont {F.}~\bibnamefont {G{\"o}hmann}}, \bibinfo {author} {\bibfnamefont {A.}~\bibnamefont {Kl{\"u}mper}},\ and\ \bibinfo {author} {\bibfnamefont {V.}~\bibnamefont {Korepin}},\ }\bibfield  {title} {\bibinfo {title} {The {O}ne-{D}imensional {H}ubbard model},\ }\href@noop {} {\bibfield  {journal} {\bibinfo  {journal} {The One-Dimensional Hubbard Model, Cambridge, UK: Cambridge University Press}\ } (\bibinfo {year} {2010})}\BibitemShut {NoStop}%
\bibitem [{\citenamefont {Baxter}(2016)}]{baxter2016exactly}%
  \BibitemOpen
  \bibfield  {author} {\bibinfo {author} {\bibfnamefont {R.}~\bibnamefont {Baxter}},\ }\href {https://books.google.de/books?id=egtcDAAAQBAJ} {\emph {\bibinfo {title} {Exactly Solved Models in Statistical Mechanics}}}\ (\bibinfo  {publisher} {Academic Press},\ \bibinfo {year} {2016})\BibitemShut {NoStop}%
\bibitem [{\citenamefont {Lieb}(1963)}]{lieb1963ExactAnalysisInteracting}%
  \BibitemOpen
  \bibfield  {author} {\bibinfo {author} {\bibfnamefont {E.~H.}\ \bibnamefont {Lieb}},\ }\bibfield  {title} {\bibinfo {title} {Exact analysis of an interacting bose gas. {{II}}. the excitation spectrum},\ }\href {https://doi.org/10.1103/PhysRev.130.1616} {\bibfield  {journal} {\bibinfo  {journal} {Phys. Rev.}\ }\textbf {\bibinfo {volume} {130}},\ \bibinfo {pages} {1616} (\bibinfo {year} {1963})}\BibitemShut {NoStop}%
\bibitem [{\citenamefont {Haldane}(1980)}]{haldaneSolidificationSolubleModel1980}%
  \BibitemOpen
  \bibfield  {author} {\bibinfo {author} {\bibfnamefont {F.~D.~M.}\ \bibnamefont {Haldane}},\ }\bibfield  {title} {\bibinfo {title} {``{{Solidification}}'' in a soluble model of bosons on a one-dimensional lattice: {{The}} ``{{Boson-Hubbard}} chain''},\ }\href {https://doi.org/10.1016/0375-9601(80)90022-5} {\bibfield  {journal} {\bibinfo  {journal} {Phys. Lett. A}\ }\textbf {\bibinfo {volume} {80}},\ \bibinfo {pages} {281} (\bibinfo {year} {1980})}\BibitemShut {NoStop}%
\bibitem [{\citenamefont {Choy}(1980)}]{choyExactResultsDegenerate1980}%
  \BibitemOpen
  \bibfield  {author} {\bibinfo {author} {\bibfnamefont {T.~C.}\ \bibnamefont {Choy}},\ }\bibfield  {title} {\bibinfo {title} {Some exact results for a degenerate {{Hubbard}} model in one dimension},\ }\href {https://doi.org/10.1016/0375-9601(80)90451-X} {\bibfield  {journal} {\bibinfo  {journal} {Phys. Lett. A}\ }\textbf {\bibinfo {volume} {80}},\ \bibinfo {pages} {49} (\bibinfo {year} {1980})}\BibitemShut {NoStop}%
\bibitem [{\citenamefont {Choy}\ and\ \citenamefont {Haldane}(1982)}]{choyFailureBetheAnsatzSolutions1982}%
  \BibitemOpen
  \bibfield  {author} {\bibinfo {author} {\bibfnamefont {T.~C.}\ \bibnamefont {Choy}}\ and\ \bibinfo {author} {\bibfnamefont {F.~D.~M.}\ \bibnamefont {Haldane}},\ }\bibfield  {title} {\bibinfo {title} {Failure of {{Bethe-Ansatz}} solutions of generalisations of the {{Hubbard}} chain to arbitrary permutation symmetry},\ }\href {https://doi.org/10.1016/0375-9601(82)90057-3} {\bibfield  {journal} {\bibinfo  {journal} {Phys. Lett. A}\ }\textbf {\bibinfo {volume} {90}},\ \bibinfo {pages} {83} (\bibinfo {year} {1982})}\BibitemShut {NoStop}%
\bibitem [{\citenamefont {Kolovsky}\ and\ \citenamefont {Buchleitner}(2004)}]{Kolovsky_2004}%
  \BibitemOpen
  \bibfield  {author} {\bibinfo {author} {\bibfnamefont {A.~R.}\ \bibnamefont {Kolovsky}}\ and\ \bibinfo {author} {\bibfnamefont {A.}~\bibnamefont {Buchleitner}},\ }\bibfield  {title} {\bibinfo {title} {Quantum chaos in the {B}ose-{H}ubbard model},\ }\href@noop {} {\bibfield  {journal} {\bibinfo  {journal} {Europhys. Lett.}\ }\textbf {\bibinfo {volume} {68}},\ \bibinfo {pages} {632} (\bibinfo {year} {2004})}\BibitemShut {NoStop}%
\bibitem [{\citenamefont {Kollath}\ \emph {et~al.}(2007)\citenamefont {Kollath}, \citenamefont {L\"auchli},\ and\ \citenamefont {Altman}}]{Kollath2007}%
  \BibitemOpen
  \bibfield  {author} {\bibinfo {author} {\bibfnamefont {C.}~\bibnamefont {Kollath}}, \bibinfo {author} {\bibfnamefont {A.~M.}\ \bibnamefont {L\"auchli}},\ and\ \bibinfo {author} {\bibfnamefont {E.}~\bibnamefont {Altman}},\ }\bibfield  {title} {\bibinfo {title} {Quench dynamics and nonequilibrium phase diagram of the {B}ose-{H}ubbard model},\ }\href {https://doi.org/10.1103/PhysRevLett.98.180601} {\bibfield  {journal} {\bibinfo  {journal} {Phys. Rev. Lett.}\ }\textbf {\bibinfo {volume} {98}},\ \bibinfo {pages} {180601} (\bibinfo {year} {2007})}\BibitemShut {NoStop}%
\bibitem [{\citenamefont {Kollath}\ \emph {et~al.}(2010)\citenamefont {Kollath}, \citenamefont {Roux}, \citenamefont {Biroli},\ and\ \citenamefont {Läuchli}}]{Kollath_2010}%
  \BibitemOpen
  \bibfield  {author} {\bibinfo {author} {\bibfnamefont {C.}~\bibnamefont {Kollath}}, \bibinfo {author} {\bibfnamefont {G.}~\bibnamefont {Roux}}, \bibinfo {author} {\bibfnamefont {G.}~\bibnamefont {Biroli}},\ and\ \bibinfo {author} {\bibfnamefont {A.~M.}\ \bibnamefont {Läuchli}},\ }\bibfield  {title} {\bibinfo {title} {Statistical properties of the spectrum of the extended {B}ose–{H}ubbard model},\ }\href {https://doi.org/10.1088/1742-5468/2010/08/P08011} {\bibfield  {journal} {\bibinfo  {journal} {J. Stat. Mech.: Theory Exp.}\ }\textbf {\bibinfo {volume} {2010}}\bibinfo  {number} { (08)},\ \bibinfo {pages} {P08011}}\BibitemShut {NoStop}%
\bibitem [{\citenamefont {de~la Cruz}\ \emph {et~al.}(2020)\citenamefont {de~la Cruz}, \citenamefont {Lerma-Hern\'andez},\ and\ \citenamefont {Hirsch}}]{Hirsch2020}%
  \BibitemOpen
\bibfield  {number} {  }\bibfield  {author} {\bibinfo {author} {\bibfnamefont {J.}~\bibnamefont {de~la Cruz}}, \bibinfo {author} {\bibfnamefont {S.}~\bibnamefont {Lerma-Hern\'andez}},\ and\ \bibinfo {author} {\bibfnamefont {J.~G.}\ \bibnamefont {Hirsch}},\ }\bibfield  {title} {\bibinfo {title} {Quantum chaos in a system with high degree of symmetries},\ }\href {https://doi.org/10.1103/PhysRevE.102.032208} {\bibfield  {journal} {\bibinfo  {journal} {Phys. Rev. E}\ }\textbf {\bibinfo {volume} {102}},\ \bibinfo {pages} {032208} (\bibinfo {year} {2020})}\BibitemShut {NoStop}%
\bibitem [{\citenamefont {Russomanno}\ \emph {et~al.}(2020)\citenamefont {Russomanno}, \citenamefont {Fava},\ and\ \citenamefont {Fazio}}]{Fazio2020}%
  \BibitemOpen
  \bibfield  {author} {\bibinfo {author} {\bibfnamefont {A.}~\bibnamefont {Russomanno}}, \bibinfo {author} {\bibfnamefont {M.}~\bibnamefont {Fava}},\ and\ \bibinfo {author} {\bibfnamefont {R.}~\bibnamefont {Fazio}},\ }\bibfield  {title} {\bibinfo {title} {Nonergodic behavior of the clean {B}ose-{H}ubbard chain},\ }\href {https://doi.org/10.1103/PhysRevB.102.144302} {\bibfield  {journal} {\bibinfo  {journal} {Phys. Rev. B}\ }\textbf {\bibinfo {volume} {102}},\ \bibinfo {pages} {144302} (\bibinfo {year} {2020})}\BibitemShut {NoStop}%
\bibitem [{\citenamefont {Links}(2021)}]{Links_2021}%
  \BibitemOpen
  \bibfield  {author} {\bibinfo {author} {\bibfnamefont {J.}~\bibnamefont {Links}},\ }\bibfield  {title} {\bibinfo {title} {The {Y}ang–{B}axter paradox},\ }\href {https://doi.org/10.1088/1751-8121/abfe48} {\bibfield  {journal} {\bibinfo  {journal} {J. Phys. A: Math. Theor.}\ }\textbf {\bibinfo {volume} {54}},\ \bibinfo {pages} {254001} (\bibinfo {year} {2021})}\BibitemShut {NoStop}%
\bibitem [{\citenamefont {Pausch}\ \emph {et~al.}(2021)\citenamefont {Pausch}, \citenamefont {Carnio}, \citenamefont {Rodr\'{\i}guez},\ and\ \citenamefont {Buchleitner}}]{Pausch2021}%
  \BibitemOpen
  \bibfield  {author} {\bibinfo {author} {\bibfnamefont {L.}~\bibnamefont {Pausch}}, \bibinfo {author} {\bibfnamefont {E.~G.}\ \bibnamefont {Carnio}}, \bibinfo {author} {\bibfnamefont {A.}~\bibnamefont {Rodr\'{\i}guez}},\ and\ \bibinfo {author} {\bibfnamefont {A.}~\bibnamefont {Buchleitner}},\ }\bibfield  {title} {\bibinfo {title} {Chaos and ergodicity across the energy spectrum of interacting bosons},\ }\href {https://doi.org/10.1103/PhysRevLett.126.150601} {\bibfield  {journal} {\bibinfo  {journal} {Phys. Rev. Lett.}\ }\textbf {\bibinfo {volume} {126}},\ \bibinfo {pages} {150601} (\bibinfo {year} {2021})}\BibitemShut {NoStop}%
\bibitem [{\citenamefont {Pausch}\ \emph {et~al.}(2022)\citenamefont {Pausch}, \citenamefont {Buchleitner}, \citenamefont {Carnio},\ and\ \citenamefont {Rodr{\'i}guez}}]{pausch2022OptimalRouteQuantum}%
  \BibitemOpen
  \bibfield  {author} {\bibinfo {author} {\bibfnamefont {L.}~\bibnamefont {Pausch}}, \bibinfo {author} {\bibfnamefont {A.}~\bibnamefont {Buchleitner}}, \bibinfo {author} {\bibfnamefont {E.~G.}\ \bibnamefont {Carnio}},\ and\ \bibinfo {author} {\bibfnamefont {A.}~\bibnamefont {Rodr{\'i}guez}},\ }\bibfield  {title} {\bibinfo {title} {Optimal route to quantum chaos in the {B}ose--{H}ubbard model},\ }\href {https://doi.org/10.1088/1751-8121/ac7e0b} {\bibfield  {journal} {\bibinfo  {journal} {J. Phys. A: Math. Theor.}\ }\textbf {\bibinfo {volume} {55}},\ \bibinfo {pages} {324002} (\bibinfo {year} {2022})}\BibitemShut {NoStop}%
\bibitem [{\citenamefont {Cazalilla}\ \emph {et~al.}(2011)\citenamefont {Cazalilla}, \citenamefont {Citro}, \citenamefont {Giamarchi}, \citenamefont {Orignac},\ and\ \citenamefont {Rigol}}]{Cazalilla2011}%
  \BibitemOpen
  \bibfield  {author} {\bibinfo {author} {\bibfnamefont {M.~A.}\ \bibnamefont {Cazalilla}}, \bibinfo {author} {\bibfnamefont {R.}~\bibnamefont {Citro}}, \bibinfo {author} {\bibfnamefont {T.}~\bibnamefont {Giamarchi}}, \bibinfo {author} {\bibfnamefont {E.}~\bibnamefont {Orignac}},\ and\ \bibinfo {author} {\bibfnamefont {M.}~\bibnamefont {Rigol}},\ }\bibfield  {title} {\bibinfo {title} {One dimensional bosons: From condensed matter systems to ultracold gases},\ }\href {https://doi.org/10.1103/RevModPhys.83.1405} {\bibfield  {journal} {\bibinfo  {journal} {Rev. Mod. Phys.}\ }\textbf {\bibinfo {volume} {83}},\ \bibinfo {pages} {1405} (\bibinfo {year} {2011})}\BibitemShut {NoStop}%
\bibitem [{\citenamefont {Pitaevskii}\ and\ \citenamefont {Stringari}(2016)}]{PitaevskiiStringariBook2016}%
  \BibitemOpen
  \bibfield  {author} {\bibinfo {author} {\bibfnamefont {L.}~\bibnamefont {Pitaevskii}}\ and\ \bibinfo {author} {\bibfnamefont {S.}~\bibnamefont {Stringari}},\ }\href {https://doi.org/10.1093/acprof:oso/9780198758884.001.0001} {\emph {\bibinfo {title} {Bose-Einstein Condensation and Superfluidity}}}\ (\bibinfo  {publisher} {Oxford University Press},\ \bibinfo {year} {2016})\BibitemShut {NoStop}%
\bibitem [{\citenamefont {Olshanii}(1998)}]{olshanii1998AtomicScatteringPresence}%
  \BibitemOpen
  \bibfield  {author} {\bibinfo {author} {\bibfnamefont {M.}~\bibnamefont {Olshanii}},\ }\bibfield  {title} {\bibinfo {title} {Atomic scattering in the presence of an external confinement and a gas of impenetrable bosons},\ }\href {https://doi.org/10.1103/PhysRevLett.81.938} {\bibfield  {journal} {\bibinfo  {journal} {Phys. Rev. Lett.}\ }\textbf {\bibinfo {volume} {81}},\ \bibinfo {pages} {938} (\bibinfo {year} {1998})}\BibitemShut {NoStop}%
\bibitem [{\citenamefont {Hofstetter}\ \emph {et~al.}(2004)\citenamefont {Hofstetter}, \citenamefont {Affleck}, \citenamefont {Nelson},\ and\ \citenamefont {Schollwöck}}]{Hofstetter2004}%
  \BibitemOpen
  \bibfield  {author} {\bibinfo {author} {\bibfnamefont {W.}~\bibnamefont {Hofstetter}}, \bibinfo {author} {\bibfnamefont {I.}~\bibnamefont {Affleck}}, \bibinfo {author} {\bibfnamefont {D.}~\bibnamefont {Nelson}},\ and\ \bibinfo {author} {\bibfnamefont {U.}~\bibnamefont {Schollwöck}},\ }\bibfield  {title} {\bibinfo {title} {Non-hermitian {L}uttinger liquids and vortex physics},\ }\href {https://doi.org/10.1209/epl/i2003-10204-2} {\bibfield  {journal} {\bibinfo  {journal} {Europhys. Lett.}\ }\textbf {\bibinfo {volume} {66}},\ \bibinfo {pages} {178} (\bibinfo {year} {2004})}\BibitemShut {NoStop}%
\bibitem [{\citenamefont {Affleck}\ \emph {et~al.}(2004)\citenamefont {Affleck}, \citenamefont {Hofstetter}, \citenamefont {Nelson},\ and\ \citenamefont {Schollwöck}}]{Affleck2004}%
  \BibitemOpen
  \bibfield  {author} {\bibinfo {author} {\bibfnamefont {I.}~\bibnamefont {Affleck}}, \bibinfo {author} {\bibfnamefont {W.}~\bibnamefont {Hofstetter}}, \bibinfo {author} {\bibfnamefont {D.~R.}\ \bibnamefont {Nelson}},\ and\ \bibinfo {author} {\bibfnamefont {U.}~\bibnamefont {Schollwöck}},\ }\bibfield  {title} {\bibinfo {title} {Non-hermitian {L}uttinger liquids and flux line pinning in planar superconductors},\ }\href {https://doi.org/10.1088/1742-5468/2004/10/P10003} {\bibfield  {journal} {\bibinfo  {journal} {J. Stat. Mech.: Theory Exp.}\ }\textbf {\bibinfo {volume} {2004}}\bibinfo  {number} { (10)},\ \bibinfo {pages} {P10003}}\BibitemShut {NoStop}%
\bibitem [{\citenamefont {Oelkers}\ and\ \citenamefont {Links}(2007)}]{oelkersGroundstatePropertiesAttractive2007}%
  \BibitemOpen
\bibfield  {number} {  }\bibfield  {author} {\bibinfo {author} {\bibfnamefont {N.}~\bibnamefont {Oelkers}}\ and\ \bibinfo {author} {\bibfnamefont {J.}~\bibnamefont {Links}},\ }\bibfield  {title} {\bibinfo {title} {Ground-state properties of the attractive one-dimensional {B}ose-{H}ubbard model},\ }\href {https://doi.org/10.1103/PhysRevB.75.115119} {\bibfield  {journal} {\bibinfo  {journal} {Phys. Rev. B}\ }\textbf {\bibinfo {volume} {75}},\ \bibinfo {pages} {115119} (\bibinfo {year} {2007})}\BibitemShut {NoStop}%
\bibitem [{\citenamefont {Valiente}\ and\ \citenamefont {Petrosyan}(2008{\natexlab{a}})}]{valienteTwoparticleStatesHubbard2008}%
  \BibitemOpen
  \bibfield  {author} {\bibinfo {author} {\bibfnamefont {M.}~\bibnamefont {Valiente}}\ and\ \bibinfo {author} {\bibfnamefont {D.}~\bibnamefont {Petrosyan}},\ }\bibfield  {title} {\bibinfo {title} {Two-particle states in the {{Hubbard}} model},\ }\href {https://doi.org/10.1088/0953-4075/41/16/161002} {\bibfield  {journal} {\bibinfo  {journal} {J. Phys. B: At. Mol. Opt. Phys.}\ }\textbf {\bibinfo {volume} {41}},\ \bibinfo {pages} {161002} (\bibinfo {year} {2008}{\natexlab{a}})}\BibitemShut {NoStop}%
\bibitem [{\citenamefont {Valiente}\ and\ \citenamefont {Petrosyan}(2009)}]{valienteScatteringResonancesTwoparticle2009}%
  \BibitemOpen
  \bibfield  {author} {\bibinfo {author} {\bibfnamefont {M.}~\bibnamefont {Valiente}}\ and\ \bibinfo {author} {\bibfnamefont {D.}~\bibnamefont {Petrosyan}},\ }\bibfield  {title} {\bibinfo {title} {Scattering resonances and two-particle bound states of the extended {{Hubbard}} model},\ }\href {https://doi.org/10.1088/0953-4075/42/12/121001} {\bibfield  {journal} {\bibinfo  {journal} {J. Phys. B: At. Mol. Opt. Phys.}\ }\textbf {\bibinfo {volume} {42}},\ \bibinfo {pages} {121001} (\bibinfo {year} {2009})}\BibitemShut {NoStop}%
\bibitem [{\citenamefont {Zhang}\ \emph {et~al.}(2013{\natexlab{a}})\citenamefont {Zhang}, \citenamefont {Braak},\ and\ \citenamefont {Kollar}}]{Kollar2013}%
  \BibitemOpen
  \bibfield  {author} {\bibinfo {author} {\bibfnamefont {J.~M.}\ \bibnamefont {Zhang}}, \bibinfo {author} {\bibfnamefont {D.}~\bibnamefont {Braak}},\ and\ \bibinfo {author} {\bibfnamefont {M.}~\bibnamefont {Kollar}},\ }\bibfield  {title} {\bibinfo {title} {Bound states in the one-dimensional two-particle {H}ubbard model with an impurity},\ }\href {https://doi.org/10.1103/PhysRevA.87.023613} {\bibfield  {journal} {\bibinfo  {journal} {Phys. Rev. A}\ }\textbf {\bibinfo {volume} {87}},\ \bibinfo {pages} {023613} (\bibinfo {year} {2013}{\natexlab{a}})}\BibitemShut {NoStop}%
\bibitem [{\citenamefont {Boschi}\ \emph {et~al.}(2014)\citenamefont {Boschi}, \citenamefont {Ercolessi}, \citenamefont {Ferrari}, \citenamefont {Naldesi}, \citenamefont {Ortolani},\ and\ \citenamefont {Taddia}}]{boschiBoundStatesExpansion2014}%
  \BibitemOpen
  \bibfield  {author} {\bibinfo {author} {\bibfnamefont {C.~D.~E.}\ \bibnamefont {Boschi}}, \bibinfo {author} {\bibfnamefont {E.}~\bibnamefont {Ercolessi}}, \bibinfo {author} {\bibfnamefont {L.}~\bibnamefont {Ferrari}}, \bibinfo {author} {\bibfnamefont {P.}~\bibnamefont {Naldesi}}, \bibinfo {author} {\bibfnamefont {F.}~\bibnamefont {Ortolani}},\ and\ \bibinfo {author} {\bibfnamefont {L.}~\bibnamefont {Taddia}},\ }\bibfield  {title} {\bibinfo {title} {Bound states and expansion dynamics of interacting bosons on a one-dimensional lattice},\ }\href {https://doi.org/10.1103/PhysRevA.90.043606} {\bibfield  {journal} {\bibinfo  {journal} {Phys. Rev. A}\ }\textbf {\bibinfo {volume} {90}},\ \bibinfo {pages} {043606} (\bibinfo {year} {2014})}\BibitemShut {NoStop}%
\bibitem [{\citenamefont {Longhi}\ and\ \citenamefont {Valle}(2013)}]{longhiTammHubbardSurface2013}%
  \BibitemOpen
  \bibfield  {author} {\bibinfo {author} {\bibfnamefont {S.}~\bibnamefont {Longhi}}\ and\ \bibinfo {author} {\bibfnamefont {G.~D.}\ \bibnamefont {Valle}},\ }\bibfield  {title} {\bibinfo {title} {Tamm--{{Hubbard}} surface states in the continuum},\ }\href {https://doi.org/10.1088/0953-8984/25/23/235601} {\bibfield  {journal} {\bibinfo  {journal} {J. Phys.: Condens. Matter}\ }\textbf {\bibinfo {volume} {25}},\ \bibinfo {pages} {235601} (\bibinfo {year} {2013})}\BibitemShut {NoStop}%
\bibitem [{\citenamefont {Polo}\ \emph {et~al.}(2020)\citenamefont {Polo}, \citenamefont {Naldesi}, \citenamefont {Minguzzi},\ and\ \citenamefont {Amico}}]{poloExactResultsPersistent2020}%
  \BibitemOpen
  \bibfield  {author} {\bibinfo {author} {\bibfnamefont {J.}~\bibnamefont {Polo}}, \bibinfo {author} {\bibfnamefont {P.}~\bibnamefont {Naldesi}}, \bibinfo {author} {\bibfnamefont {A.}~\bibnamefont {Minguzzi}},\ and\ \bibinfo {author} {\bibfnamefont {L.}~\bibnamefont {Amico}},\ }\bibfield  {title} {\bibinfo {title} {Exact results for persistent currents of two bosons in a ring lattice},\ }\href {https://doi.org/10.1103/PhysRevA.101.043418} {\bibfield  {journal} {\bibinfo  {journal} {Phys. Rev. A}\ }\textbf {\bibinfo {volume} {101}},\ \bibinfo {pages} {043418} (\bibinfo {year} {2020})}\BibitemShut {NoStop}%
\bibitem [{\citenamefont {Li}\ \emph {et~al.}(2022)\citenamefont {Li}, \citenamefont {Schneble},\ and\ \citenamefont {Wei}}]{liTwoparticleStatesOnedimensional2022}%
  \BibitemOpen
  \bibfield  {author} {\bibinfo {author} {\bibfnamefont {Y.}~\bibnamefont {Li}}, \bibinfo {author} {\bibfnamefont {D.}~\bibnamefont {Schneble}},\ and\ \bibinfo {author} {\bibfnamefont {T.-C.}\ \bibnamefont {Wei}},\ }\bibfield  {title} {\bibinfo {title} {Two-particle states in one-dimensional coupled {{Bose-Hubbard}} models},\ }\href {https://doi.org/10.1103/PhysRevA.105.053310} {\bibfield  {journal} {\bibinfo  {journal} {Phys. Rev. A}\ }\textbf {\bibinfo {volume} {105}},\ \bibinfo {pages} {053310} (\bibinfo {year} {2022})}\BibitemShut {NoStop}%
\bibitem [{\citenamefont {Zhang}\ \emph {et~al.}(2017)\citenamefont {Zhang}, \citenamefont {Greschner}, \citenamefont {Fan}, \citenamefont {Scott},\ and\ \citenamefont {Zhang}}]{zhang2017GroundstatePropertiesOnedimensional}%
  \BibitemOpen
  \bibfield  {author} {\bibinfo {author} {\bibfnamefont {W.}~\bibnamefont {Zhang}}, \bibinfo {author} {\bibfnamefont {S.}~\bibnamefont {Greschner}}, \bibinfo {author} {\bibfnamefont {E.}~\bibnamefont {Fan}}, \bibinfo {author} {\bibfnamefont {T.~C.}\ \bibnamefont {Scott}},\ and\ \bibinfo {author} {\bibfnamefont {Y.}~\bibnamefont {Zhang}},\ }\bibfield  {title} {\bibinfo {title} {Ground-state properties of the one-dimensional unconstrained pseudo-anyon {{Hubbard}} model},\ }\href {https://doi.org/10.1103/PhysRevA.95.053614} {\bibfield  {journal} {\bibinfo  {journal} {Phys. Rev. A}\ }\textbf {\bibinfo {volume} {95}},\ \bibinfo {pages} {053614} (\bibinfo {year} {2017})}\BibitemShut {NoStop}%
\bibitem [{\citenamefont {Longhi}\ and\ \citenamefont {Della~Valle}(2012)}]{longhiAnyonicBlochOscillations2012}%
  \BibitemOpen
  \bibfield  {author} {\bibinfo {author} {\bibfnamefont {S.}~\bibnamefont {Longhi}}\ and\ \bibinfo {author} {\bibfnamefont {G.}~\bibnamefont {Della~Valle}},\ }\bibfield  {title} {\bibinfo {title} {Anyonic {{Bloch}} oscillations},\ }\href {https://doi.org/10.1103/PhysRevB.85.165144} {\bibfield  {journal} {\bibinfo  {journal} {Phys. Rev. B}\ }\textbf {\bibinfo {volume} {85}},\ \bibinfo {pages} {165144} (\bibinfo {year} {2012})}\BibitemShut {NoStop}%
\bibitem [{\citenamefont {Longhi}\ and\ \citenamefont {Valle}(2012)}]{longhiAnyonsOnedimensionalLattices2012}%
  \BibitemOpen
  \bibfield  {author} {\bibinfo {author} {\bibfnamefont {S.}~\bibnamefont {Longhi}}\ and\ \bibinfo {author} {\bibfnamefont {G.~D.}\ \bibnamefont {Valle}},\ }\bibfield  {title} {\bibinfo {title} {Anyons in one-dimensional lattices: A photonic realization},\ }\href {https://doi.org/10.1364/OL.37.002160} {\bibfield  {journal} {\bibinfo  {journal} {Opt. Lett.}\ }\textbf {\bibinfo {volume} {37}},\ \bibinfo {pages} {2160} (\bibinfo {year} {2012})}\BibitemShut {NoStop}%
\bibitem [{\citenamefont {Lau}\ and\ \citenamefont {Dutta}(2022)}]{lauQuantumWalkTwo2022}%
  \BibitemOpen
  \bibfield  {author} {\bibinfo {author} {\bibfnamefont {L.~L.}\ \bibnamefont {Lau}}\ and\ \bibinfo {author} {\bibfnamefont {S.}~\bibnamefont {Dutta}},\ }\bibfield  {title} {\bibinfo {title} {Quantum walk of two anyons across a statistical boundary},\ }\href {https://doi.org/10.1103/PhysRevResearch.4.L012007} {\bibfield  {journal} {\bibinfo  {journal} {Phys. Rev. Res.}\ }\textbf {\bibinfo {volume} {4}},\ \bibinfo {pages} {L012007} (\bibinfo {year} {2022})}\BibitemShut {NoStop}%
\bibitem [{\citenamefont {Zheng}\ \emph {et~al.}(2024)\citenamefont {Zheng}, \citenamefont {Xie}, \citenamefont {Zhang}, \citenamefont {Chen},\ and\ \citenamefont {Zhang}}]{zheng2024NecessityOrthogonalBasis}%
  \BibitemOpen
  \bibfield  {author} {\bibinfo {author} {\bibfnamefont {C.}~\bibnamefont {Zheng}}, \bibinfo {author} {\bibfnamefont {J.}~\bibnamefont {Xie}}, \bibinfo {author} {\bibfnamefont {M.}~\bibnamefont {Zhang}}, \bibinfo {author} {\bibfnamefont {Y.}~\bibnamefont {Chen}},\ and\ \bibinfo {author} {\bibfnamefont {Y.}~\bibnamefont {Zhang}},\ }\bibfield  {title} {\bibinfo {title} {Necessity of orthogonal basis vectors for the two-anyon problem in a one-dimensional lattice},\ }\href {https://doi.org/10.1088/1572-9494/ad7372} {\bibfield  {journal} {\bibinfo  {journal} {Commun. Theor. Phys.}\ }\textbf {\bibinfo {volume} {76}},\ \bibinfo {pages} {125103} (\bibinfo {year} {2024})}\BibitemShut {NoStop}%
\bibitem [{\citenamefont {Bonkhoff}\ \emph {et~al.}(2023)\citenamefont {Bonkhoff}, \citenamefont {J{\"a}ger}, \citenamefont {Schneider}, \citenamefont {Pelster},\ and\ \citenamefont {Eggert}}]{bonkhoffCoherencePropertiesRepulsive2023}%
  \BibitemOpen
  \bibfield  {author} {\bibinfo {author} {\bibfnamefont {M.}~\bibnamefont {Bonkhoff}}, \bibinfo {author} {\bibfnamefont {S.~B.}\ \bibnamefont {J{\"a}ger}}, \bibinfo {author} {\bibfnamefont {I.}~\bibnamefont {Schneider}}, \bibinfo {author} {\bibfnamefont {A.}~\bibnamefont {Pelster}},\ and\ \bibinfo {author} {\bibfnamefont {S.}~\bibnamefont {Eggert}},\ }\bibfield  {title} {\bibinfo {title} {Coherence properties of the repulsive anyon-{{Hubbard}} dimer},\ }\href {https://doi.org/10.1103/PhysRevB.108.155134} {\bibfield  {journal} {\bibinfo  {journal} {Phys. Rev. B}\ }\textbf {\bibinfo {volume} {108}},\ \bibinfo {pages} {155134} (\bibinfo {year} {2023})}\BibitemShut {NoStop}%
\bibitem [{\citenamefont {Zirnbauer}(2015)}]{zirnbauerSymmetryClassesRandom2004}%
  \BibitemOpen
  \bibfield  {author} {\bibinfo {author} {\bibfnamefont {M.~R.}\ \bibnamefont {Zirnbauer}},\ }\bibfield  {title} {\bibinfo {title} {Symmetry classes},\ }in\ \href {https://doi.org/10.1093/oxfordhb/9780198744191.013.3} {\emph {\bibinfo {booktitle} {The Oxford Handbook of Random Matrix Theory}}},\ \bibinfo {editor} {edited by\ \bibinfo {editor} {\bibfnamefont {G.}~\bibnamefont {Akemann}}, \bibinfo {editor} {\bibfnamefont {J.}~\bibnamefont {Baik}},\ and\ \bibinfo {editor} {\bibfnamefont {P.}~\bibnamefont {Di~Francesco}}}\ (\bibinfo  {publisher} {Oxford University Press},\ \bibinfo {year} {2015})\BibitemShut {NoStop}%
\bibitem [{\citenamefont {Tonks}(1936)}]{tonks1936CompleteEquationState}%
  \BibitemOpen
  \bibfield  {author} {\bibinfo {author} {\bibfnamefont {L.}~\bibnamefont {Tonks}},\ }\bibfield  {title} {\bibinfo {title} {The complete equation of state of one, two and three-dimensional gases of hard elastic spheres},\ }\href {https://doi.org/10.1103/PhysRev.50.955} {\bibfield  {journal} {\bibinfo  {journal} {Phys. Rev.}\ }\textbf {\bibinfo {volume} {50}},\ \bibinfo {pages} {955} (\bibinfo {year} {1936})}\BibitemShut {NoStop}%
\bibitem [{\citenamefont {Girardeau}(1960)}]{girardeau1960RelationshipSystemsImpenetrable}%
  \BibitemOpen
  \bibfield  {author} {\bibinfo {author} {\bibfnamefont {M.}~\bibnamefont {Girardeau}},\ }\bibfield  {title} {\bibinfo {title} {Relationship between systems of impenetrable bosons and fermions in one dimension},\ }\href {https://doi.org/10.1063/1.1703687} {\bibfield  {journal} {\bibinfo  {journal} {J. Math. Phys.}\ }\textbf {\bibinfo {volume} {1}},\ \bibinfo {pages} {516} (\bibinfo {year} {1960})}\BibitemShut {NoStop}%
\bibitem [{\citenamefont {Zhang}\ \emph {et~al.}(2023{\natexlab{a}})\citenamefont {Zhang}, \citenamefont {Qian}, \citenamefont {Sun},\ and\ \citenamefont {Zhang}}]{zhangAnyonicBoundStates2023}%
  \BibitemOpen
  \bibfield  {author} {\bibinfo {author} {\bibfnamefont {W.}~\bibnamefont {Zhang}}, \bibinfo {author} {\bibfnamefont {L.}~\bibnamefont {Qian}}, \bibinfo {author} {\bibfnamefont {H.}~\bibnamefont {Sun}},\ and\ \bibinfo {author} {\bibfnamefont {X.}~\bibnamefont {Zhang}},\ }\bibfield  {title} {\bibinfo {title} {Anyonic bound states in the continuum},\ }\href {https://doi.org/10.1038/s42005-023-01245-6} {\bibfield  {journal} {\bibinfo  {journal} {Commun. Phys.}\ }\textbf {\bibinfo {volume} {6}},\ \bibinfo {pages} {1} (\bibinfo {year} {2023}{\natexlab{a}})}\BibitemShut {NoStop}%
\bibitem [{\citenamefont {Essler}\ \emph {et~al.}(2005)\citenamefont {Essler}, \citenamefont {Frahm}, \citenamefont {G{\"o}hmann}, \citenamefont {Kl{\"u}mper},\ and\ \citenamefont {Korepin}}]{essler2005OneDimensionalHubbardModel}%
  \BibitemOpen
  \bibfield  {author} {\bibinfo {author} {\bibfnamefont {F.~H.~L.}\ \bibnamefont {Essler}}, \bibinfo {author} {\bibfnamefont {H.}~\bibnamefont {Frahm}}, \bibinfo {author} {\bibfnamefont {F.}~\bibnamefont {G{\"o}hmann}}, \bibinfo {author} {\bibfnamefont {A.}~\bibnamefont {Kl{\"u}mper}},\ and\ \bibinfo {author} {\bibfnamefont {V.~E.}\ \bibnamefont {Korepin}},\ }\href@noop {} {\emph {\bibinfo {title} {The {{One-Dimensional Hubbard Model}}}}}\ (\bibinfo  {publisher} {Cambridge University Press},\ \bibinfo {year} {2005})\BibitemShut {NoStop}%
\bibitem [{\citenamefont {Batchelor}\ \emph {et~al.}(2008)\citenamefont {Batchelor}, \citenamefont {Foerster}, \citenamefont {Guan}, \citenamefont {Links},\ and\ \citenamefont {Zhou}}]{Batchelor_2008}%
  \BibitemOpen
  \bibfield  {author} {\bibinfo {author} {\bibfnamefont {M.~T.}\ \bibnamefont {Batchelor}}, \bibinfo {author} {\bibfnamefont {A.}~\bibnamefont {Foerster}}, \bibinfo {author} {\bibfnamefont {X.-W.}\ \bibnamefont {Guan}}, \bibinfo {author} {\bibfnamefont {J.}~\bibnamefont {Links}},\ and\ \bibinfo {author} {\bibfnamefont {H.-Q.}\ \bibnamefont {Zhou}},\ }\bibfield  {title} {\bibinfo {title} {The quantum inverse scattering method with anyonic grading},\ }\href {https://doi.org/10.1088/1751-8113/41/46/465201} {\bibfield  {journal} {\bibinfo  {journal} {J. Phys. A: Math. Theor.}\ }\textbf {\bibinfo {volume} {41}},\ \bibinfo {pages} {465201} (\bibinfo {year} {2008})}\BibitemShut {NoStop}%
\bibitem [{\citenamefont {Kundu}(2010)}]{kundu2010QuantumIntegrable1D}%
  \BibitemOpen
  \bibfield  {author} {\bibinfo {author} {\bibfnamefont {A.}~\bibnamefont {Kundu}},\ }\bibfield  {title} {\bibinfo {title} {Quantum {{Integrable 1D}} anyonic {{Models}}: {{Construction}} through {{Braided Yang-Baxter Equation}}},\ }\href {https://doi.org/10.3842/SIGMA.2010.080} {\bibfield  {journal} {\bibinfo  {journal} {SIGMA. Symmetry, Integrability and Geometry: Methods and Applications}\ }\textbf {\bibinfo {volume} {6}},\ \bibinfo {pages} {080} (\bibinfo {year} {2010})}\BibitemShut {NoStop}%
\bibitem [{\citenamefont {P{\^a}tu}\ \emph {et~al.}(2007)\citenamefont {P{\^a}tu}, \citenamefont {Korepin},\ and\ \citenamefont {Averin}}]{patuCorrelationFunctionsOnedimensional2007}%
  \BibitemOpen
  \bibfield  {author} {\bibinfo {author} {\bibfnamefont {O.~I.}\ \bibnamefont {P{\^a}tu}}, \bibinfo {author} {\bibfnamefont {V.~E.}\ \bibnamefont {Korepin}},\ and\ \bibinfo {author} {\bibfnamefont {D.~V.}\ \bibnamefont {Averin}},\ }\bibfield  {title} {\bibinfo {title} {Correlation functions of one-dimensional {{Lieb}}--{{Liniger}} anyons},\ }\href {https://doi.org/10.1088/1751-8113/40/50/004} {\bibfield  {journal} {\bibinfo  {journal} {J. Phys. A: Math. Theor.}\ }\textbf {\bibinfo {volume} {40}},\ \bibinfo {pages} {14963} (\bibinfo {year} {2007})}\BibitemShut {NoStop}%
\bibitem [{\citenamefont {Batchelor}\ \emph {et~al.}(2007)\citenamefont {Batchelor}, \citenamefont {Guan},\ and\ \citenamefont {He}}]{batchelorBetheAnsatz1D2007}%
  \BibitemOpen
  \bibfield  {author} {\bibinfo {author} {\bibfnamefont {M.~T.}\ \bibnamefont {Batchelor}}, \bibinfo {author} {\bibfnamefont {X.-W.}\ \bibnamefont {Guan}},\ and\ \bibinfo {author} {\bibfnamefont {J.-S.}\ \bibnamefont {He}},\ }\bibfield  {title} {\bibinfo {title} {The {{Bethe}} ansatz for {{1D}} interacting anyons},\ }\href {https://doi.org/10.1088/1742-5468/2007/03/P03007} {\bibfield  {journal} {\bibinfo  {journal} {J. Stat. Mech.: Theory Exp.}\ }\textbf {\bibinfo {volume} {2007}}\bibinfo  {number} { (03)},\ \bibinfo {pages} {P03007}}\BibitemShut {NoStop}%
\bibitem [{\citenamefont {Serbyn}\ \emph {et~al.}(2021)\citenamefont {Serbyn}, \citenamefont {Abanin},\ and\ \citenamefont {Papi{\'c}}}]{serbyn2021QuantumManybodyScars}%
  \BibitemOpen
\bibfield  {number} {  }\bibfield  {author} {\bibinfo {author} {\bibfnamefont {M.}~\bibnamefont {Serbyn}}, \bibinfo {author} {\bibfnamefont {D.~A.}\ \bibnamefont {Abanin}},\ and\ \bibinfo {author} {\bibfnamefont {Z.}~\bibnamefont {Papi{\'c}}},\ }\bibfield  {title} {\bibinfo {title} {Quantum many-body scars and weak breaking of ergodicity},\ }\href {https://doi.org/10.1038/s41567-021-01230-2} {\bibfield  {journal} {\bibinfo  {journal} {Nat. Phys.}\ }\textbf {\bibinfo {volume} {17}},\ \bibinfo {pages} {675} (\bibinfo {year} {2021})}\BibitemShut {NoStop}%
\bibitem [{\citenamefont {Banerjee}\ and\ \citenamefont {Sen}(2021)}]{banerjeeQuantumScarsZero2021}%
  \BibitemOpen
  \bibfield  {author} {\bibinfo {author} {\bibfnamefont {D.}~\bibnamefont {Banerjee}}\ and\ \bibinfo {author} {\bibfnamefont {A.}~\bibnamefont {Sen}},\ }\bibfield  {title} {\bibinfo {title} {Quantum {{Scars}} from {{Zero Modes}} in an {{Abelian Lattice Gauge Theory}} on {{Ladders}}},\ }\href {https://doi.org/10.1103/PhysRevLett.126.220601} {\bibfield  {journal} {\bibinfo  {journal} {Phys. Rev. Lett.}\ }\textbf {\bibinfo {volume} {126}},\ \bibinfo {pages} {220601} (\bibinfo {year} {2021})}\BibitemShut {NoStop}%
\bibitem [{\citenamefont {Turner}\ \emph {et~al.}(2025)\citenamefont {Turner}, \citenamefont {Szyniszewski}, \citenamefont {Mukherjee}, \citenamefont {Melendrez}, \citenamefont {Changlani},\ and\ \citenamefont {Pal}}]{turnerStableInfinitetemperatureEigenstates2025}%
  \BibitemOpen
  \bibfield  {author} {\bibinfo {author} {\bibfnamefont {C.~J.}\ \bibnamefont {Turner}}, \bibinfo {author} {\bibfnamefont {M.}~\bibnamefont {Szyniszewski}}, \bibinfo {author} {\bibfnamefont {B.}~\bibnamefont {Mukherjee}}, \bibinfo {author} {\bibfnamefont {R.}~\bibnamefont {Melendrez}}, \bibinfo {author} {\bibfnamefont {H.~J.}\ \bibnamefont {Changlani}},\ and\ \bibinfo {author} {\bibfnamefont {A.}~\bibnamefont {Pal}},\ }\href {https://doi.org/10.48550/arXiv.2407.11956} {\bibinfo {title} {Stable infinite-temperature eigenstates in {{SU}}(2)-symmetric nonintegrable models}} (\bibinfo {year} {2025}),\ \Eprint {https://arxiv.org/abs/2407.11956} {arXiv:2407.11956 [quant-ph]} \BibitemShut {NoStop}%
\bibitem [{\citenamefont {Tang}\ \emph {et~al.}(2015)\citenamefont {Tang}, \citenamefont {Eggert},\ and\ \citenamefont {Pelster}}]{tangGroundstatePropertiesAnyons2015}%
  \BibitemOpen
  \bibfield  {author} {\bibinfo {author} {\bibfnamefont {G.}~\bibnamefont {Tang}}, \bibinfo {author} {\bibfnamefont {S.}~\bibnamefont {Eggert}},\ and\ \bibinfo {author} {\bibfnamefont {A.}~\bibnamefont {Pelster}},\ }\bibfield  {title} {\bibinfo {title} {Ground-state properties of anyons in a one-dimensional lattice},\ }\href {https://doi.org/10.1088/1367-2630/17/12/123016} {\bibfield  {journal} {\bibinfo  {journal} {New J. Phys.}\ }\textbf {\bibinfo {volume} {17}},\ \bibinfo {pages} {123016} (\bibinfo {year} {2015})}\BibitemShut {NoStop}%
\bibitem [{\citenamefont {Lange}\ \emph {et~al.}(2017{\natexlab{b}})\citenamefont {Lange}, \citenamefont {Ejima},\ and\ \citenamefont {Fehske}}]{langeAnyonicHaldaneInsulator2017}%
  \BibitemOpen
  \bibfield  {author} {\bibinfo {author} {\bibfnamefont {F.}~\bibnamefont {Lange}}, \bibinfo {author} {\bibfnamefont {S.}~\bibnamefont {Ejima}},\ and\ \bibinfo {author} {\bibfnamefont {H.}~\bibnamefont {Fehske}},\ }\bibfield  {title} {\bibinfo {title} {Anyonic {{Haldane Insulator}} in {{One Dimension}}},\ }\href {https://doi.org/10.1103/PhysRevLett.118.120401} {\bibfield  {journal} {\bibinfo  {journal} {Phys. Rev. Lett.}\ }\textbf {\bibinfo {volume} {118}},\ \bibinfo {pages} {120401} (\bibinfo {year} {2017}{\natexlab{b}})}\BibitemShut {NoStop}%
\bibitem [{\citenamefont {Lieb}\ \emph {et~al.}(1961)\citenamefont {Lieb}, \citenamefont {Schultz},\ and\ \citenamefont {Mattis}}]{liebTwoSolubleModels1961}%
  \BibitemOpen
  \bibfield  {author} {\bibinfo {author} {\bibfnamefont {E.}~\bibnamefont {Lieb}}, \bibinfo {author} {\bibfnamefont {T.}~\bibnamefont {Schultz}},\ and\ \bibinfo {author} {\bibfnamefont {D.}~\bibnamefont {Mattis}},\ }\bibfield  {title} {\bibinfo {title} {Two soluble models of an antiferromagnetic chain},\ }\href {https://doi.org/10.1016/0003-4916(61)90115-4} {\bibfield  {journal} {\bibinfo  {journal} {Ann. Phys.}\ }\textbf {\bibinfo {volume} {16}},\ \bibinfo {pages} {407} (\bibinfo {year} {1961})}\BibitemShut {NoStop}%
\bibitem [{\citenamefont {Capel}\ and\ \citenamefont {Siskens}(1975)}]{capelNoteDifferenceAcyclic1975}%
  \BibitemOpen
  \bibfield  {author} {\bibinfo {author} {\bibfnamefont {H.~W.}\ \bibnamefont {Capel}}\ and\ \bibinfo {author} {\bibfnamefont {{\relax Th}.~J.}\ \bibnamefont {Siskens}},\ }\bibfield  {title} {\bibinfo {title} {Note on the difference between the a-cyclic and c-cyclic version of the {{XY}} model},\ }\href {https://doi.org/10.1016/0378-4371(75)90064-3} {\bibfield  {journal} {\bibinfo  {journal} {Physica A}\ }\textbf {\bibinfo {volume} {81}},\ \bibinfo {pages} {214} (\bibinfo {year} {1975})}\BibitemShut {NoStop}%
\bibitem [{\citenamefont {Seiberg}\ and\ \citenamefont {Shao}(2024)}]{SeibergShao2024}%
  \BibitemOpen
  \bibfield  {author} {\bibinfo {author} {\bibfnamefont {N.}~\bibnamefont {Seiberg}}\ and\ \bibinfo {author} {\bibfnamefont {S.-H.}\ \bibnamefont {Shao}},\ }\bibfield  {title} {\bibinfo {title} {{Majorana chain and Ising model - (non-invertible) translations, anomalies, and emanant symmetries}},\ }\href {https://doi.org/10.21468/SciPostPhys.16.3.064} {\bibfield  {journal} {\bibinfo  {journal} {SciPost Phys.}\ }\textbf {\bibinfo {volume} {16}},\ \bibinfo {pages} {064} (\bibinfo {year} {2024})}\BibitemShut {NoStop}%
\bibitem [{\citenamefont {{Bose}}(1924)}]{bose1924PlancksGesetzUnd}%
  \BibitemOpen
  \bibfield  {author} {\bibinfo {author} {\bibnamefont {{Bose}}},\ }\bibfield  {title} {\bibinfo {title} {{Plancks Gesetz und Lichtquantenhypothese}},\ }\href {https://doi.org/10.1007/BF01327326} {\bibfield  {journal} {\bibinfo  {journal} {Z. Physik}\ }\textbf {\bibinfo {volume} {26}},\ \bibinfo {pages} {178} (\bibinfo {year} {1924})}\BibitemShut {NoStop}%
\bibitem [{\citenamefont {Altland}\ and\ \citenamefont {Zirnbauer}(1997{\natexlab{b}})}]{altlandNonstandardSymmetryClasses1997}%
  \BibitemOpen
  \bibfield  {author} {\bibinfo {author} {\bibfnamefont {A.}~\bibnamefont {Altland}}\ and\ \bibinfo {author} {\bibfnamefont {M.~R.}\ \bibnamefont {Zirnbauer}},\ }\bibfield  {title} {\bibinfo {title} {Nonstandard symmetry classes in mesoscopic normal-superconducting hybrid structures},\ }\href {https://doi.org/10.1103/PhysRevB.55.1142} {\bibfield  {journal} {\bibinfo  {journal} {Phys. Rev. B}\ }\textbf {\bibinfo {volume} {55}},\ \bibinfo {pages} {1142} (\bibinfo {year} {1997}{\natexlab{b}})}\BibitemShut {NoStop}%
\bibitem [{\citenamefont {Hamermesh}(1989)}]{hamermeshGroupTheoryIts1989}%
  \BibitemOpen
  \bibfield  {author} {\bibinfo {author} {\bibfnamefont {M.}~\bibnamefont {Hamermesh}},\ }\href@noop {} {\emph {\bibinfo {title} {Group {{Theory}} and {{Its Application}} to {{Physical Problems}}}}},\ \bibinfo {edition} {reprint edition}\ ed.\ (\bibinfo  {publisher} {Dover Publications},\ \bibinfo {address} {New York},\ \bibinfo {year} {1989})\BibitemShut {NoStop}%
\bibitem [{\citenamefont {Ryu}\ \emph {et~al.}(2010)\citenamefont {Ryu}, \citenamefont {Schnyder}, \citenamefont {Furusaki},\ and\ \citenamefont {Ludwig}}]{ryu2010TopologicalInsulatorsSuperconductors}%
  \BibitemOpen
  \bibfield  {author} {\bibinfo {author} {\bibfnamefont {S.}~\bibnamefont {Ryu}}, \bibinfo {author} {\bibfnamefont {A.~P.}\ \bibnamefont {Schnyder}}, \bibinfo {author} {\bibfnamefont {A.}~\bibnamefont {Furusaki}},\ and\ \bibinfo {author} {\bibfnamefont {A.~W.~W.}\ \bibnamefont {Ludwig}},\ }\bibfield  {title} {\bibinfo {title} {Topological insulators and superconductors: Tenfold way and dimensional hierarchy},\ }\href {https://doi.org/10.1088/1367-2630/12/6/065010} {\bibfield  {journal} {\bibinfo  {journal} {New J. Phys.}\ }\textbf {\bibinfo {volume} {12}},\ \bibinfo {pages} {065010} (\bibinfo {year} {2010})}\BibitemShut {NoStop}%
\bibitem [{\citenamefont {Bohigas}\ and\ \citenamefont {Giannoni}(1984)}]{bohigasChaoticMotionRandom1984}%
  \BibitemOpen
  \bibfield  {author} {\bibinfo {author} {\bibfnamefont {O.}~\bibnamefont {Bohigas}}\ and\ \bibinfo {author} {\bibfnamefont {M.-J.}\ \bibnamefont {Giannoni}},\ }\bibfield  {title} {\bibinfo {title} {Chaotic motion and random matrix theories},\ }in\ \href {https://doi.org/10.1007/3-540-13392-5_1} {\emph {\bibinfo {booktitle} {Mathematical and {{Computational Methods}} in {{Nuclear Physics}}}}},\ \bibinfo {series and number} {Lecture {{Notes}} in {{Physics}}}\ (\bibinfo  {publisher} {Springer, Berlin, Heidelberg},\ \bibinfo {year} {1984})\ pp.\ \bibinfo {pages} {1--99}\BibitemShut {NoStop}%
\bibitem [{\citenamefont {Bohigas}\ \emph {et~al.}(1984)\citenamefont {Bohigas}, \citenamefont {Giannoni},\ and\ \citenamefont {Schmit}}]{bohigasCharacterizationChaoticQuantum1984}%
  \BibitemOpen
  \bibfield  {author} {\bibinfo {author} {\bibfnamefont {O.}~\bibnamefont {Bohigas}}, \bibinfo {author} {\bibfnamefont {M.~J.}\ \bibnamefont {Giannoni}},\ and\ \bibinfo {author} {\bibfnamefont {C.}~\bibnamefont {Schmit}},\ }\bibfield  {title} {\bibinfo {title} {Characterization of {{Chaotic Quantum Spectra}} and {{Universality}} of {{Level Fluctuation Laws}}},\ }\href {https://doi.org/10.1103/PhysRevLett.52.1} {\bibfield  {journal} {\bibinfo  {journal} {Phys. Rev. Lett.}\ }\textbf {\bibinfo {volume} {52}},\ \bibinfo {pages} {1} (\bibinfo {year} {1984})}\BibitemShut {NoStop}%
\bibitem [{\citenamefont {Berry}\ and\ \citenamefont {Robnik}(1984)}]{berrySemiclassicalLevelSpacings1984}%
  \BibitemOpen
  \bibfield  {author} {\bibinfo {author} {\bibfnamefont {M.~V.}\ \bibnamefont {Berry}}\ and\ \bibinfo {author} {\bibfnamefont {M.}~\bibnamefont {Robnik}},\ }\bibfield  {title} {\bibinfo {title} {Semiclassical level spacings when regular and chaotic orbits coexist},\ }\href {https://doi.org/10.1088/0305-4470/17/12/013} {\bibfield  {journal} {\bibinfo  {journal} {J. Phys. A: Math. Gen.}\ }\textbf {\bibinfo {volume} {17}},\ \bibinfo {pages} {2413} (\bibinfo {year} {1984})}\BibitemShut {NoStop}%
\bibitem [{\citenamefont {Berry}(1985)}]{berrySemiclassicalTheorySpectral1985}%
  \BibitemOpen
  \bibfield  {author} {\bibinfo {author} {\bibfnamefont {M.~V.}\ \bibnamefont {Berry}},\ }\bibfield  {title} {\bibinfo {title} {Semiclassical theory of spectral rigidity},\ }\href {https://doi.org/10.1098/rspa.1985.0078} {\bibfield  {journal} {\bibinfo  {journal} {Proc. R. Soc. Lond. A}\ }\textbf {\bibinfo {volume} {400}},\ \bibinfo {pages} {229} (\bibinfo {year} {1985})}\BibitemShut {NoStop}%
\bibitem [{\citenamefont {Berry}\ and\ \citenamefont {Robnik}(1986)}]{berryStatisticsEnergyLevels1986}%
  \BibitemOpen
  \bibfield  {author} {\bibinfo {author} {\bibfnamefont {M.~V.}\ \bibnamefont {Berry}}\ and\ \bibinfo {author} {\bibfnamefont {M.}~\bibnamefont {Robnik}},\ }\bibfield  {title} {\bibinfo {title} {Statistics of energy levels without time-reversal symmetry: {{Aharonov-Bohm}} chaotic billiards},\ }\href {https://doi.org/10.1088/0305-4470/19/5/019} {\bibfield  {journal} {\bibinfo  {journal} {J. Phys. A: Math. Gen.}\ }\textbf {\bibinfo {volume} {19}},\ \bibinfo {pages} {649} (\bibinfo {year} {1986})}\BibitemShut {NoStop}%
\bibitem [{\citenamefont {Kitaev}(2009)}]{kitaev2009PeriodicTableTopological}%
  \BibitemOpen
  \bibfield  {author} {\bibinfo {author} {\bibfnamefont {A.}~\bibnamefont {Kitaev}},\ }\bibfield  {title} {\bibinfo {title} {Periodic table for topological insulators and superconductors},\ }\href {https://doi.org/10.1063/1.3149495} {\bibfield  {journal} {\bibinfo  {journal} {AIP Conf. Proc.}\ }\textbf {\bibinfo {volume} {1134}},\ \bibinfo {pages} {22} (\bibinfo {year} {2009})}\BibitemShut {NoStop}%
\bibitem [{\citenamefont {Robnik}\ and\ \citenamefont {Berry}(1986)}]{robnikFalseTimereversalViolation1986}%
  \BibitemOpen
  \bibfield  {author} {\bibinfo {author} {\bibfnamefont {M.}~\bibnamefont {Robnik}}\ and\ \bibinfo {author} {\bibfnamefont {M.~V.}\ \bibnamefont {Berry}},\ }\bibfield  {title} {\bibinfo {title} {False time-reversal violation and energy level statistics: The role of anti-unitary symmetry},\ }\href {https://doi.org/10.1088/0305-4470/19/5/020} {\bibfield  {journal} {\bibinfo  {journal} {J. Phys. A: Math. Gen.}\ }\textbf {\bibinfo {volume} {19}},\ \bibinfo {pages} {669} (\bibinfo {year} {1986})}\BibitemShut {NoStop}%
\bibitem [{\citenamefont {Damski}\ and\ \citenamefont {Zakrzewski}(2006)}]{damskiMottinsulatorPhaseOnedimensional2006}%
  \BibitemOpen
  \bibfield  {author} {\bibinfo {author} {\bibfnamefont {B.}~\bibnamefont {Damski}}\ and\ \bibinfo {author} {\bibfnamefont {J.}~\bibnamefont {Zakrzewski}},\ }\bibfield  {title} {\bibinfo {title} {Mott-insulator phase of the one-dimensional {{Bose-Hubbard}} model: {{A}} high-order perturbative study},\ }\href {https://doi.org/10.1103/PhysRevA.74.043609} {\bibfield  {journal} {\bibinfo  {journal} {Phys. Rev. A}\ }\textbf {\bibinfo {volume} {74}},\ \bibinfo {pages} {043609} (\bibinfo {year} {2006})}\BibitemShut {NoStop}%
\bibitem [{\citenamefont {Grusdt}\ \emph {et~al.}(2013)\citenamefont {Grusdt}, \citenamefont {H{\"o}ning},\ and\ \citenamefont {Fleischhauer}}]{grusdtTopologicalEdgeStates2013}%
  \BibitemOpen
  \bibfield  {author} {\bibinfo {author} {\bibfnamefont {F.}~\bibnamefont {Grusdt}}, \bibinfo {author} {\bibfnamefont {M.}~\bibnamefont {H{\"o}ning}},\ and\ \bibinfo {author} {\bibfnamefont {M.}~\bibnamefont {Fleischhauer}},\ }\bibfield  {title} {\bibinfo {title} {Topological {{Edge States}} in the {{One-Dimensional Superlattice Bose-Hubbard Model}}},\ }\href {https://doi.org/10.1103/PhysRevLett.110.260405} {\bibfield  {journal} {\bibinfo  {journal} {Phys. Rev. Lett.}\ }\textbf {\bibinfo {volume} {110}},\ \bibinfo {pages} {260405} (\bibinfo {year} {2013})}\BibitemShut {NoStop}%
\bibitem [{\citenamefont {Yu}\ \emph {et~al.}(2017)\citenamefont {Yu}, \citenamefont {Sun},\ and\ \citenamefont {Zhai}}]{yuSymmetryProtectedDynamical2017}%
  \BibitemOpen
  \bibfield  {author} {\bibinfo {author} {\bibfnamefont {J.}~\bibnamefont {Yu}}, \bibinfo {author} {\bibfnamefont {N.}~\bibnamefont {Sun}},\ and\ \bibinfo {author} {\bibfnamefont {H.}~\bibnamefont {Zhai}},\ }\bibfield  {title} {\bibinfo {title} {Symmetry {{Protected Dynamical Symmetry}} in the {{Generalized Hubbard Models}}},\ }\href {https://doi.org/10.1103/PhysRevLett.119.225302} {\bibfield  {journal} {\bibinfo  {journal} {Phys. Rev. Lett.}\ }\textbf {\bibinfo {volume} {119}},\ \bibinfo {pages} {225302} (\bibinfo {year} {2017})}\BibitemShut {NoStop}%
\bibitem [{\citenamefont {Sutherland}(1986{\natexlab{a}})}]{sutherlandLocalizationElectronicWave1986}%
  \BibitemOpen
  \bibfield  {author} {\bibinfo {author} {\bibfnamefont {B.}~\bibnamefont {Sutherland}},\ }\bibfield  {title} {\bibinfo {title} {Localization of electronic wave functions due to local topology},\ }\href {https://doi.org/10.1103/PhysRevB.34.5208} {\bibfield  {journal} {\bibinfo  {journal} {Phys. Rev. B}\ }\textbf {\bibinfo {volume} {34}},\ \bibinfo {pages} {5208} (\bibinfo {year} {1986}{\natexlab{a}})}\BibitemShut {NoStop}%
\bibitem [{\citenamefont {Lieb}(1989)}]{liebTwoTheoremsHubbard1989}%
  \BibitemOpen
  \bibfield  {author} {\bibinfo {author} {\bibfnamefont {E.~H.}\ \bibnamefont {Lieb}},\ }\bibfield  {title} {\bibinfo {title} {Two theorems on the {{Hubbard}} model},\ }\href {https://doi.org/10.1103/PhysRevLett.62.1201} {\bibfield  {journal} {\bibinfo  {journal} {Phys. Rev. Lett.}\ }\textbf {\bibinfo {volume} {62}},\ \bibinfo {pages} {1201} (\bibinfo {year} {1989})}\BibitemShut {NoStop}%
\bibitem [{\citenamefont {Heilmann}\ and\ \citenamefont {Lieb}(1971)}]{heilmann1971ViolationNoncrossingRule}%
  \BibitemOpen
  \bibfield  {author} {\bibinfo {author} {\bibfnamefont {O.~J.}\ \bibnamefont {Heilmann}}\ and\ \bibinfo {author} {\bibfnamefont {E.~H.}\ \bibnamefont {Lieb}},\ }\bibfield  {title} {\bibinfo {title} {Violation of the {{Noncrossing Rule}}: {{The Hubbard Hamiltonian}} for {{Benzene}}},\ }\href {https://doi.org/10.1111/j.1749-6632.1971.tb34956.x} {\bibfield  {journal} {\bibinfo  {journal} {Ann. N. Y. Acad. Sci.}\ }\textbf {\bibinfo {volume} {172}},\ \bibinfo {pages} {584} (\bibinfo {year} {1971})}\BibitemShut {NoStop}%
\bibitem [{\citenamefont {Sriram~Shastry}(1988)}]{sriramshastryDecoratedStartriangleRelations1988}%
  \BibitemOpen
  \bibfield  {author} {\bibinfo {author} {\bibfnamefont {B.}~\bibnamefont {Sriram~Shastry}},\ }\bibfield  {title} {\bibinfo {title} {Decorated star-triangle relations and exact integrability of the one-dimensional {{Hubbard}} model},\ }\href {https://doi.org/10.1007/BF01022987} {\bibfield  {journal} {\bibinfo  {journal} {J. Stat. Phys.}\ }\textbf {\bibinfo {volume} {50}},\ \bibinfo {pages} {57} (\bibinfo {year} {1988})}\BibitemShut {NoStop}%
\bibitem [{\citenamefont {Grosse}(1989)}]{grosseSymmetryHubbardModel1989}%
  \BibitemOpen
  \bibfield  {author} {\bibinfo {author} {\bibfnamefont {H.}~\bibnamefont {Grosse}},\ }\bibfield  {title} {\bibinfo {title} {The symmetry of the {{Hubbard}} model},\ }\href {https://doi.org/10.1007/BF00401869} {\bibfield  {journal} {\bibinfo  {journal} {Lett. Math. Phys.}\ }\textbf {\bibinfo {volume} {18}},\ \bibinfo {pages} {151} (\bibinfo {year} {1989})}\BibitemShut {NoStop}%
\bibitem [{\citenamefont {Nicolau}\ \emph {et~al.}(2026)\citenamefont {Nicolau}, \citenamefont {Ljubotina},\ and\ \citenamefont {Serbyn}}]{nicolau2026FragmentationZeroModes}%
  \BibitemOpen
  \bibfield  {author} {\bibinfo {author} {\bibfnamefont {E.}~\bibnamefont {Nicolau}}, \bibinfo {author} {\bibfnamefont {M.}~\bibnamefont {Ljubotina}},\ and\ \bibinfo {author} {\bibfnamefont {M.}~\bibnamefont {Serbyn}},\ }\bibfield  {title} {\bibinfo {title} {Fragmentation, {{Zero Modes}}, and {{Collective Bound States}} in {{Constrained Models}}},\ }\href {https://doi.org/10.1103/sl79-1xgb} {\bibfield  {journal} {\bibinfo  {journal} {PRX Quantum}\ }\textbf {\bibinfo {volume} {7}},\ \bibinfo {pages} {010352} (\bibinfo {year} {2026})}\BibitemShut {NoStop}%
\bibitem [{\citenamefont {Von~Neumann}\ and\ \citenamefont {Wigner}(1929)}]{Neumann1929}%
  \BibitemOpen
  \bibfield  {author} {\bibinfo {author} {\bibfnamefont {J.}~\bibnamefont {Von~Neumann}}\ and\ \bibinfo {author} {\bibfnamefont {E.}~\bibnamefont {Wigner}},\ }\bibfield  {title} {\bibinfo {title} {On some peculiar discrete eigenvalues},\ }\href@noop {} {\bibfield  {journal} {\bibinfo  {journal} {Phys. Z}\ }\textbf {\bibinfo {volume} {30}},\ \bibinfo {pages} {465} (\bibinfo {year} {1929})}\BibitemShut {NoStop}%
\bibitem [{\citenamefont {Wigner}\ and\ \citenamefont {Von~Neumann}(1929)}]{Wigner1929}%
  \BibitemOpen
  \bibfield  {author} {\bibinfo {author} {\bibfnamefont {E.}~\bibnamefont {Wigner}}\ and\ \bibinfo {author} {\bibfnamefont {J.}~\bibnamefont {Von~Neumann}},\ }\bibfield  {title} {\bibinfo {title} {Uber das verhalten von eigenwerten bei adiabatischen prozessen},\ }\href@noop {} {\bibfield  {journal} {\bibinfo  {journal} {Phys. Zeit}\ }\textbf {\bibinfo {volume} {30}},\ \bibinfo {pages} {467} (\bibinfo {year} {1929})}\BibitemShut {NoStop}%
\bibitem [{\citenamefont {Stepanov}\ \emph {et~al.}(2008)\citenamefont {Stepanov}, \citenamefont {M\"uller},\ and\ \citenamefont {Stolze}}]{Stepanov2008}%
  \BibitemOpen
  \bibfield  {author} {\bibinfo {author} {\bibfnamefont {V.~V.}\ \bibnamefont {Stepanov}}, \bibinfo {author} {\bibfnamefont {G.}~\bibnamefont {M\"uller}},\ and\ \bibinfo {author} {\bibfnamefont {J.}~\bibnamefont {Stolze}},\ }\bibfield  {title} {\bibinfo {title} {Quantum integrability and nonintegrability in the spin-boson model},\ }\href {https://doi.org/10.1103/PhysRevE.77.066202} {\bibfield  {journal} {\bibinfo  {journal} {Phys. Rev. E}\ }\textbf {\bibinfo {volume} {77}},\ \bibinfo {pages} {066202} (\bibinfo {year} {2008})}\BibitemShut {NoStop}%
\bibitem [{\citenamefont {Braak}\ \emph {et~al.}(2014)\citenamefont {Braak}, \citenamefont {Zhang},\ and\ \citenamefont {Kollar}}]{braakIntegrabilityWeakDiffraction2014}%
  \BibitemOpen
  \bibfield  {author} {\bibinfo {author} {\bibfnamefont {D.}~\bibnamefont {Braak}}, \bibinfo {author} {\bibfnamefont {J.~M.}\ \bibnamefont {Zhang}},\ and\ \bibinfo {author} {\bibfnamefont {M.}~\bibnamefont {Kollar}},\ }\bibfield  {title} {\bibinfo {title} {Integrability and weak diffraction in a two-particle {B}ose-{H}ubbard model},\ }\href {https://doi.org/10.1088/1751-8113/47/46/465303} {\bibfield  {journal} {\bibinfo  {journal} {J. Phys. A: Math. Theor.}\ }\textbf {\bibinfo {volume} {47}},\ \bibinfo {pages} {465303} (\bibinfo {year} {2014})}\BibitemShut {NoStop}%
\bibitem [{\citenamefont {Zhang}\ \emph {et~al.}(2012)\citenamefont {Zhang}, \citenamefont {Braak},\ and\ \citenamefont {Kollar}}]{Kollar2012}%
  \BibitemOpen
  \bibfield  {author} {\bibinfo {author} {\bibfnamefont {J.~M.}\ \bibnamefont {Zhang}}, \bibinfo {author} {\bibfnamefont {D.}~\bibnamefont {Braak}},\ and\ \bibinfo {author} {\bibfnamefont {M.}~\bibnamefont {Kollar}},\ }\bibfield  {title} {\bibinfo {title} {Bound states in the continuum realized in the one-dimensional two-particle hubbard model with an impurity},\ }\href {https://doi.org/10.1103/PhysRevLett.109.116405} {\bibfield  {journal} {\bibinfo  {journal} {Phys. Rev. Lett.}\ }\textbf {\bibinfo {volume} {109}},\ \bibinfo {pages} {116405} (\bibinfo {year} {2012})}\BibitemShut {NoStop}%
\bibitem [{\citenamefont {Mehta}(1991)}]{mehta1991random}%
  \BibitemOpen
  \bibfield  {author} {\bibinfo {author} {\bibfnamefont {M.}~\bibnamefont {Mehta}},\ }\href {https://books.google.de/books?id=-sloQgAACAAJ} {\emph {\bibinfo {title} {Random Matrices}}}\ (\bibinfo  {publisher} {Academic Press},\ \bibinfo {year} {1991})\BibitemShut {NoStop}%
\bibitem [{\citenamefont {Guhr}\ \emph {et~al.}(1998)\citenamefont {Guhr}, \citenamefont {Müller–Groeling},\ and\ \citenamefont {Weidenmüller}}]{GUHR1998189}%
  \BibitemOpen
  \bibfield  {author} {\bibinfo {author} {\bibfnamefont {T.}~\bibnamefont {Guhr}}, \bibinfo {author} {\bibfnamefont {A.}~\bibnamefont {Müller–Groeling}},\ and\ \bibinfo {author} {\bibfnamefont {H.~A.}\ \bibnamefont {Weidenmüller}},\ }\bibfield  {title} {\bibinfo {title} {Random-matrix theories in quantum physics: common concepts},\ }\href {https://doi.org/https://doi.org/10.1016/S0370-1573(97)00088-4} {\bibfield  {journal} {\bibinfo  {journal} {Phys. Rep.}\ }\textbf {\bibinfo {volume} {299}},\ \bibinfo {pages} {189} (\bibinfo {year} {1998})}\BibitemShut {NoStop}%
\bibitem [{\citenamefont {Rabson}\ \emph {et~al.}(2004)\citenamefont {Rabson}, \citenamefont {Narozhny},\ and\ \citenamefont {Millis}}]{Rabson2004}%
  \BibitemOpen
  \bibfield  {author} {\bibinfo {author} {\bibfnamefont {D.~A.}\ \bibnamefont {Rabson}}, \bibinfo {author} {\bibfnamefont {B.~N.}\ \bibnamefont {Narozhny}},\ and\ \bibinfo {author} {\bibfnamefont {A.~J.}\ \bibnamefont {Millis}},\ }\bibfield  {title} {\bibinfo {title} {Crossover from poisson to wigner-dyson level statistics in spin chains with integrability breaking},\ }\href {https://doi.org/10.1103/PhysRevB.69.054403} {\bibfield  {journal} {\bibinfo  {journal} {Phys. Rev. B}\ }\textbf {\bibinfo {volume} {69}},\ \bibinfo {pages} {054403} (\bibinfo {year} {2004})}\BibitemShut {NoStop}%
\bibitem [{\citenamefont {Brody}\ \emph {et~al.}(1981)\citenamefont {Brody}, \citenamefont {Flores}, \citenamefont {French}, \citenamefont {Mello}, \citenamefont {Pandey},\ and\ \citenamefont {Wong}}]{brody1981randommatrix}%
  \BibitemOpen
  \bibfield  {author} {\bibinfo {author} {\bibfnamefont {T.~A.}\ \bibnamefont {Brody}}, \bibinfo {author} {\bibfnamefont {J.}~\bibnamefont {Flores}}, \bibinfo {author} {\bibfnamefont {J.~B.}\ \bibnamefont {French}}, \bibinfo {author} {\bibfnamefont {P.~A.}\ \bibnamefont {Mello}}, \bibinfo {author} {\bibfnamefont {A.}~\bibnamefont {Pandey}},\ and\ \bibinfo {author} {\bibfnamefont {S.~S.~M.}\ \bibnamefont {Wong}},\ }\bibfield  {title} {\bibinfo {title} {Random-matrix physics: spectrum and strength fluctuations},\ }\href {https://doi.org/10.1103/RevModPhys.53.385} {\bibfield  {journal} {\bibinfo  {journal} {Rev. Mod. Phys.}\ }\textbf {\bibinfo {volume} {53}},\ \bibinfo {pages} {385} (\bibinfo {year} {1981})}\BibitemShut {NoStop}%
\bibitem [{\citenamefont {Poilblanc}\ \emph {et~al.}(1993)\citenamefont {Poilblanc}, \citenamefont {Ziman}, \citenamefont {Bellissard}, \citenamefont {Mila},\ and\ \citenamefont {Montambaux}}]{poilblancPoissonVsGOE1993}%
  \BibitemOpen
  \bibfield  {author} {\bibinfo {author} {\bibfnamefont {D.}~\bibnamefont {Poilblanc}}, \bibinfo {author} {\bibfnamefont {T.}~\bibnamefont {Ziman}}, \bibinfo {author} {\bibfnamefont {J.}~\bibnamefont {Bellissard}}, \bibinfo {author} {\bibfnamefont {F.}~\bibnamefont {Mila}},\ and\ \bibinfo {author} {\bibfnamefont {G.}~\bibnamefont {Montambaux}},\ }\bibfield  {title} {\bibinfo {title} {Poisson vs. {{GOE Statistics}} in {{Integrable}} and {{Non-Integrable Quantum Hamiltonians}}},\ }\href {https://doi.org/10.1209/0295-5075/22/7/010} {\bibfield  {journal} {\bibinfo  {journal} {Europhys. Lett.}\ }\textbf {\bibinfo {volume} {22}},\ \bibinfo {pages} {537} (\bibinfo {year} {1993})}\BibitemShut {NoStop}%
\bibitem [{\citenamefont {Kieburg}(2026)}]{kieburg2026quantumchaoticsystemsrandommatrix}%
  \BibitemOpen
  \bibfield  {author} {\bibinfo {author} {\bibfnamefont {M.}~\bibnamefont {Kieburg}},\ }\href {https://arxiv.org/abs/2604.12141} {\bibinfo {title} {Quantum chaotic systems: a random-matrix approach}} (\bibinfo {year} {2026}),\ \Eprint {https://arxiv.org/abs/2604.12141} {arXiv:2604.12141 [quant-ph]} \BibitemShut {NoStop}%
\bibitem [{\citenamefont {Takahashi}\ and\ \citenamefont {Iida}(2001)}]{takahasi_energy-level_2001}%
  \BibitemOpen
  \bibfield  {author} {\bibinfo {author} {\bibfnamefont {K.}~\bibnamefont {Takahashi}}\ and\ \bibinfo {author} {\bibfnamefont {S.}~\bibnamefont {Iida}},\ }\bibfield  {title} {\bibinfo {title} {Energy-level correlations in chiral symmetric disordered systems: Corrections to the universal results},\ }\href {https://doi.org/10.1103/PhysRevB.63.214201} {\bibfield  {journal} {\bibinfo  {journal} {Phys. Rev. B}\ }\textbf {\bibinfo {volume} {63}},\ \bibinfo {pages} {214201} (\bibinfo {year} {2001})}\BibitemShut {NoStop}%
\bibitem [{\citenamefont {Evangelou}\ and\ \citenamefont {Katsanos}(2003)}]{Evangelou_spectral_2003}%
  \BibitemOpen
  \bibfield  {author} {\bibinfo {author} {\bibfnamefont {S.~N.}\ \bibnamefont {Evangelou}}\ and\ \bibinfo {author} {\bibfnamefont {D.~E.}\ \bibnamefont {Katsanos}},\ }\bibfield  {title} {\bibinfo {title} {Spectral statistics in chiral-orthogonal disordered systems},\ }\href {https://doi.org/10.1088/0305-4470/36/12/322} {\bibfield  {journal} {\bibinfo  {journal} {J. Phys. A: Math. Gen.}\ }\textbf {\bibinfo {volume} {36}},\ \bibinfo {pages} {3237} (\bibinfo {year} {2003})}\BibitemShut {NoStop}%
\bibitem [{\citenamefont {Rehemanjiang}\ \emph {et~al.}(2020)\citenamefont {Rehemanjiang}, \citenamefont {Richter}, \citenamefont {Kuhl},\ and\ \citenamefont {St\"ockmann}}]{rehemanjiang_microwave_2020}%
  \BibitemOpen
  \bibfield  {author} {\bibinfo {author} {\bibfnamefont {A.}~\bibnamefont {Rehemanjiang}}, \bibinfo {author} {\bibfnamefont {M.}~\bibnamefont {Richter}}, \bibinfo {author} {\bibfnamefont {U.}~\bibnamefont {Kuhl}},\ and\ \bibinfo {author} {\bibfnamefont {H.-J.}\ \bibnamefont {St\"ockmann}},\ }\bibfield  {title} {\bibinfo {title} {Microwave realization of the chiral orthogonal, unitary, and symplectic ensembles},\ }\href {https://doi.org/10.1103/PhysRevLett.124.116801} {\bibfield  {journal} {\bibinfo  {journal} {Phys. Rev. Lett.}\ }\textbf {\bibinfo {volume} {124}},\ \bibinfo {pages} {116801} (\bibinfo {year} {2020})}\BibitemShut {NoStop}%
\bibitem [{\citenamefont {Fogarty}\ \emph {et~al.}(2021)\citenamefont {Fogarty}, \citenamefont {Garc{\'{i}}a-March}, \citenamefont {Santos},\ and\ \citenamefont {Harshman}}]{Fogarty2021}%
  \BibitemOpen
  \bibfield  {author} {\bibinfo {author} {\bibfnamefont {T.}~\bibnamefont {Fogarty}}, \bibinfo {author} {\bibfnamefont {M.~{\'{A}}.}\ \bibnamefont {Garc{\'{i}}a-March}}, \bibinfo {author} {\bibfnamefont {L.~F.}\ \bibnamefont {Santos}},\ and\ \bibinfo {author} {\bibfnamefont {N.~L.}\ \bibnamefont {Harshman}},\ }\bibfield  {title} {\bibinfo {title} {Probing the edge between integrability and quantum chaos in interacting few-atom systems},\ }\href@noop {} {\bibfield  {journal} {\bibinfo  {journal} {{Quantum}}\ }\textbf {\bibinfo {volume} {5}},\ \bibinfo {pages} {486} (\bibinfo {year} {2021})}\BibitemShut {NoStop}%
\bibitem [{\citenamefont {Berry}\ and\ \citenamefont {Tabor}(1977)}]{Berry1977}%
  \BibitemOpen
  \bibfield  {author} {\bibinfo {author} {\bibfnamefont {M.~V.}\ \bibnamefont {Berry}}\ and\ \bibinfo {author} {\bibfnamefont {M.}~\bibnamefont {Tabor}},\ }\bibfield  {title} {\bibinfo {title} {Level clustering in the regular spectrum},\ }\href {https://doi.org/10.1103/PhysRevResearch.5.023128} {\bibfield  {journal} {\bibinfo  {journal} {Proc. R. Soc. Lond.}\ }\textbf {\bibinfo {volume} {356}},\ \bibinfo {pages} {375–394} (\bibinfo {year} {1977})}\BibitemShut {NoStop}%
\bibitem [{\citenamefont {Garcia-March}\ \emph {et~al.}(2018)\citenamefont {Garcia-March}, \citenamefont {Frank}, \citenamefont {Bonneau}, \citenamefont {Schmiedmayer}, \citenamefont {Lewenstein},\ and\ \citenamefont {Santos}}]{Garcia-March_2018}%
  \BibitemOpen
  \bibfield  {author} {\bibinfo {author} {\bibfnamefont {M.~A.}\ \bibnamefont {Garcia-March}}, \bibinfo {author} {\bibfnamefont {S.~v.}\ \bibnamefont {Frank}}, \bibinfo {author} {\bibfnamefont {M.}~\bibnamefont {Bonneau}}, \bibinfo {author} {\bibfnamefont {J.}~\bibnamefont {Schmiedmayer}}, \bibinfo {author} {\bibfnamefont {M.}~\bibnamefont {Lewenstein}},\ and\ \bibinfo {author} {\bibfnamefont {L.~F.}\ \bibnamefont {Santos}},\ }\bibfield  {title} {\bibinfo {title} {Relaxation, chaos, and thermalization in a three-mode model of a {B}ose–{E}instein condensate},\ }\href {https://doi.org/10.1088/1367-2630/aaed68} {\bibfield  {journal} {\bibinfo  {journal} {New J. Phys.}\ }\textbf {\bibinfo {volume} {20}},\ \bibinfo {pages} {113039} (\bibinfo {year} {2018})}\BibitemShut {NoStop}%
\bibitem [{\citenamefont {Pandey}\ and\ \citenamefont {Ramaswamy}(1991)}]{Pandey1991}%
  \BibitemOpen
  \bibfield  {author} {\bibinfo {author} {\bibfnamefont {A.}~\bibnamefont {Pandey}}\ and\ \bibinfo {author} {\bibfnamefont {R.}~\bibnamefont {Ramaswamy}},\ }\bibfield  {title} {\bibinfo {title} {Level spacings for harmonic-oscillator systems},\ }\href {https://doi.org/10.1103/PhysRevA.43.4237} {\bibfield  {journal} {\bibinfo  {journal} {Phys. Rev. A}\ }\textbf {\bibinfo {volume} {43}},\ \bibinfo {pages} {4237} (\bibinfo {year} {1991})}\BibitemShut {NoStop}%
\bibitem [{\citenamefont {Pecci}\ \emph {et~al.}(2023)\citenamefont {Pecci}, \citenamefont {Aupetit-Diallo}, \citenamefont {Albert}, \citenamefont {Vignolo},\ and\ \citenamefont {Minguzzi}}]{Minguzzi2023}%
  \BibitemOpen
  \bibfield  {author} {\bibinfo {author} {\bibfnamefont {G.}~\bibnamefont {Pecci}}, \bibinfo {author} {\bibfnamefont {G.}~\bibnamefont {Aupetit-Diallo}}, \bibinfo {author} {\bibfnamefont {M.}~\bibnamefont {Albert}}, \bibinfo {author} {\bibfnamefont {P.}~\bibnamefont {Vignolo}},\ and\ \bibinfo {author} {\bibfnamefont {A.}~\bibnamefont {Minguzzi}},\ }\bibfield  {title} {\bibinfo {title} {Persistent currents in a strongly interacting multicomponent {Bose} gas on a ring},\ }\href {https://doi.org/10.5802/crphys.157} {\bibfield  {journal} {\bibinfo  {journal} {C. R. Phys.. Physique}\ }\textbf {\bibinfo {volume} {24}},\ \bibinfo {pages} {87} (\bibinfo {year} {2023})}\BibitemShut {NoStop}%
\bibitem [{\citenamefont {Izergin}\ and\ \citenamefont {Korepin}(1982)}]{izergin1982PauliPrincipleOnedimensional}%
  \BibitemOpen
  \bibfield  {author} {\bibinfo {author} {\bibfnamefont {A.~G.}\ \bibnamefont {Izergin}}\ and\ \bibinfo {author} {\bibfnamefont {V.~E.}\ \bibnamefont {Korepin}},\ }\bibfield  {title} {\bibinfo {title} {Pauli principle for one-dimensional bosons and the algebraic bethe ansatz},\ }\href {https://doi.org/10.1007/BF00400323} {\bibfield  {journal} {\bibinfo  {journal} {Lett. Mat. Phys.}\ }\textbf {\bibinfo {volume} {6}},\ \bibinfo {pages} {283} (\bibinfo {year} {1982})}\BibitemShut {NoStop}%
\bibitem [{\citenamefont {Alba}\ \emph {et~al.}(2013)\citenamefont {Alba}, \citenamefont {Saha},\ and\ \citenamefont {Haque}}]{Alba_2013}%
  \BibitemOpen
  \bibfield  {author} {\bibinfo {author} {\bibfnamefont {V.}~\bibnamefont {Alba}}, \bibinfo {author} {\bibfnamefont {K.}~\bibnamefont {Saha}},\ and\ \bibinfo {author} {\bibfnamefont {M.}~\bibnamefont {Haque}},\ }\bibfield  {title} {\bibinfo {title} {Bethe ansatz description of edge-localization in the open-boundary xxz spin chain},\ }\href {https://doi.org/10.1088/1742-5468/2013/10/P10018} {\bibfield  {journal} {\bibinfo  {journal} {J. Stat. Mech.: Theory Exp.}\ }\textbf {\bibinfo {volume} {2013}}\bibinfo  {number} { (10)},\ \bibinfo {pages} {P10018}}\BibitemShut {NoStop}%
\bibitem [{\citenamefont {Batchelor}\ \emph {et~al.}(2005)\citenamefont {Batchelor}, \citenamefont {Guan}, \citenamefont {Oelkers},\ and\ \citenamefont {Lee}}]{batchelor20051DInteractingBose}%
  \BibitemOpen
\bibfield  {number} {  }\bibfield  {author} {\bibinfo {author} {\bibfnamefont {M.~T.}\ \bibnamefont {Batchelor}}, \bibinfo {author} {\bibfnamefont {X.~W.}\ \bibnamefont {Guan}}, \bibinfo {author} {\bibfnamefont {N.}~\bibnamefont {Oelkers}},\ and\ \bibinfo {author} {\bibfnamefont {C.}~\bibnamefont {Lee}},\ }\bibfield  {title} {\bibinfo {title} {The {{1D}} interacting {{Bose}} gas in a hard wall box},\ }\href {https://doi.org/10.1088/0305-4470/38/36/001} {\bibfield  {journal} {\bibinfo  {journal} {J. Phys. A: Math. Gen.}\ }\textbf {\bibinfo {volume} {38}},\ \bibinfo {pages} {7787} (\bibinfo {year} {2005})}\BibitemShut {NoStop}%
\bibitem [{\citenamefont {Zhang}\ \emph {et~al.}(2013{\natexlab{b}})\citenamefont {Zhang}, \citenamefont {Braak},\ and\ \citenamefont {Kollar}}]{zhang2013BoundStatesOnedimensional}%
  \BibitemOpen
  \bibfield  {author} {\bibinfo {author} {\bibfnamefont {J.~M.}\ \bibnamefont {Zhang}}, \bibinfo {author} {\bibfnamefont {D.}~\bibnamefont {Braak}},\ and\ \bibinfo {author} {\bibfnamefont {M.}~\bibnamefont {Kollar}},\ }\bibfield  {title} {\bibinfo {title} {Bound states in the one-dimensional two-particle {{Hubbard}} model with an impurity},\ }\href {https://doi.org/10.1103/PhysRevA.87.023613} {\bibfield  {journal} {\bibinfo  {journal} {Phys. Rev. A}\ }\textbf {\bibinfo {volume} {87}},\ \bibinfo {pages} {023613} (\bibinfo {year} {2013}{\natexlab{b}})}\BibitemShut {NoStop}%
\bibitem [{\citenamefont {Zhang}\ \emph {et~al.}(2023{\natexlab{b}})\citenamefont {Zhang}, \citenamefont {Qian}, \citenamefont {Sun},\ and\ \citenamefont {Zhang}}]{zhang2023AnyonicBoundStates}%
  \BibitemOpen
  \bibfield  {author} {\bibinfo {author} {\bibfnamefont {W.}~\bibnamefont {Zhang}}, \bibinfo {author} {\bibfnamefont {L.}~\bibnamefont {Qian}}, \bibinfo {author} {\bibfnamefont {H.}~\bibnamefont {Sun}},\ and\ \bibinfo {author} {\bibfnamefont {X.}~\bibnamefont {Zhang}},\ }\bibfield  {title} {\bibinfo {title} {Anyonic bound states in the continuum},\ }\href {https://doi.org/10.1038/s42005-023-01245-6} {\bibfield  {journal} {\bibinfo  {journal} {Commun. Phys.}\ }\textbf {\bibinfo {volume} {6}},\ \bibinfo {pages} {1} (\bibinfo {year} {2023}{\natexlab{b}})}\BibitemShut {NoStop}%
\bibitem [{\citenamefont {Valiente}\ and\ \citenamefont {Petrosyan}(2008{\natexlab{b}})}]{valiente2008TwoparticleStatesHubbard}%
  \BibitemOpen
  \bibfield  {author} {\bibinfo {author} {\bibfnamefont {M.}~\bibnamefont {Valiente}}\ and\ \bibinfo {author} {\bibfnamefont {D.}~\bibnamefont {Petrosyan}},\ }\bibfield  {title} {\bibinfo {title} {Two-particle states in the {{Hubbard}} model},\ }\href {https://doi.org/10.1088/0953-4075/41/16/161002} {\bibfield  {journal} {\bibinfo  {journal} {J. Phys. B: At. Mol. Opt. Phys.}\ }\textbf {\bibinfo {volume} {41}},\ \bibinfo {pages} {161002} (\bibinfo {year} {2008}{\natexlab{b}})}\BibitemShut {NoStop}%
\bibitem [{\citenamefont {Ma}\ \emph {et~al.}(2010)\citenamefont {Ma}, \citenamefont {Chen}, \citenamefont {Fan},\ and\ \citenamefont {Liu}}]{ma2010AbelianNonAbelianQuantum}%
  \BibitemOpen
  \bibfield  {author} {\bibinfo {author} {\bibfnamefont {Y.-Q.}\ \bibnamefont {Ma}}, \bibinfo {author} {\bibfnamefont {S.}~\bibnamefont {Chen}}, \bibinfo {author} {\bibfnamefont {H.}~\bibnamefont {Fan}},\ and\ \bibinfo {author} {\bibfnamefont {W.-M.}\ \bibnamefont {Liu}},\ }\bibfield  {title} {\bibinfo {title} {Abelian and non-{{Abelian}} quantum geometric tensor},\ }\href {https://doi.org/10.1103/PhysRevB.81.245129} {\bibfield  {journal} {\bibinfo  {journal} {Phys. Rev. B}\ }\textbf {\bibinfo {volume} {81}},\ \bibinfo {pages} {245129} (\bibinfo {year} {2010})}\BibitemShut {NoStop}%
\bibitem [{\citenamefont {Moudgalya}\ \emph {et~al.}(2022)\citenamefont {Moudgalya}, \citenamefont {Bernevig},\ and\ \citenamefont {Regnault}}]{Moudgalya_2022}%
  \BibitemOpen
  \bibfield  {author} {\bibinfo {author} {\bibfnamefont {S.}~\bibnamefont {Moudgalya}}, \bibinfo {author} {\bibfnamefont {B.~A.}\ \bibnamefont {Bernevig}},\ and\ \bibinfo {author} {\bibfnamefont {N.}~\bibnamefont {Regnault}},\ }\bibfield  {title} {\bibinfo {title} {Quantum many-body scars and hilbert space fragmentation: a review of exact results},\ }\href {https://doi.org/10.1088/1361-6633/ac73a0} {\bibfield  {journal} {\bibinfo  {journal} {Rep. Prog. Phys.}\ }\textbf {\bibinfo {volume} {85}},\ \bibinfo {pages} {086501} (\bibinfo {year} {2022})}\BibitemShut {NoStop}%
\bibitem [{\citenamefont {Brighi}\ \emph {et~al.}(2023)\citenamefont {Brighi}, \citenamefont {Ljubotina},\ and\ \citenamefont {Serbyn}}]{Serbyn2023}%
  \BibitemOpen
  \bibfield  {author} {\bibinfo {author} {\bibfnamefont {P.}~\bibnamefont {Brighi}}, \bibinfo {author} {\bibfnamefont {M.}~\bibnamefont {Ljubotina}},\ and\ \bibinfo {author} {\bibfnamefont {M.}~\bibnamefont {Serbyn}},\ }\bibfield  {title} {\bibinfo {title} {{Hilbert space fragmentation and slow dynamics in particle-conserving quantum East models}},\ }\href {https://doi.org/10.21468/SciPostPhys.15.3.093} {\bibfield  {journal} {\bibinfo  {journal} {SciPost Phys.}\ }\textbf {\bibinfo {volume} {15}},\ \bibinfo {pages} {093} (\bibinfo {year} {2023})}\BibitemShut {NoStop}%
\bibitem [{\citenamefont {Kaneko}\ \emph {et~al.}(2024)\citenamefont {Kaneko}, \citenamefont {Kunimi},\ and\ \citenamefont {Danshita}}]{kaneko2024QuantumManybodyScars}%
  \BibitemOpen
  \bibfield  {author} {\bibinfo {author} {\bibfnamefont {R.}~\bibnamefont {Kaneko}}, \bibinfo {author} {\bibfnamefont {M.}~\bibnamefont {Kunimi}},\ and\ \bibinfo {author} {\bibfnamefont {I.}~\bibnamefont {Danshita}},\ }\bibfield  {title} {\bibinfo {title} {Quantum many-body scars in the {B}ose-{H}ubbard model with a three-body constraint},\ }\href {https://doi.org/10.1103/PhysRevA.109.L011301} {\bibfield  {journal} {\bibinfo  {journal} {Phys. Rev. A}\ }\textbf {\bibinfo {volume} {109}},\ \bibinfo {pages} {L011301} (\bibinfo {year} {2024})}\BibitemShut {NoStop}%
\bibitem [{\citenamefont {Ronzheimer}\ \emph {et~al.}(2013)\citenamefont {Ronzheimer}, \citenamefont {Schreiber}, \citenamefont {Braun}, \citenamefont {Hodgman}, \citenamefont {Langer}, \citenamefont {McCulloch}, \citenamefont {Heidrich-Meisner}, \citenamefont {Bloch},\ and\ \citenamefont {Schneider}}]{Bloch2013}%
  \BibitemOpen
  \bibfield  {author} {\bibinfo {author} {\bibfnamefont {J.~P.}\ \bibnamefont {Ronzheimer}}, \bibinfo {author} {\bibfnamefont {M.}~\bibnamefont {Schreiber}}, \bibinfo {author} {\bibfnamefont {S.}~\bibnamefont {Braun}}, \bibinfo {author} {\bibfnamefont {S.~S.}\ \bibnamefont {Hodgman}}, \bibinfo {author} {\bibfnamefont {S.}~\bibnamefont {Langer}}, \bibinfo {author} {\bibfnamefont {I.~P.}\ \bibnamefont {McCulloch}}, \bibinfo {author} {\bibfnamefont {F.}~\bibnamefont {Heidrich-Meisner}}, \bibinfo {author} {\bibfnamefont {I.}~\bibnamefont {Bloch}},\ and\ \bibinfo {author} {\bibfnamefont {U.}~\bibnamefont {Schneider}},\ }\bibfield  {title} {\bibinfo {title} {Expansion dynamics of interacting bosons in homogeneous lattices in one and two dimensions},\ }\href {https://doi.org/10.1103/PhysRevLett.110.205301} {\bibfield  {journal} {\bibinfo  {journal} {Phys. Rev. Lett.}\ }\textbf {\bibinfo {volume} {110}},\ \bibinfo {pages} {205301} (\bibinfo {year} {2013})}\BibitemShut {NoStop}%
\bibitem [{\citenamefont {Liu}\ \emph {et~al.}(2018)\citenamefont {Liu}, \citenamefont {Garrison}, \citenamefont {Deng}, \citenamefont {Gong},\ and\ \citenamefont {Gorshkov}}]{liuAsymmetricParticleTransport2018}%
  \BibitemOpen
  \bibfield  {author} {\bibinfo {author} {\bibfnamefont {F.}~\bibnamefont {Liu}}, \bibinfo {author} {\bibfnamefont {J.~R.}\ \bibnamefont {Garrison}}, \bibinfo {author} {\bibfnamefont {D.-L.}\ \bibnamefont {Deng}}, \bibinfo {author} {\bibfnamefont {Z.-X.}\ \bibnamefont {Gong}},\ and\ \bibinfo {author} {\bibfnamefont {A.~V.}\ \bibnamefont {Gorshkov}},\ }\bibfield  {title} {\bibinfo {title} {Asymmetric {{Particle Transport}} and {{Light-Cone Dynamics Induced}} by {{Anyonic Statistics}}},\ }\href {https://doi.org/10.1103/PhysRevLett.121.250404} {\bibfield  {journal} {\bibinfo  {journal} {Phys. Rev. Lett.}\ }\textbf {\bibinfo {volume} {121}},\ \bibinfo {pages} {250404} (\bibinfo {year} {2018})}\BibitemShut {NoStop}%
\bibitem [{\citenamefont {Lin}\ \emph {et~al.}(2023)\citenamefont {Lin}, \citenamefont {Ke},\ and\ \citenamefont {Lee}}]{linTopologicalInvariantsInteracting2023}%
  \BibitemOpen
  \bibfield  {author} {\bibinfo {author} {\bibfnamefont {L.}~\bibnamefont {Lin}}, \bibinfo {author} {\bibfnamefont {Y.}~\bibnamefont {Ke}},\ and\ \bibinfo {author} {\bibfnamefont {C.}~\bibnamefont {Lee}},\ }\bibfield  {title} {\bibinfo {title} {Topological invariants for interacting systems: {{From}} twisted boundary conditions to center-of-mass momentum},\ }\href {https://doi.org/10.1103/PhysRevB.107.125161} {\bibfield  {journal} {\bibinfo  {journal} {Phys. Rev. B}\ }\textbf {\bibinfo {volume} {107}},\ \bibinfo {pages} {125161} (\bibinfo {year} {2023})}\BibitemShut {NoStop}%
\bibitem [{\citenamefont {Reiner}\ \emph {et~al.}(2004)\citenamefont {Reiner}, \citenamefont {Stanton},\ and\ \citenamefont {White}}]{REINER200417}%
  \BibitemOpen
  \bibfield  {author} {\bibinfo {author} {\bibfnamefont {V.}~\bibnamefont {Reiner}}, \bibinfo {author} {\bibfnamefont {D.}~\bibnamefont {Stanton}},\ and\ \bibinfo {author} {\bibfnamefont {D.}~\bibnamefont {White}},\ }\bibfield  {title} {\bibinfo {title} {The cyclic sieving phenomenon},\ }\href {https://doi.org/https://doi.org/10.1016/j.jcta.2004.04.009} {\bibfield  {journal} {\bibinfo  {journal} {J. Comb. Theory Ser. Ay, Series A}\ }\textbf {\bibinfo {volume} {108}},\ \bibinfo {pages} {17} (\bibinfo {year} {2004})}\BibitemShut {NoStop}%
\bibitem [{\citenamefont {Sutherland}(1986{\natexlab{b}})}]{Sutherland1986}%
  \BibitemOpen
  \bibfield  {author} {\bibinfo {author} {\bibfnamefont {B.}~\bibnamefont {Sutherland}},\ }\bibfield  {title} {\bibinfo {title} {Localization of electronic wave functions due to local topology},\ }\href {https://doi.org/10.1103/PhysRevB.34.5208} {\bibfield  {journal} {\bibinfo  {journal} {Phys. Rev. B}\ }\textbf {\bibinfo {volume} {34}},\ \bibinfo {pages} {5208} (\bibinfo {year} {1986}{\natexlab{b}})}\BibitemShut {NoStop}%
\bibitem [{\citenamefont {Maimaiti}\ \emph {et~al.}(2017)\citenamefont {Maimaiti}, \citenamefont {Andreanov}, \citenamefont {Park}, \citenamefont {Gendelman},\ and\ \citenamefont {Flach}}]{Flach2017}%
  \BibitemOpen
  \bibfield  {author} {\bibinfo {author} {\bibfnamefont {W.}~\bibnamefont {Maimaiti}}, \bibinfo {author} {\bibfnamefont {A.}~\bibnamefont {Andreanov}}, \bibinfo {author} {\bibfnamefont {H.~C.}\ \bibnamefont {Park}}, \bibinfo {author} {\bibfnamefont {O.}~\bibnamefont {Gendelman}},\ and\ \bibinfo {author} {\bibfnamefont {S.}~\bibnamefont {Flach}},\ }\bibfield  {title} {\bibinfo {title} {Compact localized states and flat-band generators in one dimension},\ }\href {https://doi.org/10.1103/PhysRevB.95.115135} {\bibfield  {journal} {\bibinfo  {journal} {Phys. Rev. B}\ }\textbf {\bibinfo {volume} {95}},\ \bibinfo {pages} {115135} (\bibinfo {year} {2017})}\BibitemShut {NoStop}%
\bibitem [{\citenamefont {Chen}\ \emph {et~al.}(2020)\citenamefont {Chen}, \citenamefont {Szyniszewski},\ and\ \citenamefont {Schomerus}}]{Schomerus2020}%
  \BibitemOpen
  \bibfield  {author} {\bibinfo {author} {\bibfnamefont {C.~P.}\ \bibnamefont {Chen}}, \bibinfo {author} {\bibfnamefont {M.}~\bibnamefont {Szyniszewski}},\ and\ \bibinfo {author} {\bibfnamefont {H.}~\bibnamefont {Schomerus}},\ }\bibfield  {title} {\bibinfo {title} {Many-body localization of zero modes},\ }\href {https://doi.org/10.1103/PhysRevResearch.2.023118} {\bibfield  {journal} {\bibinfo  {journal} {Phys. Rev. Res.}\ }\textbf {\bibinfo {volume} {2}},\ \bibinfo {pages} {023118} (\bibinfo {year} {2020})}\BibitemShut {NoStop}%
\end{thebibliography}
\end{document}